\documentclass[review]{elsarticle}

\usepackage[colorlinks,citecolor=blue,linktoc=all,linkcolor=cyan]{hyperref}
\usepackage{graphicx}

\usepackage[T1]{fontenc}
\usepackage{mathrsfs}             
\usepackage{slashed}              
\usepackage{amsmath}
\usepackage{amssymb}
\usepackage{amsbsy}
\usepackage{amsfonts}

\numberwithin{equation}{section}
\numberwithin{table}{section}
\numberwithin{figure}{section}

\journal{Progress in Particle and Nuclear Physics}

\usepackage{titlesec}
\usepackage{sectsty}
\titleformat{\section}{\normalfont\Large\bfseries}{\thesection}{1em}{}
\titleformat{\subsection}{\normalfont\large\bfseries}{\thesubsection}{1em}{}
\titleformat{\subsubsection}{\normalfont\normalsize\bfseries}{\thesubsubsection}{1em}{}

\newcommand{\ket}[1]{\left|#1\right>}

\newcommand{\f}[1]{\mbox{\boldmath$#1$}}
\newcommand{\fk}[1]{\mbox{\boldmath$\scriptstyle#1$}}
\newcommand{\vau}{\mbox{\boldmath$v$}}
\newcommand{\na}{\mbox{\boldmath$\nabla$}}
\newcommand{\bea}{\begin{eqnarray}}
\newcommand{\ea}{\end{eqnarray}}
\newcommand{\eea}{\end{eqnarray}}
\newcommand{\ord}{\,{\cal O}}

\newcommand{\rs}[1]{{#1}}

\begin{document}

\begin{frontmatter}

\title{Ultra-cold atoms as quantum simulators for relativistic phenomena}

\author[mymainaddress,mysecondaryaddress]{Ralf Sch\"utzhold\corref{mycorrespondingauthor}}
\cortext[mycorrespondingauthor]{Corresponding author}
\ead{r.schuetzhold@hzdr.de}

\address[mymainaddress]
{Helmholtz-Zentrum Dresden-Rossendorf, Bautzner Landstra{\ss}e 400, 01328 Dresden, Germany}
\address[mysecondaryaddress]
{Institut f\"ur Theoretische Physik, Technische Universit\"at Dresden, 01062 Dresden, Germany}

\begin{abstract}
The goal of this article is to review developments regarding the use of ultra-cold atoms as 
quantum simulators.
Special emphasis is placed on relativistic quantum phenomena, which are presumably most 
interesting for the audience of this journal. 
After a brief introduction into the main idea of quantum simulators and the basic physics of 
ultra-cold atoms, relativistic quantum phenomena of linear fields are discussed, including 
Hawking radiation, the Unruh effect, cosmological particle creation, the Gibbons-Hawking and 
Ginzburg effects, super-radiance, Sauter-Schwinger and Breit-Wheeler pair creation,
as well as the dynamical Casimir effect. 
After that, the focus is shifted to phenomena of non-linear fields, such as the sine-Gordon 
model, the Kibble-Zurek mechanism, false-vacuum decay, and quantum back-reaction.  
%
In order to place everything into proper context,
the basic underlying mechanisms of these phenomena are briefly recapitulated
before their simulators are discussed. 
Even though effort is made to provide a review as fair as possible, there can be co claim 
of completeness and the selection as well as the relative weights of the topics may well 
reflect \rs{the personal view and taste of the author.} 
\end{abstract}


\end{frontmatter}

\newpage

\thispagestyle{empty}
\tableofcontents


\newpage

\section{Introduction}\label{Introduction}


Let us start by setting the stage. 
Unless explicitly noted for illustrative purposes, natural units with 
\bea
\hbar=c=\varepsilon_0=\mu_0=1 
\ea
will be used throughout. 
Note that the elementary charge $q$ could then be expressed in terms of a dimensionless number 
via the QED fine structure constant 
$\alpha_{\rm QED}=q^2/(4\pi\varepsilon_0\hbar c)\approx1/137$ but we shall not do this here and 
keep the charge $q$ explicitly. 
Furthermore, since we are dealing with ultra-cold atoms instead of 
ions\footnote{This would open up a completely new field  \cite{Blatt:2012chk}, 
see, e.g., 
\cite{Alsing:2005dno,Lamata:2007qq,Schutzhold:2007mx,Horstmann:2009yh,Menicucci:2010xs,
Horstmann:2010xd,Wittemer:2019agm,Joshi:2023rvd} 
for a few examples out of many more.}, 
for example, the real charge $q$ will typically be replaced by an effective charge $q_{\rm eff}$ 
for the quantum simulators. 
Similarly, the real speed of light $c$ will also be replaced by an effective speed of light 
$c_{\rm eff}$, typically the propagation velocity of the relevant quasi-particles, in most of the 
quantum simulators. 
The fact that this effective speed of light $c_{\rm eff}$ is often much smaller than $c$
renders the experimental realization of the quantum phenomena under consideration typically 
far easier. 

As another interesting option, one could consider replacing the Planck constant 
$\hbar$ by an effective value $\hbar_{\rm eff}$.
This could correspond to modeling 
the quantum fluctuations (governed by $\hbar$)
by classical fluctuations, for example, such that $\hbar_{\rm eff}$ is bigger than $\hbar$.
However, such a replacement would probably be no longer considered a quantum simulator and thus 
we do not discuss this option here. 

Four vectors such as $k_\mu$ are defined with respect to the Minkoswki metric $g_{\mu\nu}^{\rm Min}$ 
in flat space-time (or a general metric $g_{\mu\nu}$ in curved space-time) for which we use the  
signature in the ``particle-physics convention'' $(+1,-1,-1,-1)$.
The cases where we go to Euclidean space will be mentioned explicitly. 

\subsection{Quantum simulators}\label{Quantum simulators}

Before turning to ultra-cold atoms, it is probably helpful to specify the term ``quantum simulators'' 
and the main ideas behind it, see, e.g., 
\cite{Altman:2019vbv,Buluta:2009vjg,Lewenstein:2012afn,Georgescu:2013oza}. 
To this end, let us take a small detour into the field of quantum computation, see, e.g., 
\cite{Nielsen:2012yss}, 
where we start with the standard scenario of a sequential quantum algorithm, see also 
Fig.~\ref{fig:sequential}.

\subsubsection{Sequential quantum algorithms}\label{Sequential quantum algorithms}

\begin{figure}[tbp]
\centering
\includegraphics[width=.7\textwidth]{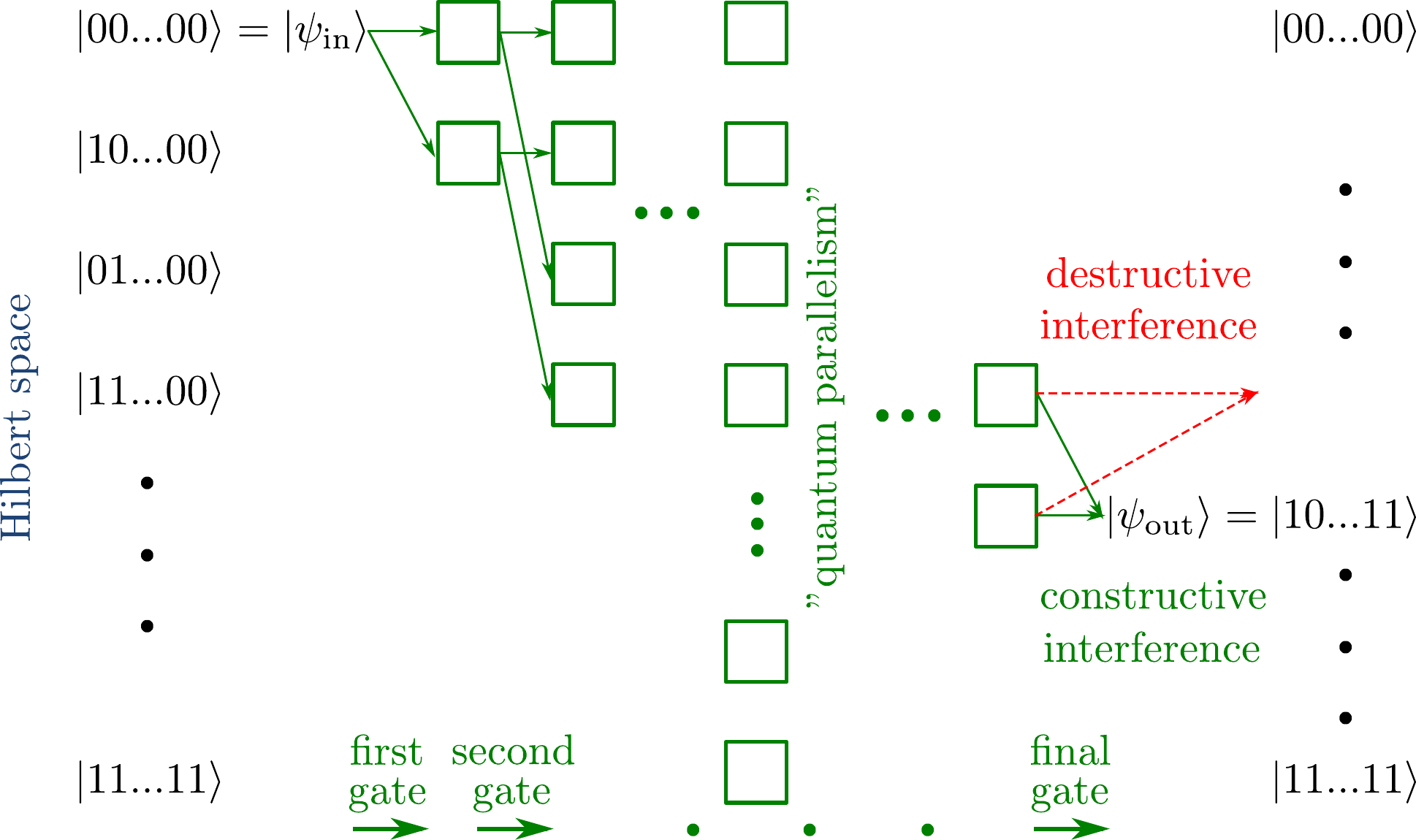}
\caption{Sketch of a sequential quantum algorithm.}
\label{fig:sequential}
\end{figure}

Without loss of generality, we may assume that the initial state $\ket{\psi_{\rm in}}$
is one of the computational basis states, say $\ket{\psi_{\rm in}}=\ket{00\dots00}$. 
Then, in order to execute the quantum algorithm, we apply the a series of gate operations to the
state, where the first gate could be a Hadamard gate ${\mathfrak H}_1$ acting on the first qubit
\bea
\ket{\psi_{\rm in}}=\ket{00\dots00}\,\to\, 
{\mathfrak H}_1\ket{00\dots00}=\frac{\ket{00\dots00}+\ket{10\dots00}}{\sqrt{2}}
\,.
\ea
Then we apply the second gate (for example, a Hadamard gate ${\mathfrak H}_2$
acting on the second qubit) and so on, see Fig.~\ref{fig:sequential}.
After applying a number of gates comparable to the number $n$ of qubits, the quantum state 
is turned into a superposition of a huge number (possibly even all $N=2^n$) 
of computational basis states (``quantum parallelism''). 
If the quantum state would only contain a small number of computational basis states,
one could simulate the quantum algorithm efficiently on a classical computer, i.e., 
there would be no real quantum advantage. 

Towards the end of the quantum algorithm, however, this superposition of a huge number 
of computational basis states should be focused into a smaller set of possible solutions states. 
For example, in the case of the Grover search algorithm \cite{Grover:1997fa}
with only one possible solution,
the final state $\ket{\psi_{\rm out}}$ would be (to a very good approximation) one of the 
computational basis states, such as $\ket{\psi_{\rm out}}=\ket{10\dots11}$. 
The way this works is that we have constructive interference for the paths leading to this 
desired final state $\ket{\psi_{\rm out}}$ whereas the paths leading to all other 
(unwanted) output states interfere destructively and thus their amplitudes are suppressed,
as indicated in Fig.~\ref{fig:sequential}.

Obviously, the interference necessary for such a quantum algorithm to work is prone to 
decoherence and other imperfections. 
Especially the relative phase between contributions from computational basis states
which differ in many qubits is quite vulnerable since the environment could induce a 
small phase shift (or other error) at each qubit. 
Furthermore, the quantum algorithm contains a significant number of quantum gates 
(typically scaling polynomially with the number $n$ of qubits) such that the errors 
can accumulate. 
Thus, making such a quantum algorithm work for a significant number of qubits 
(such that it can solve problems which are intractable on a classical computer) 
would require significant error correction -- which in turn, necessitates very low 
decoherence rates for the gates and qubits. 
This is probably the main obstacle for actually realizing such quantum computers.  

\subsubsection{Quantum ground-state algorithms}\label{Quantum ground-state algorithms}

One idea to avoid or at least reduce the problem of decoherence mentioned above could be to 
replace the scheme of sequential quantum algorithms by another approach where one constructs 
a Hamiltonian $\hat H_{\rm problem}$ such that the ground state of this Hamiltonian 
$\hat H_{\rm problem}$ encodes the solution to the problem one is interested in. 
This ground state should then be quite robust to decoherence -- especially if there is 
a sizable gap to the excited state(s).
Actually, constructing such a Hamiltonian $\hat H_{\rm problem}$ is possible for the problems 
(e.g., factoring large numbers \cite{Shor:2006sdo}) 
usually discussed in connection with quantum computing 
-- and even for some more problems (e.g., the traveling salesman problem) for which no 
sequential quantum algorithm offering a strong speed-up is known. 

However, even after constructing such a Hamiltonian $\hat H_{\rm problem}$, there is still the 
task of actually finding the ground state.
For problem Hamiltonians $\hat H_{\rm problem}$ which are supposed to answer interesting questions, 
the potential landscape typically contains many local minima (like a glassy system) and thus 
going to the global minimum is difficult since the system is likely to get stuck in one of the 
local minima. 
Thus, just starting in some initial state $\ket{\psi_{\rm in}}$ and cooling the system does 
probably not work within a reasonable time. 
One idea to go down in energy is to use annealing techniques where thermal or quantum fluctuations 
are used to overcome the potential barriers and thus to ``rescue'' the system from a local minimum,
see, e.g., \cite{Das:2008zzd}. 

\begin{figure}[tbp]
\centering
\includegraphics[width=.7\textwidth]{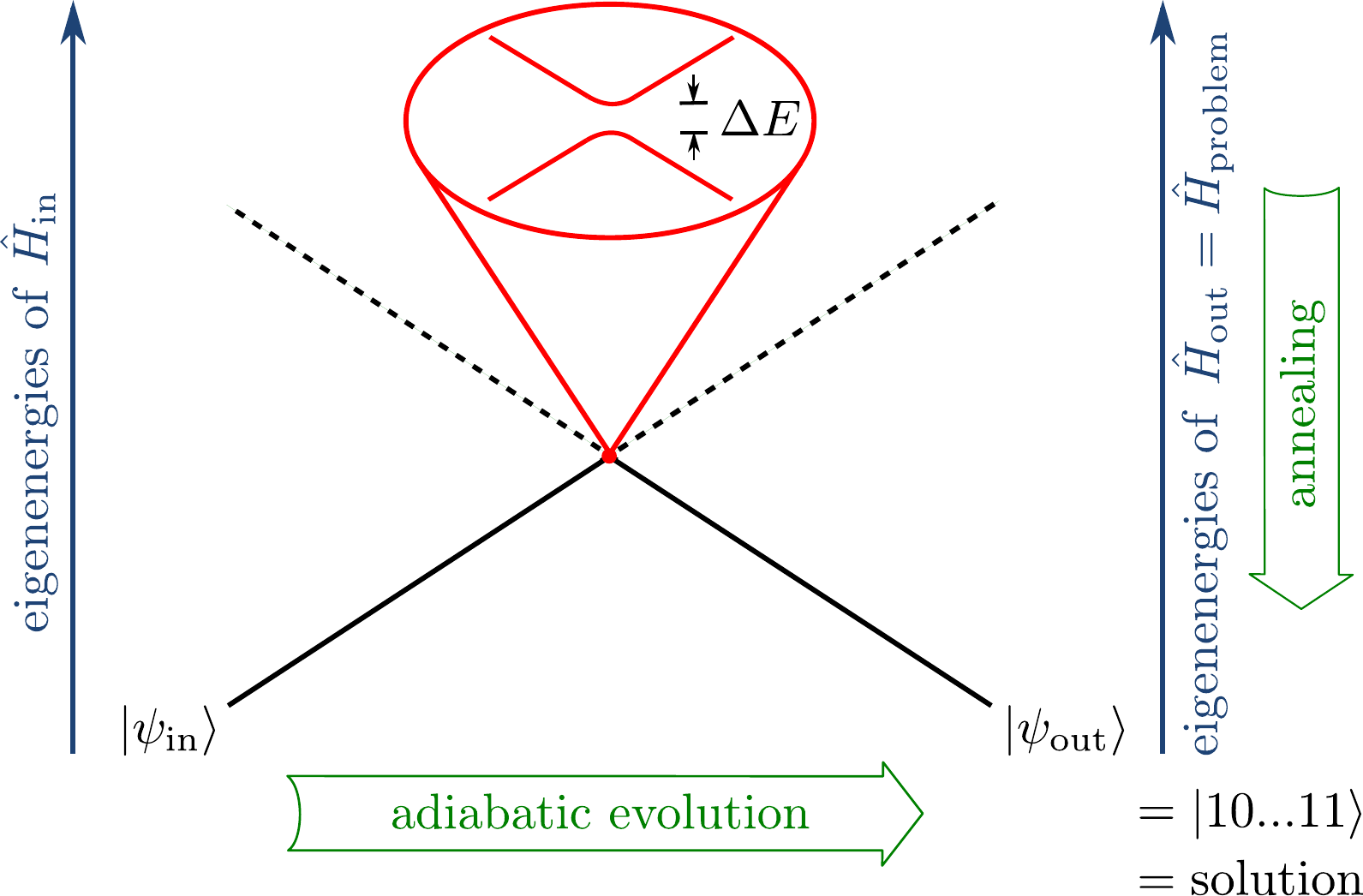}
\caption{Sketch of a quantum ground-state (e.g., adiabatic) algorithm.}
\label{fig:adiabatic}
\end{figure}

As another idea, one could start with an initial Hamiltonian $\hat H_{\rm in}$ which can easily 
be realized experimentally and whose ground state $\ket{\psi_{\rm in}}$ is known, has a 
sufficiently large energy gap to the excited states, and can be prepared experimentally. 
Then the initial Hamiltonian $\hat H_{\rm in}$ is slowly deformed into the final problem 
Hamiltonian $\hat H_{\rm problem}$, see, e.g., \cite{Farhi:2000ikn,Farhi:2001ltl}. 
The most simple way would be a linear interpolation scheme with a run-time $T$ 
\bea
\label{run-time}
\hat H(t)=\frac{T-t}{T}\,\hat H_{\rm in} + \frac{t}{T}\,\hat H_{\rm problem}
\,,
\ea
but more involved schemes are conceivable as well, see, e.g., \cite{Roland:2002huo}. 
If this deformation is performed slowly enough (i.e., $T$ is large enough), 
the adiabatic theorem demands that we indeed end up in the desired ground state of 
$\hat H_{\rm problem}$.
Thus, this scheme is referred to as adiabatic quantum computing. 
How slow is slow enough is then mainly determined by the energy gap $\Delta E(t)$
as each instant of time $t$ between the instantaneous ground state of $\hat H(t)$
and the first excited state(s), see, e.g.,  \cite{Schaller+Mostame}. 

\begin{figure}[tbp]
\centering
\includegraphics[width=.7\textwidth]{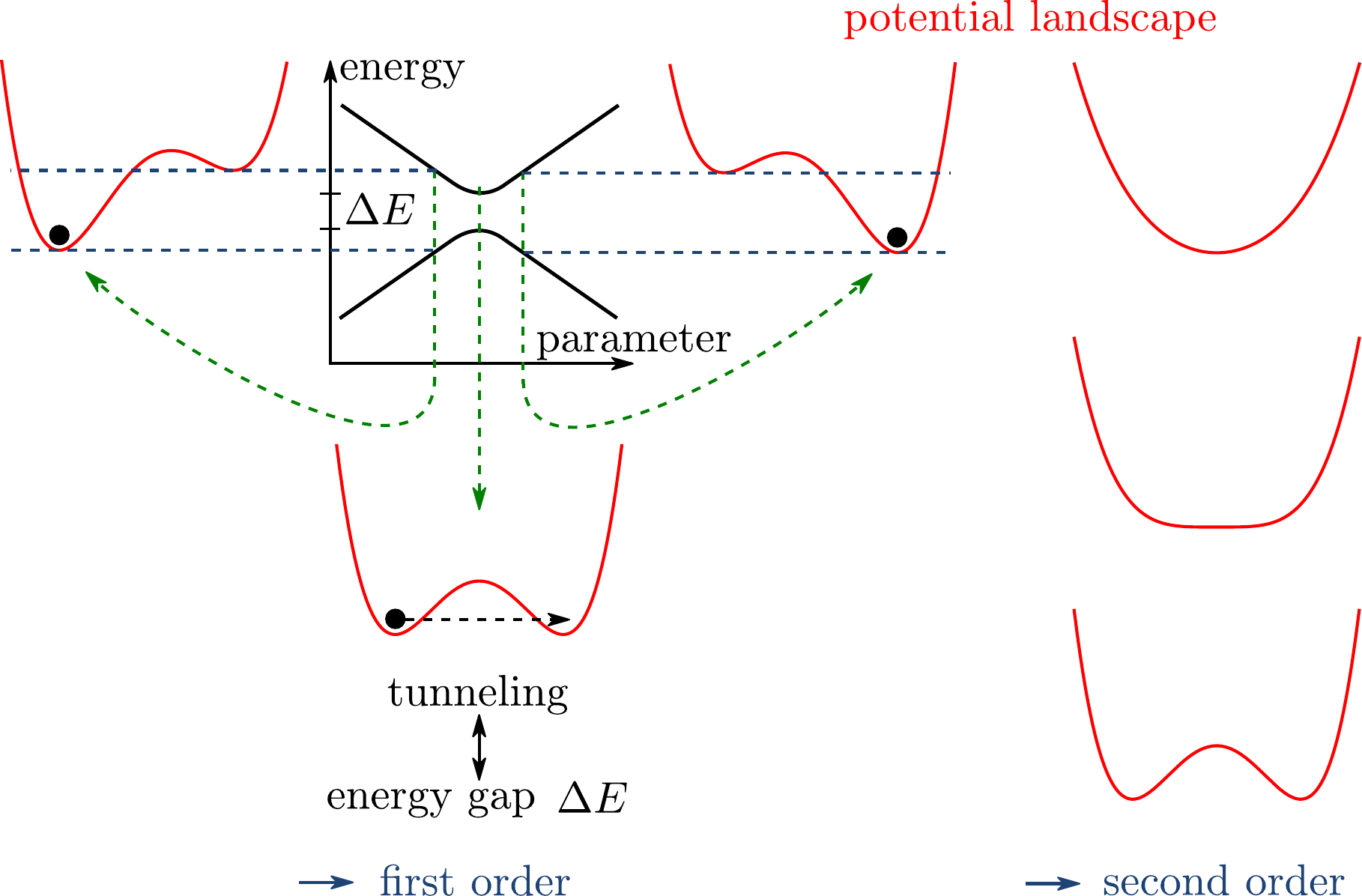}
\caption{Left: Landau-Zener transition and associated effective potential landscape, 
which corresponds to a first-order phase transition. 
Right: effective potential landscape for a symmetry-breaking (top to bottom) or 
symmetry-restoring (bottom to top) second-order phase transition. 
Such a symmetry-breaking phase transition will be relevant for the Kibble-Zurek mechanism 
discussed in Sec.~\ref{Kibble-Zurek mechanism}. 
The potential landscape on the right-hand side, on the other hand, is analogous to the 
false-vacuum decay discussed in Sec.~\ref{False-vacuum decay}.}
\label{fig:first-second}
\end{figure}

Unfortunately, the ``law of the conservation of difficulty'' strikes here again, 
since this energy gap $\Delta E(t)$ typically becomes extremely small at some time $t$
(or even at multiple times), see Fig.~\ref{fig:adiabatic}. 
As an intuitive picture, one can use the analogy to a Landau-Zener transition where the 
system has to tunnel from the initial minimum of the potential landscape to the final 
minimum (quite analogous to false vacuum decay discussed in Sec.~\ref{False-vacuum decay} 
below) which takes some time and the small tunneling rate is related to the 
energy gap $\Delta E$, see Fig.~\ref{fig:first-second}. 
This point where the energy gap $\Delta E$ becomes very small corresponds to the 
analogue of a first-order phase transition (for a large but finite system).
Carrying this analogy a bit further, one could expect that a second-order phase 
transition (where the system does not have to tunnel from one minimum to another)
might actually be advantageous, see also \cite{Schaller:2006,Schaller:2010uou} 
and Fig.~\ref{fig:first-second}. 

Apart from the long run-time $T$ required by the adiabatic theorem, such an avoided level 
crossing with a small energy gap $\Delta E$ may also be especially prone to decoherence,
see, e.g., \cite{Childs:2001ge,Tiersch:2007,Mostame:2010}. 
In summary, although adiabatic quantum algorithms or, more generally, quantum ground-state 
algorithms might offer some advantages in comparison to sequential quantum algorithms, 
their potential for solving real computational problems is still an open question.  
However, as it became evident above, we see more and more analogies to physical systems, 
such as glassy behaviour and many local minima in the potential landscape, first-order and 
second-order phase transitions, and Landau-Zener tunneling etc. 

\subsubsection{Quantum simulators for physical systems}
\label{Quantum simulators for physical systems}

Apart from solving classical problems such as factoring large numbers, quantum computers 
should also be able to simulate quantum systems much better than classical computers. 
Actually, this was one of the main motivations for 
Feynman \cite{Feynman:1981tf,Feynman:1984bi}
who referred to quantum computers as ``universal quantum simulators.'' 
A typical question here is the following: given an initial state $\ket{\psi_{\rm in}}$ 
and a possibly time-dependent physical Hamiltonian $\hat H_{\rm phys}(t)$, what is the 
resulting
final state $\ket{\psi_{\rm out}}=\hat U(T)\ket{\psi_{\rm in}}$ after a period of time $T$.
More precisely, quantum computers do not provide the full information about this 
quantum state $\ket{\psi_{\rm out}}$ (such as all the amplitudes in front of the basis states)
but properties of this state, i.e., observables. 
This is related to the problem of how to measure relevant observables in such a quantum 
simulator -- which will be discussed below in Sec.~\ref{Measurement schemes}. 

Provided that this Hamiltonian $\hat H_{\rm phys}(t)$ has reasonable properties
(see, e.g., \cite{Nielsen:2012yss}),
the above problem can be mapped to a sequential quantum algorithm 
as described above, e.g., by discretizing the time $t$ and approximating the small 
time steps of the quantum evolution $\hat U(t)$ by the Trotter formula.
%
Depending on the structure of $\hat H_{\rm phys}(t)$, there are quite efficient schemes 
for doing that. 
However, since the problem of decoherence described above prevented us from constructing 
such universal quantum computers (which allow the implementation of arbitrary sequential
quantum algorithms) with many qubits, the direct application of this scheme is not possible
with present technology -- at least for large numbers of qubits, see also 
\cite{Fauseweh:2024cow}. 

As a way out, one could instead construct a specific quantum system (i.e., a quantum simulator) 
such that its Hamiltonian $\hat H_{\rm sim}(t)$ can be mapped to the physical Hamiltonian 
$\hat H_{\rm phys}(t)$ to be simulated, in the sense that it reproduces its essential features. 
Examples of this idea will be discussed below -- basically, in the whole rest of this review. 

Now, one could object that the problem of decoherence (which was a huge problem for the ideas 
above) would also spoil this approach. 
One of the main points is that this objection is not necessarily correct:
Provided that the similarity between $\hat H_{\rm sim}(t)$ and $\hat H_{\rm phys}(t)$
is close enough, one would expect that all decoherence channels and small imperfections 
which could apply to the quantum simulator $\hat H_{\rm sim}(t)$ should also be possible 
for the real physical system $\hat H_{\rm phys}(t)$.
Thus, one would expect that these decoherence channels do not pose a very serious problem 
-- or, if they do, one would also like to know about it. 
For example, if the real physical system $\hat H_{\rm phys}(t)$ displays some properties 
such as Hawking radiation (see Sec.~\ref{Hawking radiation}) 
or high-temperature superconductivity (see Sec.~\ref{Hubbard Hamiltonians}), 
but small imperfections or perturbations in $\hat H_{\rm sim}(t)$ destroy these features,
then one would expect that analogous imperfections or perturbations in the real physical 
system $\hat H_{\rm phys}(t)$ would have a similar effect. 
Turning this argument around, the robustness of the phenomena under consideration 
(e.g., Hawking radiation or high-temperature superconductivity) is one of the important 
points one can study using these quantum simulators (see also Sec.~\ref{Summary and outlook}).
Quite generally, one would expect that physical phenomena which we can observe should be 
robust against small enough imperfections or perturbations because all Hamiltonians 
$\hat H_{\rm phys}(t)$ we do calculations with are approximations. 
Thus, such physical phenomena which are robust should also be observable in a suitable 
quantum simulator.  

So far, we have mainly discussed quantum simulators in analogy to sequential quantum
algorithms in Sec.~\ref{Sequential quantum algorithms}, 
but very similar arguments apply to the ground-state case discussed in 
Sec.~\ref{Quantum ground-state algorithms}. 
If the Hamiltonian $\hat H_{\rm phys}$ (or, equivalently, $\hat H_{\rm sim}$ 
provided that they are similar enough) has many local minima in its potential 
landscape where the system is likely to get stuck then it would be important 
to know this. 
Moreover, in such a case the precise structure of the true ground state may not 
be so important for our physical understanding (since it is never reached anyway), 
instead the typical structure of the low-lying local minima would determine the 
characteristic (e.g., glassy) behaviour of this system.  
Similar arguments apply to cases where the ground state is very vulnerable to 
small imperfections or perturbations.

For the transition scenario as in Eq.~\eqref{run-time} between two physically relevant 
Hamiltonians, it would also be interesting to know whether there is a point where the 
energy gap becomes extremely small (i.e., the analogue of a phase transition) and where 
that critical point is. 
As explained above, traversing this point can easily result in non-adiabatic behaviour,
which is basically the underlying mechanism for the Kibble-Zurek effect discussed in 
Sec.~\ref{Kibble-Zurek mechanism}. 

Here, we follow the route described above and consider quantum simulators for specific 
physical systems instead of universal quantum computers. 
More explicitly, the term ``quantum simulators'' is meant to refer to laboratory systems 
(in our case ultra-cold atoms) whose relevant degrees of freedom can be described 
by the Hamiltonian $\hat H_{\rm sim}(t)$ which reproduces essential features of the 
Hamiltonian $\hat H_{\rm phys}(t)$ of another physical quantum system we want to simulate.
Clearly, this is most relevant for investigating (or reproducing) quantum effects --  
but in several cases it can also be interesting to study their classical analogues.
It should be mentioned here that this notation is by no means unique.
For example, sometimes also the terms ``analogue quantum simulators'' 
or ``analogue quantum simulations'' are used, where ``analogue'' 
could refer to an analogy or to the opposite of ``digital'' 
(i.e., based on bits or qubits, see, e.g., \cite{Fauseweh:2024cow}) or both.  

\subsection{Ultra-cold atoms}\label{Ultracold atoms}

Ultra-cold atoms are very promising candidates for quantum simulators since our experimental 
capabilities for manipulation, control and read out are very advanced, see, e.g., 
\cite{Jaksch:2005,
Bloch:2005she,
Bloch:2012uep,
Weiss:2017kzz, 
Gross:2017ehn,
Schaefer:2020,
Yang:2020arq, 
Kaufman:2021kwf, 
Pause:2023pao}. 
One can combine different species of atoms (e.g., bosonic and fermionic) with the 
desired properties or exploit their internal states. 
The associated energy scales are quite low and thus the resulting time scales are rather long 
(in the millisecond regime or even longer) such that it is easy to manipulate ultra-cold atoms 
fast enough to reach non-equilibrium conditions. 
Characteristic length scales in the micrometer regime and above are also much longer than those 
in other areas (such as solid state or particle physics) and thus can be resolved by optical means.  
It is possible to tune the strength and the geometry of the external potential felt be the atoms 
as well as their interaction among each other. 
%
%
Furthermore, these atoms can be well isolated from the environment and 
cooled down to extremely low temperatures below one nano-Kelvin, which is the lowest 
temperature reachable in the laboratory so far -- allowing us to reach quantum degeneracy.  
Actually, in view of the temperature of the cosmic microwave background radiation of about 
2.7~Kelvin\footnote{Even lower temperatures of approximately one Kelvin might exist in the 
Boomerang Nebula, which indicates that this system has not equilibrated 
with its surroundings yet. 
However, this is still far above the temperature of the ultra-cold atoms.},  
these systems have been called ``the coldest systems in the Universe.'' 
Note, however, that this low temperature is -- strictly speaking -- only an effective 
temperature since ultra-cold atomic vapor is a meta-stable state and not the true ground state.
For typical examples such as Rubidium or Sodium atoms, the real ground state would be a metallic 
solid.  
Their vaporized state can live quite long if they are very dilute. 
In this case, almost all collisions are two-body collisions, where the atoms cannot 
``clump together'' and form molecules due to energy-momentum conservation. 
Molecule formation is thus only possible in three-body collisions where the third atom 
carries away the excess energy. 
However, in dilute systems (possibly in combination with suitable potentials such as 
optical lattices) these three-body collisions are very rare events. 
Nevertheless, these three-body losses results in decoherence -- even at arbitrary low 
temperatures and for nearly perfect isolation from any environment -- which pose ultimate 
limits on the quantum properties of these simulators, see, e.g., 
\cite{Jack:2002,Raetzel:2021}.  

\subsubsection{Many-body Hamiltonian}\label{Many-body Hamiltonian}

Neglecting these three-body losses, ultra-cold atoms can be described by the general 
Hamiltonian 
%
\bea
\label{Hamiltonian-general}
\hat H=\int d^3r
\left(
\sum_a\frac{(\na\hat\psi^\dagger_a)\cdot(\na\hat\psi_a)}{2m_a}
+
\sum_{ab}
V_{ab}
\hat\psi_a^\dagger\hat\psi_b
\right) 
+
\int d^3r\,d^3r'\,
\sum_{abcd}
W_{abcd}(\f{r},\f{r}')
\hat\psi_a^\dagger(\f{r})\hat\psi_b^\dagger(\f{r}')
\hat\psi_c(\f{r}')\hat\psi_d(\f{r})
\,.
\ea
Since the ultra-cold atoms are naturally quite slow, the non-relativistic description above 
is typically sufficient. 
The many-particle field operators $\hat\psi_a$ and $\hat\psi^\dagger_a$ annihilate or create 
atoms, which could be bosons or fermions or a mixture, where $a$ labels the different atomic 
species (different atoms or isotopes, spin states or internal states etc.) 
which could also have different masses $m_a$. 

The potential term $V_{ab}\hat\psi_a^\dagger\hat\psi_b$ is typically generated by external 
electromagnetic fields. 
For example, off-resonant optical laser fields, blue or red detuned, may generate 
diagonal terms $V_{aa}\hat\psi_a^\dagger\hat\psi_a$ corresponding to 
repulsive or attractive potentials for the atoms, respectively. 
Resonant or near-resonant optical laser fields could induce transitions corresponding to 
off-diagonal terms $V_{ab}\hat\psi_a^\dagger\hat\psi_b$ with $a\neq b$ if the labels 
$a$ and $b$ distinguish the internal atomic states. 
As another possibility, a static magnetic field $B$ in $x$-direction can induce such an 
off-diagonal term $\hat\psi_\uparrow^\dagger\hat\psi_\downarrow$ if the labels $a$ and $b$ 
correspond to the spin of the fermionic atoms (with $s=1/2$) in $z$-direction. 
These potentials can be used to confine the atomic cloud or to form an optical lattice 
(or a combination of both).

As another interesting possibility, one could depart from the treatment of the photon  
field as an external field and take the photonic degrees of freedom as dynamical variables
into account -- for example, after enhancing the coupling to certain photon modes via  
enclosing the ultra-cold atoms by a cavity, see, e.g., \cite{Mivehvar:2021lpo}. 

Finally, $W_{abcd}(\f{r},\f{r}')$ describes the interactions between the atoms. 
As explained above, we neglect processes involving three or more atoms which would 
correspond to the product of six or more many-particle field operators $\hat\psi_a$ 
and $\hat\psi^\dagger_a$.
For ultra-cold atoms, mainly three interaction types are relevant. 
Interactions or collisions with ranges much shorter than all other relevant length 
scales can be approximated by contact interactions 
$W_{abcd}(\f{r},\f{r}')\approx g_{abcd}\,\delta^3(\f{r}-\f{r}')$ with the coupling 
constants or scattering cross sections being encoded in $g_{abcd}$. 
In most cases, they will be the same for all atomic species 
$g_{abcd}=g\,\delta_{ad}\delta_{bc}$ but one can obtain additional interesting effects 
if the inter-species and the intra-species coupling differ, see also
Sec.~\ref{Cosmological particle creation}.
As a further step, one could also imagine interaction-induced transitions which would 
correspond to off-diagonal elements, but we shall not consider them here.
Other important examples are the (magnetic) dipole-dipole interaction which is of
finite range and behaves as $W_{abcd}(\f{r},\f{r}')\sim1/|\f{r}-\f{r}'|^3$ for fixed
dipole moments, see, e.g., \cite{Lahaye:2009qqr,Chomaz:2022cgi}.
If the dipole moments are not fixed but fluctuating, one can still obtain an interaction 
of the van-der-Waals type, which behaves as $1/|\f{r}-\f{r}'|^6$. 
These interactions can be amplified strongly using Rydberg atoms, i.e., atoms in 
high-lying excited states, see, e.g., 
\cite{Weimer:2010tfn,
Saffman:2010ocs, 
Bernien:2017ubn, 
Nguyen:2018, 
Keesling:2018ish, 
Surace:2019dtp, 
Browaeys:2020kzz, 
Ebadi:2020ldi, 
Cohen:2021axm, 
Malz:2022pof}.
This option does also offer more knobs for manipulation and control because the 
interaction properties of those Rydberg atoms strongly depend on their quantum numbers.


\subsubsection{Measurement schemes}\label{Measurement schemes}

Of course, realizing the Hamiltonian which allows us to simulate the phenomenon under consideration
is not the end of the story, we also need a read-out scheme in order to measure the observables
we are interested in.
For ultra-cold atoms, the majority of the schemes relies on measuring the positions of the 
atoms\footnote{However, alternative ideas are conceivable as well, see, e.g.,
\cite{tenBrinke:2015,tenBrinke:2016}.}
or the position-dependent number or density of atoms in some way, e.g.,
by taking pictures or by sending the atoms into an atom detector, see, e.g.,
\cite{Jaskula:2012ab}. 
Obviously, the former method can be less destructive than the latter and thus could be repeated
several times during the course of evolution.
However, one should keep in mind that such a non-destructive measurement is never
``interaction-free''.
According to the laws of quantum theory, such a measurement changes the quantum state of the
system (unless it happens to be in an atom number eigen-state) and thus perturbs it.
This perturbation \rs{or back-action} can be small (e.g., in the case of weak measurements), 
but it cannot be avoided completely.

In some cases, the observables under consideration are directly related to the number of atoms
depending on their position, such as the various lattice sites
(see, e.g., \cite{Tarruell:2018ikc}),
or the density profile of the atoms or their density-density correlations
(see, e.g., 
\cite{Lahav:2009wx,Steinhauer:2015ava,Steinhauer:2015saa,MunozdeNova:2018fxv,Kolobov:2019qfs}) 
which can be extracted by repeating the experiments many times.
In these cases, the detection scheme is usually referred to as {\em in situ} and can be
non-destructive (as discussed above).
In other cases, the trick is to translate the observables of interest to the number or
density of atoms.
As one example, if we consider different atomic species (e.g., hyperfine states)
and start with atoms of one species only, suitable light-induced resonant transitions
could be used to translate the number of phonons (i.e., excitation quanta)
to the number of atoms in a different species.
Then, with resonant light which strongly couples to the latter species but basically
not to the former, one can measure the number of atoms in that species and thereby
infer the original number of phonons \cite{Schutzhold:2006}. 

For measuring observables which are canonically conjugate to the atom numbers or densities,
one can apply methods and ideas from quantum state tomography:
After the actual quantum simulation itself, one could add periods of suitable unitary
evolutions of varying delay times which rotate the canonically conjugate variables into each
other.
By measuring the atom numbers or densities after varying time delays, one can also infer
their canonically conjugate variables (i.e., phases), see, e.g., \cite{Viermann:2022wgw}. 
A related concept is the {\em time-of-flight} technique, where the trap confining the
atoms is switched off after the quantum simulation.
If the interactions between the atoms are sufficiently weak (dilute-gas limit) and the
(initial) confining potential is sufficiently strong, the atoms basically start to move
according to their momenta at the switch-off time.
Then, letting the atomic cloud expand until it is much larger than its initial size,
a snap-shot of the atomic cloud allows us to infer the momentum distribution at the
switch-off time -- in contrast to an {\em in situ} measurement of the position distribution.
As another difference, such a {\em time-of-flight} measurement is obviously destructive
while an {\em in situ} measurement can be non-destructive.

\subsubsection{Hubbard Hamiltonians}\label{Hubbard Hamiltonians}

\rs{Even though the main focus of this review lies on relativistic particle physics,
whereas Hubbard Hamiltonians belong more to condensed-matter or solid-state physics,
let us briefly discuss them -- possible relations will hopefully 
become more evident below.}
%
%
Focusing on bosons for a moment, we may expand the field operators $\hat\psi_a(\f{r})$ into a 
complete set of orthonormal basis functions $f_a^I(\f{r})$ via
\bea
\label{complete-set}
\hat\psi_a(\f{r})=\sum_I\hat b_I f_a^I(\f{r})
\,.
\ea
Insertion of this expansion into the general form~\eqref{Hamiltonian-general} yields 
\bea
\label{pre-Bose-Hubbard}
\hat H=-\sum_{IJ} T_{IJ} \hat b_I^\dagger\hat b_J 
+ \sum_{IJKL} 
U_{IJKL}
\hat b_I^\dagger\hat b_J^\dagger\hat b_K \hat b_L 
\,,
\ea
where the $T_{IJ}$ are determined by the matrix elements of the differential operator 
in Eq.~\eqref{Hamiltonian-general} between the basis functions $f_a^I(\f{r})$ and 
$f_b^J(\f{r})$. 
Similarly, the interaction terms $U_{IJKL}$ represent convolutions of the interaction kernel 
$W_{abcd}(\f{r},\f{r}')$ in Eq.~\eqref{Hamiltonian-general} with four basis functions. 
The diagonal terms $T_{II}$ can be identified with the single-particle 
energies per mode $I$ and the off-diagonal terms $T_{IJ}$ as the tunneling or 
hopping matrix between the modes $I$ and $J$. 
%
%
This form~\eqref{pre-Bose-Hubbard} is still quite general, but let us now apply 
some specifications. 
First, we assume a periodic potential 
such as an optical lattice and 
choose the basis set $f_a^I(\f{r})$ to be the associated Wannier functions. 
Second, this lattice potential is supposed to be sufficiently deep such that 
the different bands are well separated in energy and we may focus on the lowest 
Wannier band.  
In addition, for deep lattice potentials, the Wannier states in this lowest band 
are strongly localized. 
Third, we consider only one atomic species and short-range interactions.  
With these assumptions and approximations, we arrive at the 
Bose-Hubbard Hamiltonian \cite{Jaksch:1998zz} 
with the number operator $\hat n_I=\hat b_I^\dagger\hat b_I$
for the lattice site $I$ 
\bea
\label{Bose-Hubbard Hamiltonian}
\hat H
\approx 
-\sum_{IJ} T_{IJ} \hat b_I^\dagger\hat b_J 
+ U \sum_I
\hat n_I(\hat n_I-1)
\,.
\ea
Due to the strong localization of the Wannier states (for deep lattice potentials),
it is sufficient if we keep the tunneling matrix elements $T_{IJ}$ between neighboring 
lattice sites $I$ and $J$ only. 

A renowned phenomenon of the Bose-Hubbard Hamiltonian~\eqref{Bose-Hubbard Hamiltonian} 
is the super-fluid--Mott transition, see, e.g., \cite{Greiner:2002gqo}. 
For small interactions $U$, the ground state is a super-fluid 
(i.e., a Bose-Einstein condensate) as characterized by a long-range order parameter 
$\langle\hat b_I^\dagger\hat b_J\rangle$.
For large on-site repulsion $U$, however, the character of the ground state changes. 
Assuming unit filling $\langle\hat n_I\rangle=1$, we have essentially 
one atom per lattice site and the large energy penalty $U$ prevents them from hopping to 
the next lattice site.
As a result, the ground state is insulating and separated by a finite gap 
(the Mott gap) from the excited states. 
In this case, the coherence $\langle\hat b_I^\dagger\hat b_J\rangle$ essentially vanishes  
for $I\neq J$. 

Applying the same approximations for fermionic atoms in optical lattices, we arrive at the 
Fermi-Hubbard Hamiltonian 
\bea
\label{Fermi-Hubbard}
\hat H
\approx 
-\sum_{IJs} T_{IJ} \hat c_{Is}^\dagger\hat c_{Js}  
+ U \sum_I
\hat n_I^\uparrow\hat n_I^\downarrow
\,,
\ea
which is even more famous than the bosonic version~\eqref{Bose-Hubbard Hamiltonian}. 
Naturally, this system has also been realized experimentally, see, e.g., 
\cite{Tarruell:2018ikc,Jordens:2008qmo,Bakr:2009lrk,Mazurenko:2017asv}
as well as
\cite{Yang:2020yer,Su:2023urk}.
Due to the additional spin degree \rs{of freedom} 
$s\in\{\uparrow,\downarrow\}$, its dynamics is more 
involved and less well understood. 
Nevertheless, for large $U$, the ground state does also become a Mott insulator in the 
case of half filling $\langle\hat n_I^\uparrow\rangle=\langle\hat n_I^\downarrow\rangle=1/2$.
For small $U$, on the other hand, the state would be metallic (in more than one dimension),
i.e., conducting.
In the Mott insulator state, conduction requires the generation of quasi-particle excitations 
in the form of doublons (i.e., doubly occupied lattice sites) or holons 
(i.e., empty lattice sites).
The proper description of their dynamics, including the interactions among each other 
(see, e.g., \cite{IJTP} and references therein) is an open question and might also be 
relevant for our understanding of high-temperature superconductivity.  




\subsubsection{Dirac Hamiltonian}\label{Dirac Hamiltonian}

After this detour to the non-relativistic Hubbard Hamiltonians, 
let us now turn our attention to the Dirac Hamiltonian 
\bea
\label{Dirac-Hamiltonian-3+1}
\hat H
=
\int d^3r\,\hat\Psi^\dagger\left(-i\f{\alpha}\cdot\na+m\beta\right)\hat\Psi 
=
\int d^3r\,\hat\Psi^\dagger\left(-i\gamma^0\f{\gamma}\cdot\na+m\gamma^0\right)\hat\Psi 
\,,
\ea
\rs{which} we first consider in 3+1 dimensions.  
For the $\gamma^\mu$ matrices (or, equivalently, for the $\f{\alpha}=\gamma^0\f{\gamma}$ 
and $\beta=\gamma^0$ matrices), we use the Dirac basis 
\bea
\label{Dirac-basis}
\gamma^0=
\left(
\begin{array}{cc}
\f{1} & 0 \\
0 & -\f{1} 
\end{array}
\right)
\,,\;
\gamma^{1}=
\left(
\begin{array}{cc}
0 & \sigma_x \\
-\sigma_x & 0
\end{array}
\right)
\,,\;
\gamma^{2}=
\left(
\begin{array}{cc}
0 & \sigma_y \\
-\sigma_y & 0
\end{array}
\right)
\,,\;
\gamma^{3}=
\left(
\begin{array}{cc}
0 & \sigma_z \\
-\sigma_z & 0
\end{array}
\right)
\,,
\ea
where $\sigma_x$, $\sigma_y$ and $\sigma_z$ are the usual Pauli spin matrices.  
Other basis sets, such as the Weyl basis or the Majorana basis 
(with purely imaginary Dirac matrices) can also be used. 
The bi-spinor field operators $\hat\Psi$ and $\hat\Psi^\dagger$ have four 
components encoding the two spin degrees of freedom as well as the particle and 
anti-particle components.  

After spatial discretization $\hat\Psi(\f{r})\to\hat c_{Is}$, the above 
Hamiltonian~\eqref{Dirac-Hamiltonian-3+1} can in principle be cast into a form 
$T_{IJ}^{ss'} \hat c_{Is}^\dagger\hat c_{Js'}$ similar to the 
Hubbard Hamiltonian~\eqref{Fermi-Hubbard} with vanishing interaction $U=0$, 
but realizing such a hopping matrix $T_{IJ}^{ss'}$ experimentally may be 
a bit involved, see also \cite{Mazza:2011kf}. 

Fortunately, many phenomena such as the Sauter-Schwinger effect discussed in 
Sec.~\ref{Sauter-Schwinger and Breit-Wheeler effect} do not necessarily 
require 3+1 dimensions but can already be observed in 2+1 or even 1+1 dimensions.
Thus, let us simplify the problem and consider the Dirac Hamiltonian in 1+1 
dimensions (again $s=1/2$) with $\gamma^0=\sigma_z$ and $\gamma^1=i\sigma_x$
\bea
\label{Dirac-Hamiltonian-1+1}
\hat H=\int dx\,\hat\Psi^\dagger\left(i\sigma_y\partial_x+m\sigma_z\right)\hat\Psi 
\,.
\ea
Here, $2\times2$ Dirac matrices are sufficient and thus we only have two components 
for $\hat\Psi$ and $\hat\Psi^\dagger$. 
Note that these two components describe the particle and anti-particle components, 
but not the spin degree of freedom. 

As it turns out, this form~\eqref{Dirac-Hamiltonian-1+1} admits a quite simple lattice 
representation, see, e.g., \cite{IgnacioCirac:2010us,Szpak:2011pra,Szpak:2011jj}. 
For example, we may encode the upper and lower components of $\hat\Psi$ in the even 
and odd lattice sites $I$. 
(Corresponding to the remark above, we consider the spin of the fermionic atoms to be fixed, 
e.g., by a strong magnetic field.)  
Then, the spatial derivative $\partial_x\hat\Psi$ at the position of the lattice site 
$I$ can be represented by $(\hat c_{I+1}-\hat c_{I-1})/(2\Delta x)$ where $\Delta x$ 
is the distance between the lattice sites. 
As a result, the kinetic term in Eq.~\eqref{Dirac-Hamiltonian-1+1} becomes a sum of 
simple hopping terms $\pm(\hat c_I^\dagger\hat c_{I\pm1}+{\rm h.c.})$.
The alternating signs $\pm$ in front of these hopping terms can be eliminated by local 
phase shifts.
For example, we may go through the one-dimensional lattice from left to right, i.e., 
with increasing $I$.
Wherever such a hopping term $(\hat c_I^\dagger\hat c_{I+1}+{\rm h.c.})$ comes with 
the ``wrong'' sign, we may ``repair'' this by applying a phase shift 
$\hat c_{I+1}\to-\hat c_{I+1}$ to the right lattice site. 

Note that this ability to ``repair'' the phases (which would also work for more general 
complex phases $e^{i\varphi_I}$) is a peculiarity of a one-dimensional lattice, 
which has the same number of lattice sites and links (between neighboring lattice sites). 
In general higher-dimensional lattices, such as a square lattice in two dimensions, 
the number of lattice sites is smaller than the number of lattice links 
(even if only links between between neighboring lattice sites are considered) 
and thus the number of possible phase shifts (i.e., lattice sites) is not sufficient to 
``repair'' all the possible relative phases (at the links). 
This problem has to be taken into consideration when trying to realize lattice versions 
of higher-dimensional theories, such as the Hamiltonian~\eqref{Dirac-Hamiltonian-3+1} 
with the matrices~\eqref{Dirac-basis}.  
The issue becomes even more relevant for the implementation of gauge fields discussed 
in Sec.~\ref{Gauge fields} where phase accumulated along a closed path (i.e., a Wilson loop)
can be non-trivial, e.g., if a magnetic field is present. 

\begin{figure}[tbp]
\centering
\includegraphics[width=.7\textwidth]{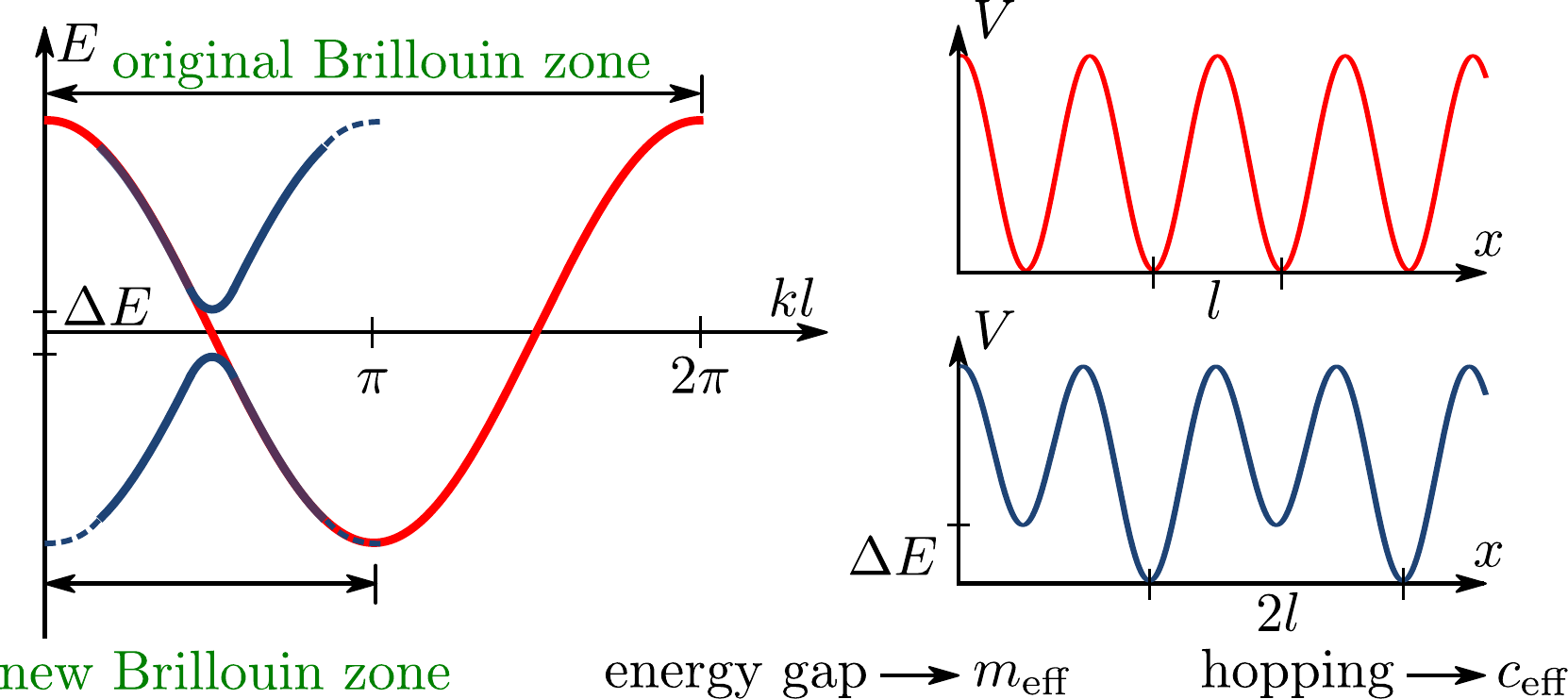}
\caption{Sketch of the dispersion relations (left) for the original (top) and the 
bi-chromatic (bottom) optical lattices (right).}
\label{fig:dirac-dispersion}
\end{figure}

Coming back to our 1+1 dimensional system~\eqref{Dirac-Hamiltonian-1+1} and its lattice 
realization, the hopping strength determines the effective speed of light $c_{\rm eff}$ which 
is naturally much slower than the real speed of light $c$, see Sec.~\ref{Introduction}. 
The mass term $m\sigma_z$ corresponds to a local (on-site) potential shift which is 
alternating, i.e., positive for the upper component of $\hat\Psi$ on the even sites 
and negative for the lower component of $\hat\Psi$ on the odd sites.
As a result, we get a bi-chromatic lattice where the even sites have a higher energy 
than the odd sites, see Fig.~\ref{fig:dirac-dispersion}. 

The emergence of quasi-relativistic behaviour in such a lattice can also be understood 
on the level of the dispersion relations, see Fig.~\ref{fig:dirac-dispersion}. 
Without the off-set between the even on add lattice sites representing the mass term 
$m\sigma_z$, the lattice reproduces the standard cosine dispersion relation if we focus 
on the lowest Wannier band. 
Since we are going to distinguish odd and even sites by the off-set, we take our 
elementary cell to contain two lattice sites, such that our Brillouin zone in the 
reciprocal lattice is cut in half. 
Thus, we have to fold the usual cosine dispersion relation which can be done by 
taking both signs into account, i.e., mirroring it at the $k$-axis, see
Fig.~\ref{fig:dirac-dispersion}. 
Now, adding a small off-set basically turns the crossing of the two branches of the 
cosine dispersion relation at its zero into an avoided level crossing.
In the vicinity of this avoided level crossing, the dispersion relation now has the 
quasi-relativistic form $\pm\sqrt{c_{\rm eff}^2k^2+m_{\rm eff}^2c_{\rm eff}^4}$.

To conclude this section, let us briefly discuss the Dirac Hamiltonian in 2+1 dimensions.
Using the Dirac matrices $\gamma^0=\sigma_z$, $\gamma^1=i\sigma_x$ and $\gamma^2=i\sigma_y$,
it can also be realized via a two-component spinor $\hat\Psi$.
On the one hand, this is a great simplification when it comes to constructing a quantum 
simulator, but, on the other hand, this implies that some of the concepts known from the 
3+1 dimensional case~\eqref{Dirac-Hamiltonian-3+1} and \eqref{Dirac-basis} cannot be applied 
here. 
Apart from the fact that the spin degree of freedom is not captured (as already mentioned above),
there is no $\gamma^5$ matrix (as an independent matrix) here, i.e., one cannot apply the 
concept of chirality in the same way as in 3+1 dimensions\footnote{Obviously, one can have 
circular motion in this case, but this is not really the same as chirality -- even though 
the two concepts are sometimes mixed up a bit.}. 
One of the simplest ways to actually realize a lattice version of this 2+1 dimensional 
Dirac Hamiltonian is to have fermionic atoms in a hexagonal or honeycomb lattice, i.e., 
the same structure as graphene.
In this case, a quasi-relativistic dispersion relation 
$\omega\approx c_{\rm eff}|\f{k}-\f{k}_{\rm Dirac}|$ 
without a mass gap arises naturally in the vicinity of the two Dirac points
$\f{k}_{\rm Dirac}$ where $\omega=0$, see, e.g., \cite{CastroNeto:2007fxn}. 
The fact that we obtain two inequivalent Dirac points (further Dirac points in the Brillouin
zone are related via symmetries to these two) is related to the fermion doubling problem:
When directly discretizing the Dirac Hamiltonian on a lattice, one typically obtains multiple 
copies of the original fermions in the low-energy sector. 
%
This should also be taken into account when constructing quantum simulators, 
e.g., by focusing on (or coupling to) one of these copies only -- or by by actually 
using these different copies for representing different fermion species of the 
original problem. 

\subsubsection{Scalar field}\label{Scalar field}

In the previous section, the original atomic field operators $\hat\psi_a(\f{r})$ 
and the quasi-relativistic Dirac field operators $\hat\Psi(\f{r})$ to be simulated 
we related by a simple linear transformation such as Eq.~\eqref{complete-set}
plus the discretization on a lattice.
In this section, let us consider a different scheme where the motional degrees of freedom 
of the atoms are mapped to the quasi-relativistic field to be simulated, in this case 
the scalar field $\hat\phi$.  
An alternative approach based on the phonon field in Bose-Einstein condensates will be 
discussed in Sec.~\ref{Analogue gravity}. 

\begin{figure}[tbp]
\centering
\includegraphics[width=.7\textwidth]{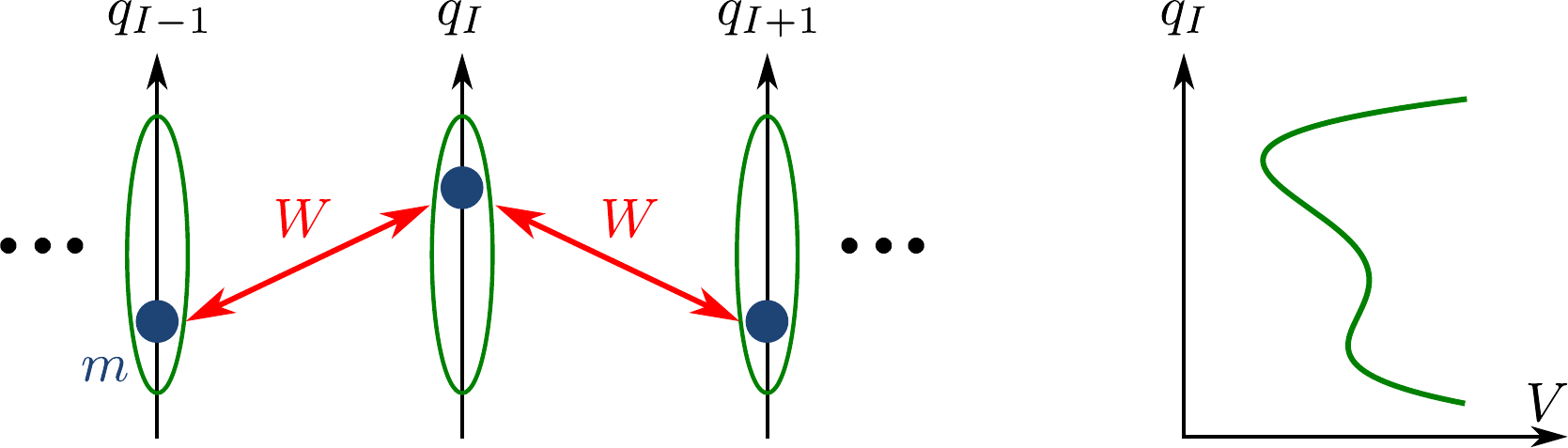}
\caption{Sketch of the simulator for the scalar-field Lagrangian~\eqref{scalar-field+potential}.
\rs{On the left-hand side, the atoms at the positions $q_I$ are displayed as blue dots together 
with their interactions $W$ between nearest neighbors.
On the right-hand side, the local on-site potentials $V(q_I)$ are depicted, where the 
axes are rotated in order to align the $q_I$ direction with that on the left-hand side.}}
\label{fig:scalar} 
\end{figure}

To this end, we consider a chain (or array) of atoms with spin zero which are strongly
localized in the $x$ and $y$ directions by suitable potentials (e.g., a deep  optical lattice)
such than they stay in the lowest quantum well states in those directions,
see Fig.~\ref{fig:scalar}.
Thus, we may focus on their motion in the remaining $z$ direction where we assume that we
have one atom in each lattice site $I$.
Denoting the position of the atom with mass $m$ in the lattice site $I$ by $q_I$,
this motion in $z$ direction is influenced by the potential $V(q_I)$ which is
supposed to be 
much weaker than the confining potentials in the $x$ and $y$ directions.
Furthermore, we assume that the lattice sites are close enough that we may have a
strong coupling (e.g., by having Rydberg atoms) between the lattice sites
$W(q_{I+1}-q_I)$ where we neglect to the coupling between next to nearest neighbors, 
see Fig.~\ref{fig:scalar}. 
Together with the usual kinetic \rs{energy} term, the Lagrangian for this system reads 
\bea
L=\sum_I\left(\frac{m}{2}\,\dot q_I^2-V(q_I)-W(q_{I+1}-q_I)\right) 
\,.
\ea
If we assume that the interaction $W(q_{I+1}-q_I)$ is quite strong and attractive,
it tends to align the atoms, i.e., to make the difference $q_{I+1}-q_I$ quite small
(e.g., much smaller than the lattice spacing).
Thus, we may Taylor expand this interaction around its minimum at $q_{I+1}=q_I$
with the curvature $\kappa$ (effective spring constant)
\bea
L\approx\sum_I\left(\frac{m}{2}\,\dot q_I^2-V(q_I)-\frac{\kappa}{2}(q_{I+1}-q_I)^2\right)
\,.
\ea
In addition, again using that the difference $q_{I+1}-q_I$ is quite small,
we may go to the continuum limit with the identification $q_I(t)\to\phi(t,x)$
where the lattice site $I$ is replaced by the coordinate $x=I\Delta x$
with the lattice spacing $\Delta x$.
In this limit, the difference $q_{I+1}-q_I$ becomes the spatial derivative
$(q_{I+1}-q_I)/\Delta x\to\partial_x\phi$ and the sum over all lattice sites $I$
turns into an integral $\Delta x\sum_I\dots\to\int dx\dots$.
As a result, we arrive at the Lagrangian density for a scalar field
\bea
\label{scalar-field+potential}
{\cal L}=\frac{1}{2}\,\dot\phi^2-\frac{c_{\rm eff}^2}{2}\,(\partial_x\phi)^2-{\cal V}(\phi) 
\,,
\ea
where the effective speed of light is set by the spring constant $\kappa$,
see also Sec.~\ref{Introduction}.
The on-site potential ${\cal V}(\phi)$ can be adjusted by suitably shaped laser beams,
for example.
For a purely parabolic potential ${\cal V}(\phi)$, we would simulate a free massive
scalar field, but more involved potentials could be used to study
the sine-Gordon model, see Sec.~\ref{Sine-Gordon model},
the Kibble-Zurek mechanism, see Sec.~\ref{Kibble-Zurek mechanism}, or
the analogue of false-vacuum decay, see Sec.~\ref{False-vacuum decay}.
%
Note that the experimentally available potentials and interaction strengths
(e.g., using Rydberg atoms) are large enough to simulate the ground state properties
of these theories at nano-Kelvin temperatures.
Obviously, the above concept can be generalized to 2+1 dimensions in a quite straightforward
manner, while a 3+1 dimensional version requires more effort.

\subsubsection{Gauge fields}\label{Gauge fields} 

After having discussed the Dirac field~\eqref{Dirac-Hamiltonian-1+1} and the scalar 
field~\eqref{scalar-field+potential}, let us now turn to vector fields $A_\mu$. 
Here, it is important to carefully distinguish several levels of the incorporation of 
those fields $A_\mu$, see also \cite{Wiese:2013uua}. 
At the lowest level, one can consider external fields $A_\mu$ which can depend on space and 
time, but are fixed, i.e., treated as classical background fields.
For example, one could have the quantum Dirac field $\hat\Psi$ 
as in Eq.~\eqref{Dirac-Hamiltonian-1+1}, but propagating within the background of an external 
electromagnetic field $A_\mu$. 
In 1+1 dimensions, the vector potential $A_\mu=(A_0,A_1)$ can be simplified by a
gauge transformation $A_\mu\to A_\mu + \partial_\mu\chi$, e.g., by gauging
the spatial part away $A_1=0$ such that only the potential $A_0=\Phi$ remains.
Note that this gauge transformation is related to the phase shift discussed above in 
Sec.~\ref{Dirac Hamiltonian} since the associated gauge transformation for the Dirac field 
just reads $\Psi\to\exp\{iq\chi\}\Psi$. 
After this gauge transformation, the interaction Hamiltonian reads
\bea
\hat H_{\rm int}
=
q\int dx\,\hat\Psi^\dagger\hat\Psi\,A_0
=
q\int dx\,\hat\Psi^\dagger\hat\Psi\,\Phi
\,. 
\ea
In the lattice representation in Fig.~\ref{fig:dirac-dispersion}, such a term can simply 
be modeled by an additional optical laser with suitably shaped intensity profile
(which is determined by $\Phi$). 
As another option, a spatially homogeneous force $F(t)$ (representing a purely time-dependent 
electric field) can also be applied by accelerating (e.g., shaking) the lattice. 

Going to higher dimensions -- as also explained above in Sec.~\ref{Dirac Hamiltonian} --
the situation becomes more complicated, i.e., more interesting. 
In addition to the homogeneous force $F(t)$ generated by accelerating the lattice,
one could also imagine a rotating lattice such that the Coriolis force corresponds 
to the Lorentz force in the presence of a magnetic field. 
More involved gauge fields can be modeled by suitable phase imprints on the hopping 
matrix elements $T_{IJ}$ such that a particle hopping from one lattice site to another 
acquires a phase. 
For ultra-cold atoms in optical lattices, such phases can be induced in driven systems,
e.g., by applying suitable time-dependent electromagnetic fields. 
In this way, the atoms evolve (in the continuum limit) as if they were under the influence 
of an effective field $A_\mu$, which is referred to as a synthetic gauge field or 
an artificial gauge field, see, e.g., 
\cite{Surace:2019dtp,
Jaksch:2003opb, 
Ruseckas:2005vgb,
Gerbier:2010zz,
Zhang:2012tia,
Goldman:2013xka,
Celi:2013gma, 
Aidelsburger:2017qlh,
Hey:2018tzy,
Gorg:2018xyc, 
Galitski:2019yff}. 

\begin{figure}[tbp]
\centering
\includegraphics[width=.7\textwidth]{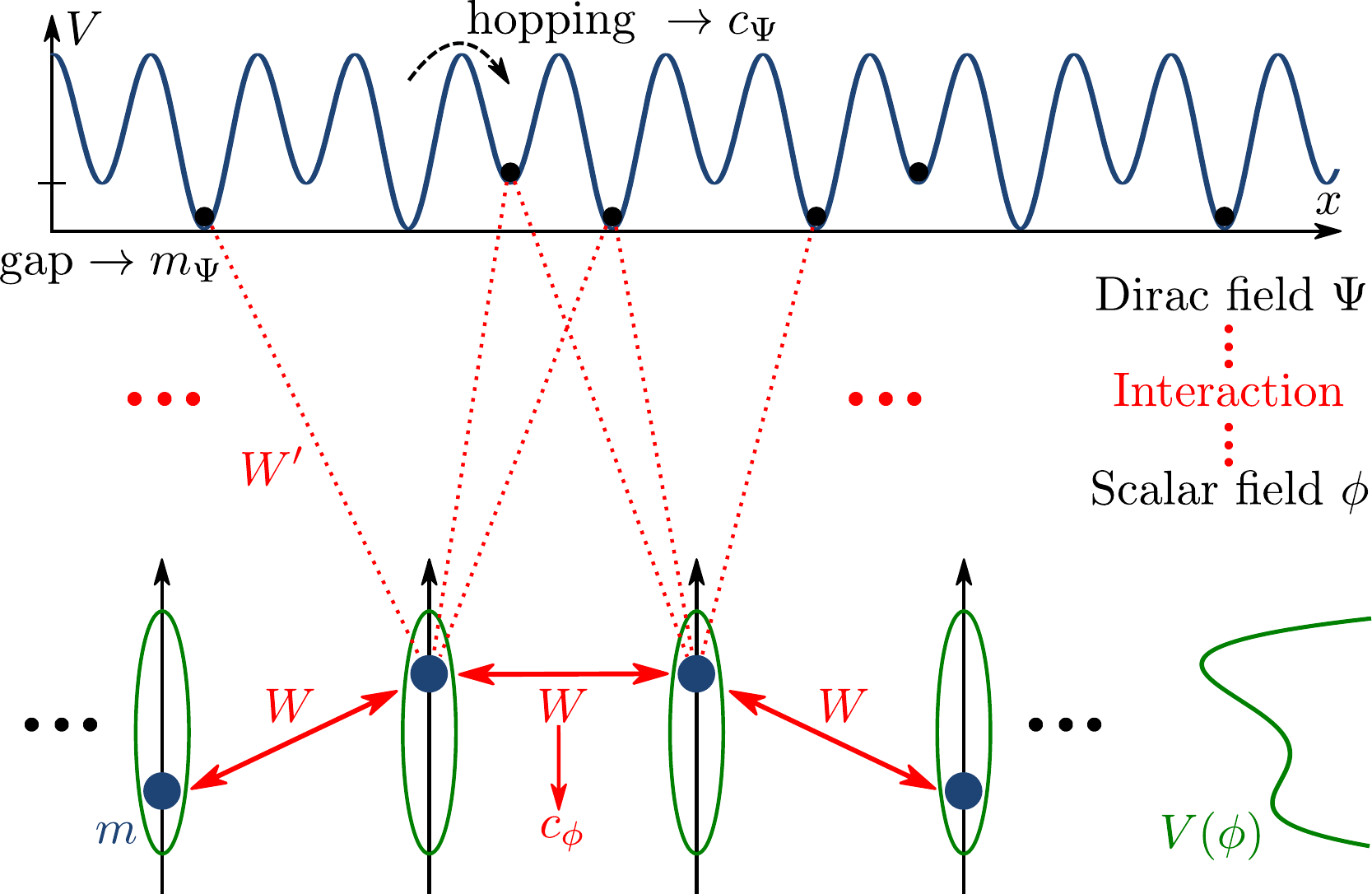}
\caption{Sketch of the simulator for the Dirac field~\eqref{Dirac-Hamiltonian-1+1}
coupled to the scalar-field~\eqref{scalar-field+potential}.}
\label{fig:dirac+scalar} 
\end{figure}

Going one step beyond external fields, one could also include the interaction between
the atoms.
However, since the dipolar or van-der-Waals interactions between the atoms display a 
distance dependence which is 
different\footnote{This would not be the case for ions, but they are not considered here.} 
from the Coulomb interactions between 
Dirac particles such as electrons and positrons, for example, one would typically 
obtain a qualitative instead of a quantitative analogy. 
Still, such a qualitative analogy would allow us to tackle interesting questions, 
such as the back-reaction (due to the Coulomb interaction between electrons and positrons)
onto the Sauter-Schwinger effect, see Sec.~\ref{Sauter-Schwinger and Breit-Wheeler effect}. 
On the qualitative level, these investigations could also shed light on other phenomena,
such as string breaking in Quantum Chromodynamics (QCD). 

The next level would be a fully dynamical treatment of the gauge fields 
(instead of dipolar or van-der-Waals interactions where the interaction kernels  
$\propto1/|\f{r}-\f{r}'|^3$ or $\propto1/|\f{r}-\f{r}'|^6$ are fixed). 
In this case, one should also have quantized degrees of freedom modeling those gauge fields, 
see, e.g., 
\cite{Yang:2020yer,
Zohar:2011cw,
Zohar:2012ay,
Zohar:2012xf, 
Zohar:2013zla,
Zohar:2015hwa, 
Gonzalez-Cuadra:2017lvz, 
Barbiero:2018wui,
Schweizer:2019lwx}.
As a very simple illustration, one could consider combining the simulator for 
the Dirac field~\eqref{Dirac-Hamiltonian-1+1} in Fig.~\ref{fig:dirac-dispersion} 
with that for the scalar field~\eqref{scalar-field+potential} in Fig.~\ref{fig:scalar},
which is sketched in Fig.~\ref{fig:dirac+scalar}. 
This simulator is realized via a one-dimensional bi-chromatic lattice half filled with 
fermionic atoms which represent the Dirac field~\eqref{Dirac-Hamiltonian-1+1}.
They do not interact with each other but can hop from one lattice site to the next, 
such that the hopping strength determines the effective speed of light $c_{\Psi}$ 
for the Dirac field. 
The effective mass $m_\Psi$ in the Dirac equation is generated by the energy difference 
between the even and odd sites (as in Fig.~\ref{fig:dirac-dispersion}). 
The simulator for the scalar field as in Fig.~\ref{fig:scalar} is then placed parallel 
to this one-dimensional fermionic lattice in the same plane, i.e., such that the motion 
$q_I$ of those atoms changes their distance to the ferimonic atoms, i.e., impacts their 
interaction $W'$, e.g., via having Rydberg atoms in the simulator for the scalar field. 
They do interact among each other and the associated interaction $W$ generates the 
effective speed of light $c_\phi$ for the scalar field, as in Fig.~\ref{fig:scalar}. 
In the continuum limit, this combination gives rise to the effective Lagrangian 
\bea
\label{dirac+scalar-field}
{\cal L}
=
\Psi^\dagger\left(
i\partial_t-i c_\Psi\sigma_y\partial_x-m_\Psi c_\Psi^2\sigma_z+
g_{\rm eff}\phi 
\right) 
\Psi
+
\frac{1}{2}\,\dot\phi^2-\frac{c_\phi^2}{2}\,(\partial_x\phi)^2-{\cal V}(\phi) 
\,,
\ea
where the effective coupling strength $g_{\rm eff}$ between the Dirac and the scalar field is
determined by the interaction $W'$ between the fermionic and the Rydberg atoms. 
As a result, although the fermionic atoms do not directly interact with each other, 
they acquire an indirect interaction, which is mediated via the dynamical and 
quantized scalar field $\hat\phi$.
For example, if we assume a parabolic potential ${\cal V}(\phi)\propto\phi^2$ 
for the scalar field (i.e., an effective mass term $m_\phi$), this field 
would mediate a one-dimensional Yukawa type interaction 
$\propto\exp\{-m_\phi c_\phi|x-x'|\}$ in the static limit. 
For large masses $m_\phi$, this interaction would become very short range and 
could be approximated by an effective local four-Fermion interaction 
$\propto(\Psi^\dagger\Psi)^2$, similar to the $U$ term in the 
Fermi-Hubbard model~\eqref{Fermi-Hubbard} or the Thirring model discussed in 
Sec.~\ref{Sine-Gordon model}. 

Already this simple example~\eqref{dirac+scalar-field} indicates the complexity 
required for quantum simulators of dynamical gauge fields (e.g., the QCD Lagrangian),  
see, e.g., 
\cite{Yang:2020yer,
Zohar:2011cw,
Zohar:2012ay,
Zohar:2012xf, 
Zohar:2013zla,
Zohar:2015hwa, 
Gonzalez-Cuadra:2017lvz, 
Barbiero:2018wui,
Schweizer:2019lwx},
especially when they should also be coupled to fermionic fields 
(e.g., representing the quarks).
Nevertheless, one can make progress using results of lattice gauge field theory,
e.g., by using the plaquette formulation.  


\newpage
\section{Linear fields}\label{Linear fields}

To start our discussion, let us first focus on linear fields, i.e., quantum fields which obey 
linear evolution equations (to the level of approximation we are considering). 
Still, such fields can lead to highly non-trivial phenomena if we study their evolution under 
the influence of external (i.e., classical) background fields, such as gravitational or 
electromagnetic fields. 

In most of these cases, it is sufficient to consider the simple example of a mass-less and 
minimally coupled scalar field $\phi$ propagating in a curved space-time with metric 
$g^{\mu\nu}$ as described by the Klein-Fock-Gordon equation 
\bea
\label{minimally-coupled-scalar}
\Box\phi
=
\nabla_\mu\nabla^\mu\phi 
=
\frac{1}{\sqrt{-g}}\,\partial_\mu\left(\sqrt{-g}\,g^{\mu\nu}\partial_\nu\phi\right) 
=0
\,,
\ea
where $g$ is the determinant of the metric $g_{\mu\nu}$. 
Later on, we will also study the Dirac field in the presence of an external 
electromagnetic field in Sec.~\ref{Sauter-Schwinger and Breit-Wheeler effect}.
Of course, the phenomena discussed below (such as Hawking radiation or cosmological
particle creation \cite{Birrell:1982ix}) do also occur for other fields,
such as the quantized electromagnetic
(i.e., photon) field $\hat A_\mu$. 

\subsection{Curved space-time analogues (a.k.a.\ analogue gravity)}\label{Analogue gravity}

As we have seen in Sec.~\ref{Scalar field}, one option to simulate a scalar field is to arrange 
atoms with finite-range interactions in an optical lattice.
In the following, let us take a different route and consider an atomic cloud in the form of a 
Bose-Einstein condensate, where we assume only one species of atoms with contact interactions.  
Then, the evolution equation for the scalar (non-relativistic)
field operator $\hat\psi$ (in the Heisenberg picture)
following from the Hamiltonian~\eqref{Hamiltonian-general} reads 
\bea
\label{pre-gp}
i\hbar\,\frac{\partial\hat\psi}{\partial t}
=
\left(-\frac{\hbar^2}{2m}\,\na^2+V+g_s\hat\psi^\dagger\hat\psi\right)\hat\psi
\,,
\ea
where $g_s$ denotes the contact ($s$-wave scattering) interaction strength.
In a Bose-Einstein condensate, almost all atoms occupy the same quantum state as 
described by condensate wave function $\psi_{\rm c}(t,\f{r})$ which can be time-dependent 
in general, see, e.g., \cite{Dalfovo:1999zz}. 
Thus, we may apply the mean-field approximation $\hat\psi(t,\f{r})\approx\psi_{\rm c}(t,\f{r})$
where the many-particle field operator $\hat\psi(t,\f{r})$ is effectively replaced by a 
c-number valued function $\psi_{\rm c}(t,\f{r})$. 
Note that this approximation $\hat\psi(t,\f{r})\approx\psi_{\rm c}(t,\f{r})$ 
is a strongly simplified picture.  
For example, 
\rs{considering}
a state with a definite number of atoms 
\bea
\label{number}
\hat N=\int d^3r\,\hat\psi^\dagger\hat\psi=\int d^3r\,\hat\rho
\,,
\ea
\rs{and taking the quantum expectation value of the field operator $\hat\psi$,
we would find that it vanishes} 
$\langle\hat\psi\rangle=0$ which contradicts the 
mean-field approximation $\hat\psi(t,\f{r})\approx\psi_{\rm c}(t,\f{r})$. 
This contradiction can be repaired by a refined mean-field approximation 
$\hat\psi(t,\f{r})\approx\hat A\hat N^{-1/2}\psi_{\rm c}(t,\f{r})$
where $\hat A^\dagger$ and $\hat A$ are suitable creation and annihilation
operators for the total number of atoms in the cloud 
$\hat N=\hat A^\dagger\hat A$, see, e.g.,
\cite{Gardiner:1997yk,Girardeau:1998,Girardeau:1959zz}.  
More refined expansion schemes such as 
$\hat\psi(t,\f{r})\approx\hat A\hat N^{-1/2}[\psi_{\rm c}(t,\f{r})+\hat\chi(t,\f{r})]$
can also include the quantum or thermal corrections $\hat\chi$ or fluctuations around 
this mean field or even higher-order corrections, see, e.g., \cite{Mean-field-expansion}.  

Inserting the mean-field approximation into the evolution equation~\eqref{pre-gp} 
yields the Gross-Pitaevskii equation 
\bea
\label{Gross-Pitaevskii}
i\hbar\,\frac{\partial\psi_{\rm c}}{\partial t}
=
\left(-\frac{\hbar^2}{2m}\,\na^2+V+g_s|\psi_{\rm c}|^2\right)\psi_{\rm c}
\,.
\ea
To interpret this equation, let us apply the Madelung split of the condensate 
wave function $\psi_{\rm c}$ into the condensate density $\rho_{\rm c}=|\psi_{\rm c}|^2$ 
and a remaining phase (in analogy to the eikonal ansatz)
\bea
\label{Madelung}
\psi_{\rm c}=\sqrt{\rho_{\rm c}}\exp\{iS/\hbar\}
\,.
\ea
After inserting this split~\eqref{Madelung} into the Gross-Pitaevskii 
equation~\eqref{Gross-Pitaevskii}, the real part yields the evolution equation for 
the phase $S$ which is very similar to the Hamilton-Jacobi equation in classical mechanics 
\bea
\label{Hamilton-Jacobi}
\dot S+V+g_s\rho_{\rm c}+\frac{(\na S)^2}{2m}
=
\frac{\hbar^2}{2m}\,\frac{\na^2\sqrt{\rho_{\rm c}}}{\sqrt{\rho_{\rm c}}}
\,,
\ea
apart from the additional ``quantum pressure'' term or ``quantum potential'' 
on the right-hand side.
Originally, the term ``quantum potential'' stems from the Bohm formulation of quantum 
mechanics, but we may just use it to refer to the quantum corrections to the 
classical Hamilton-Jacobi equation~\eqref{Hamilton-Jacobi} on the left-hand side. 

The imaginary part of the Gross-Pitaevskii equation~\eqref{Gross-Pitaevskii} yields the 
equation of continuity $\dot\rho_{\rm c}+\na\cdot(\rho_{\rm c}\vau_{\rm c})=0$ 
with the condensate velocity $\vau_{\rm c}=\na S/m$.
Thus, the phase $S$ is proportional to the velocity potential with 
$\vau_{\rm c}=\na S/m=\na\phi_{\rm c}$ and the above equation~\eqref{Hamilton-Jacobi}
turns into the Bernoulli equation $\dot\phi_{\rm c}+V/m+g_s\rho_{\rm c}/m+\vau_{\rm c}^2/2=0$ 
if we neglect the ``quantum pressure'' term on right-hand side. 
From this Bernoulli equation, we may read off the internal pressure $p=g\rho_{\rm c}^2/(2m)$ 
due to the contact interaction $g_s$ between the atoms.
As a result, the condensate supports sound waves, which we may describe by linearizing 
the Bernoulli and continuity equations around a given background profile 
$\phi_{\rm c}=\phi_{\rm c}^0+\delta\phi_{\rm c}$ and 
$\rho_{\rm c}=\rho_{\rm c}^0+\delta\rho_{\rm c}$. 
Combining these two first-order (in $\partial_t$) equations into one second-order equation, 
we find the wave equation for those sound waves 
\bea 
\label{wave-equation-sound}
\left[
\left(\partial_t+\na\cdot\vau_{\rm c}^0\right) 
\frac{m}{g_s}
\left(\partial_t+\vau_{\rm c}^0\cdot\na\right) 
-
\na\cdot\left(\rho_{\rm c}^0\, 
\na\right) 
\right]\delta\phi_{\rm c}=0
\,,
\ea
where all differential operators act on everything on their right.
Here we may directly read off the
speed of sound
$c_{\rm s}=\sqrt{g_s\rho_{\rm c}^0/m}$, typically on the order of millimeter per second.
As explained in Sec.~\ref{Introduction}, this is the analogue to the speed of light.
Note that this speed of sound can be manipulated by modifying the condensate density 
$\rho_{\rm c}^0$ or the interaction strength $g_s$ 
(e.g., via Feshbach resonances \cite{Chin:2010crf})
and thus may depend on space and time $c_{\rm s}=c_{\rm s}(t,\f{r})$.

Now we may follow the seminal paper by Unruh \cite{Unruh:1980cg} and find that the wave 
equation~\eqref{wave-equation-sound} has exactly the same form as that for a 
mass-less and minimally coupled scalar field in a curved 
space-time~\eqref{minimally-coupled-scalar} 
provided that we insert the effective sonic or acoustic metric 
\bea 
\label{sonic-metric} 
g_{\mu\nu}^{\rm eff}
=
\left(
\frac{\rho_{\rm c}^0}{c_{\rm s}}
\right)^{2/(D-1)}
\left(
\begin{array}{cc}
c_{\rm s}^2 -(\vau_{\rm c}^0)^2 & \vau_{\rm c}^0 \\
\vau_{\rm c}^0 & -\f{1}
\end{array}
\right) 
\,,
\ea
where $D$ is the number of spatial dimensions to be simulated. 
In one spatial dimension, the mass-less and minimally coupled scalar field 
is conformally invariant and thus the identification works -- strictly speaking -- 
only if $\rho_{\rm c}^0/c_{\rm s}$ is constant. 
Note that the above metric~\eqref{sonic-metric} is only defined up to a constant factor. 

This analogy~\eqref{sonic-metric} between sound waves in Bose-Einstein 
condensates~\eqref{wave-equation-sound} on the one hand and mass-less and minimally 
coupled scalar fields in curved space-times~\eqref{minimally-coupled-scalar} on the 
other hand allows us to simulate many effects such as Hawking radiation 
(see the next section) in the laboratory. 
However, as one might already expect from counting the degrees of freedom entering the 
effective metric~\eqref{sonic-metric} -- three for $\vau_{\rm c}^0$, one for $c_{\rm s}$
and possibly one more for the factor in front of the metric~\eqref{sonic-metric} -- 
it is not possible to simulate all possible curved space-times. 
For example, the full Kerr metric of a rotating black hole or the metric describing 
gravitational waves cannot be modelled with~\eqref{sonic-metric}, see also 
\cite{Baak:2023zjf}. 
%
%
Nevertheless, the effective metric~\eqref{sonic-metric} is flexible enough to simulate many 
interesting phenomena -- which will be the subject of the next sections. 

Note that, for predicting quantum effect such as Hawking radiation, the classical equation 
of motion~\eqref{wave-equation-sound} is not sufficient, one should also consider the 
commutation relations of the field operators, see also \cite{Unruh:2003ss}. 
For the original field operators $\hat\psi^\dagger$ and $\hat\psi$ of the atoms,
they read 
\bea
\left[\hat\psi(t,\f{r}),\hat\psi^\dagger(t,\f{r}')\right]=\delta^3(\f{r}-\f{r}')
\;,\quad 
\left[\hat\psi(t,\f{r}),\hat\psi(t,\f{r}')\right]=
\left[\hat\psi^\dagger(t,\f{r}),\hat\psi^\dagger(t,\f{r}')\right]=0
\,.
\ea
Now inserting the simplified mean-field split 
$\hat\psi(t,\f{r})\approx\psi_{\rm c}+\hat\chi(t,\f{r})$, we find the same commutation 
relations for the Bogoliubov-de~Gennes fluctuations $\hat\chi^\dagger$ and $\hat\chi$.
For the more precise ansatz 
$\hat\psi(t,\f{r})\approx\hat A\hat N^{-1/2}[\psi_{\rm c}(t,\f{r})+\hat\chi(t,\f{r})]$,
this remains correct for those Bogoliubov-de~Gennes fluctuations $\hat\chi^\dagger$ and 
$\hat\chi$ which are orthogonal to the spatial mode function $\psi_{\rm c}(t,\f{r})$ 
of the condensate itself. 
Now the linear combinations $\hat\chi^\dagger\pm\hat\chi$ of the 
Bogoliubov-de~Gennes fluctuations yield the density $\delta\hat\rho_{\rm c}$ 
and phase $\delta\hat\phi_{\rm c}$ such that they indeed obey the correct commutation
relations expected for a scalar quantum field -- i.e., the analogy does also hold on the 
quantum level. 

As another important point, sound waves in Bose-Einstein condensates allow us to study 
deviations from the wave equations~\eqref{sonic-metric} and thus~\eqref{minimally-coupled-scalar}
at short length scales, i.e., large wave numbers $k$.
In this regime, the dispersion relation of the sound waves differs from the usual linear 
dispersion $\omega=c_{\rm s}k$ which is valid at low $k$. 
In order to derive these deviations, one could linearize Eq.~\eqref{Hamilton-Jacobi} 
including the ``quantum pressure'' term (together with the continuity equation) or 
insert the mean-field expansion 
$\hat\psi(t,\f{r})\approx\hat A\hat N^{-1/2}[\psi_{\rm c}+\hat\chi(t,\f{r})]$ 
into Eq.~\eqref{pre-gp} which yields the Bogoliubov-de~Gennes equations for the 
fluctuations $\hat\chi$ and $\hat\chi^\dagger$.
Both approaches are equivalent and give the modified dispersion relation 
\bea
\label{BEC-disperion}
\left(\omega-\vau_{\rm c}^0\cdot\f{k}\right)^2
=
c_{\rm s}^2\f{k}^2\left(1+\frac{\xi^2\f{k}^2}{4}\right)
\,,
\ea
where the healing\footnote{\rs{The healing length corresponds to the distance over 
which the condensate wave function (or order parameter) can heal, i.e., recover 
from a local perturbation, such as a vortex.
It can be determined by balancing the kinetic energy density 
$\hbar^2|\na\psi_{\rm c}|^2/(2m)$ and the internal interaction energy density 
$ g_s|\psi_{\rm c}|^4/2$.}} 
length $\xi=1/\sqrt{mg_s\rho_{\rm c}^0}$ can be used to separate the 
two regimes. 
For $\xi^2\f{k}^2\ll1$ we have the usual linear dispersion relation for sound 
$(\omega-\vau_{\rm c}^0\cdot\f{k})^2=c_{\rm s}^2\f{k}^2$ while for 
$\xi^2\f{k}^2\gg1$, the dispersion relation becomes quadratic and approaches the 
free-particle form $(\omega-\vau_{\rm c}^0\cdot\f{k})^2=[\hbar^2\f{k}^2/(2m)]^2$.  
As a result, phonons with large $k$ can move faster than $c_{\rm s}$, which will be 
important for the discussions in the next section. 

A a final remark, the analogy between Eqs.~\eqref{sonic-metric} 
and~\eqref{minimally-coupled-scalar} applies to the {\em kinematics} of a scalar field 
propagating in a given curved space-time, the {\em dynamics} of the space-time itself, 
i.e., the Einstein equations, are not reproduced. 
Instead, the flow background is determined by the Gross-Pitaevskii 
equation~\eqref{Gross-Pitaevskii}. 
Thus, the widely used term ``analogue gravity'' might actually be a bit misleading -- 
which was the reason for the deviating title of this section. 
Still, even though the analogy does ``only'' apply on the kinematical level, 
it allows us to study several phenomena (see the next sections) and thus it has 
developed into its own research field by now, which is still growing. 
Hence, it is impossible to provide a complete list of references here, see, e.g.,
the overview \cite{Barcelo:2005fc} for a recently updated list.

\subsection{Hawking radiation}\label{Hawking radiation}

\begin{figure}[tbp]
\centering
\includegraphics[width=.7\textwidth]{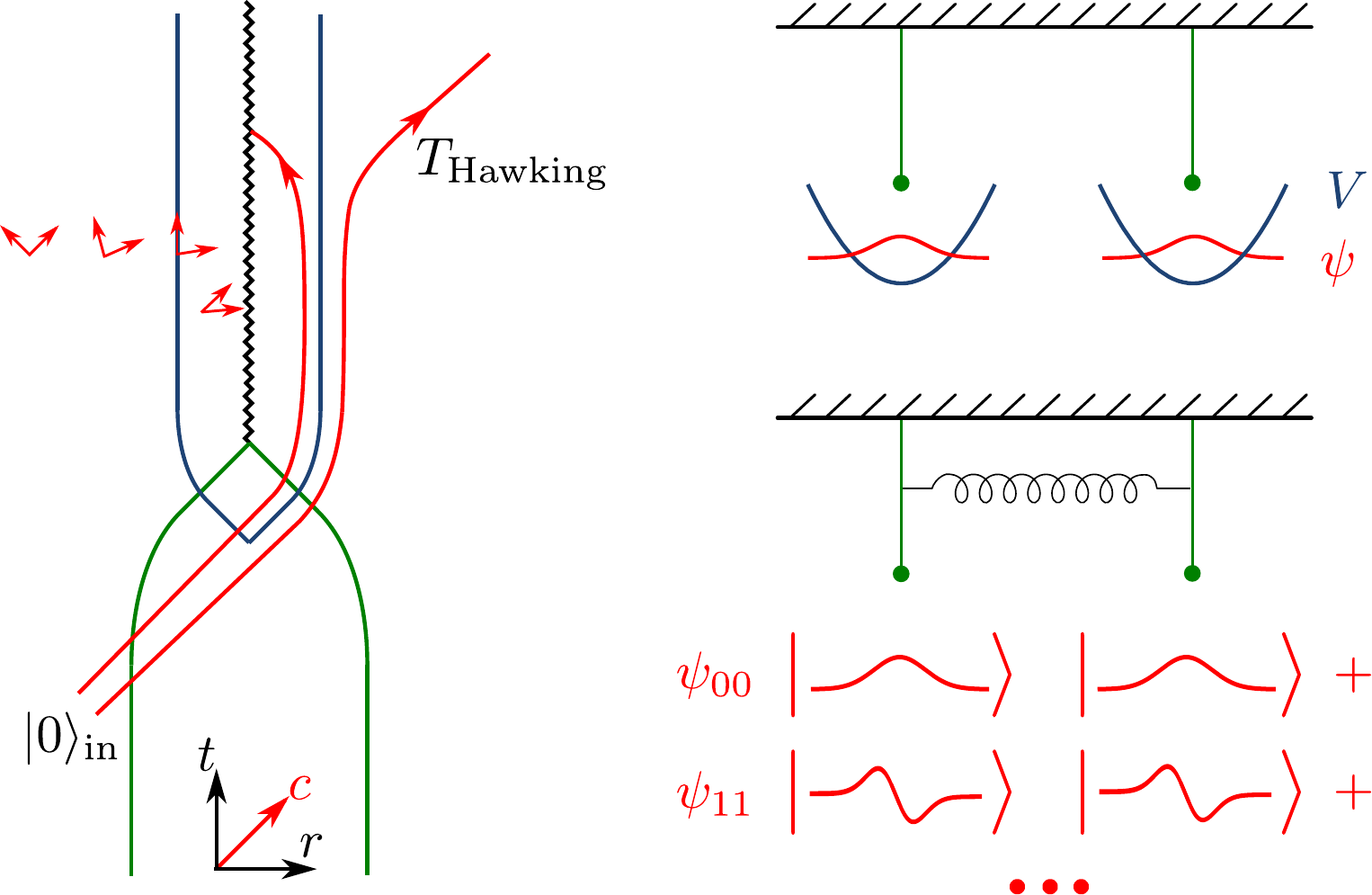}
\caption{Sketch of the space-time diagram of black-hole formation (left) and the 
intuitive picture for Hawking radiation based on the two harmonic oscillators (right).}
\label{fig:hawking} 
\end{figure}

Let us start with a brief explanation of the mechanism behind Hawking radiation
\cite{Hawking:1974rv,Hawking:1975vcx}. 
To this end, we consider the space-time diagram of a star collapsing to a black 
hole\footnote{In contrast to some claims in the literature, space-time curvature 
alone is not sufficient for predicting Hawking radiation. 
As a counter-example, one may consider a neutron star described by a static metric 
which is regular everywhere. 
Then, the ground state (with respect to this static frame) of the quantum field is 
regular and there is no lasting particle creation such as Hawking radiation.
%
%
Thus, the existence of the black-hole horizon is essential for Hawking radiation,
see also \cite{Schutzhold:2024kfg}.},
first on the classical level, see Fig.~\ref{fig:hawking}. 
As usual, time $t$ goes up in this diagram, i.e., on the ordinate, while the abscissa
represents the radial coordinate $r$, assuming spherical symmetry.
Note that due to $r>0$ the left and right-hand sides of the space-time diagram should be 
identified, but we display both of them separately for better visibility. 
The green curves display the surface of the collapsing star which end up in the
central singularity at $r=0$ (black zigzag line). 
The red curves and arrows represent light rays, which are tilted due to the strong 
gravitational 
field\footnote{For simplicity, we neglect back-scattering effects by the effective potential 
(angular momentum and curvature).}.  
The event horizon of the black hole is indicated by the blue curve and marks the 
border between the last light ray which is able to escape to $r\to\infty$ and the 
first light ray which is trapped by the strong gravitational field and eventually 
hits the central singularity at $r=0$.

Now what Hawking did was to consider quantum fields propagating on this classical 
background space-time. 
Strikingly, he found that even when these fields start off in their initial vacuum state
$\ket{0}_{\rm in}$,
the final state of those fields which are able to escape to infinity is not
the vacuum but instead a thermal state with the famous Hawking temperature 
\bea
\label{Hawking}
T_{\rm Hawking}=\frac{\hbar c^3}{8\pi G_{\rm N}k_{\rm B} M}
\,,
\ea
where $M$ is the mass of the black hole. 
As a visualization of the order of magnitude, the typical wavelength of photons 
emitted at such a temperature is comparable to the Schwarzschild radius 
$R_{\rm S}=2G_{\rm N}M/c^2$ which marks the location of the horizon
(blue curve in Fig.~\ref{fig:hawking} left).
Even though this temperature is extremely small for typical astronomical black holes, 
it is considered to be very important for our fundamental understanding of nature 
and has many ramifications such as the concept of black-hole entropy etc. 
\rs{Hawking's discovery \cite{Hawking:1974rv,Hawking:1975vcx} sparked an 
enormous amount of research activity and can be considered one of the most 
influential findings of modern fundamental physics.}

As an intuitive explanation for Hawking radiation, often the picture based on 
particle--anti-particle pairs is used.
In this picture, the vacuum is visualized as containing many ``virtual'' 
particle--anti-particle pairs which are constantly created and annihilated. 
If one of them (the in-falling partner particle, see also \cite{Hotta:2015yla}) 
is then trapped behind the horizon,
the other one is free to escape to $r\to\infty$, i.e., it becomes ``real''   
and constitutes the Hawking radiation. 
This intuitive picture may be appropriate for visualizing the quantum vacuum 
fluctuations and their entanglement at finite spatial distances (which are important 
ingredients for Hawking radiation), but one should be careful and not take it too 
seriously. 
For example, taking this picture too seriously, one has to entertain the possibility 
that a photon and an ''anti-photon'' annihilate in vacuum without creating any new particles. 
Furthermore, the concept of classical trajectories for both, the outgoing Hawking radiation 
and the in-falling partner particle, which are supposed to originate from the same space-time 
point, does not quite work, \rs{see also Fig.~\ref{fig:hawking}}.

Thus, let us discuss a more refined picture for the origin of Hawking radiation based on two 
harmonic oscillators, see Fig.~\ref{fig:hawking}. 
If these two harmonic oscillators are not coupled, the ground state of the total quantum 
system is just the product of the two ground states of the two individual oscillators,
including the ground-state quantum fluctuations (i.e., Gaussian wave functions).  
In case of a coupling\footnote{Using the analogy sketched in Fig.~\ref{fig:scalar}, 
the oscillators could represent the degrees of freedom at two neighboring spatial 
positions. 
The coupling between the two (i.e., the spring) then allows the quantum field to propagate
from one position to another.}
between the two oscillators 
(in Fig.~\ref{fig:hawking} depicted by a spring between them)
the ground state of the total quantum system is an entangled state where the quantum fluctuations 
are correlated (it is a two-mode squeezed state). 
For example, if the position of one oscillator fluctuates a bit to the right, the position 
of the other oscillator is more likely to fluctuate to the right too. 
An analogous correlation can be found in the occupation numbers of the two oscillators 
when considered separately. 
If the occupation number of one oscillator is zero then the occupation number of the
other oscillator vanishes as well. 
However, in the ground state of the total quantum system (including the coupling by the spring)
there is also a finite amplitude for the occupation number of one oscillator being unity -- 
in which case the occupation number of the other oscillator is also unity, and so on. 

Now, if we were to switch off the coupling between the two oscillators slowly, i.e., 
adiabatically, this entangled ground state would evolve into the factorizing ground state 
of the two individual oscillators discussed above. 
However, if the coupling is switched off rapidly, i.e., non-adiabatically 
(say suddenly), the quantum state has not enough time to react and the final state 
retains some or even all of the initial entanglement, i.e., it deviates from the ground state.
In the case of Hawking radiation, this non-adiabatic change arises from the tearing apart 
of the waves at the horizon, where one part of the wave (represented by one oscillator) 
falls into the black hole while the other part of the wave 
(represented by the other oscillator) escapes to infinity. 
This tearing apart of waves as a non-adiabatic process causes a deviation from the ground 
state and its rapidity is set by the surface gravity of the black hole, which in turn 
determines the temperature~\eqref{Hawking}.
Even though the combined quantum state of the two oscillators is a pure state, 
the reduced state of the oscillator which escaped to infinity is a thermal, i.e., mixed state. 
In this picture, the two oscillators initially represent the quantum vacuum fluctuations 
at two different positions (which are also correlated \cite{Schutzhold:2010ig}) 
which then evolve on trajectories 
on either side of the horizon, see Fig.~\ref{fig:hawking}. 
Due to the tearing apart by the gravitational field of the black hole, 
one of them escapes to infinity and represents the outgoing Hawking radiation 
(in a thermal state) while to other one corresponds to the in-falling partner 
\cite{Hotta:2015yla}. 
For real black hole, the correlation between the outgoing Hawking particles 
and their in-falling partners can obviously not be observed, but for the 
black-hole analogues discussed below, this is an important point for the 
experimental observations, see also \cite{Balbinot:2007de,Carusotto:2008ep}.

Note that for the derivation of Hawking radiation, the regularity at the horizon 
is a crucial point. 
While for freely falling observers, the vicinity of the horizon is locally 
indistinguishable from vacuum (i.e., a regular\footnote{This regularity is 
usually described by the Hadamard condition.} state),
static observers far away measure thermal radiation. 
Actually, this argument can even be used to derive Hawking radiation is a rigorous
manner, see, e.g., \cite{Fredenhagen:1989kr}. 
It has been proposed to abandon the regularity at the horizon 
(e.g., in ``firewall'' scenarios \cite{Almheiri:2012rt}) 
in order to address some open questions 
such as the black-hole information puzzle. 
However, by abandoning the regularity at the horizon one also abandons the
main reason for the existence of Hawking radiation in the first place. 
For example, one could consider the quantum field to be in the Boulware state
\cite{Boulware:1974dm} 
which is the local ground state of all static observers outside the black hole. 
This state does not contain Hawking radiation and is a well-defined state 
outside the black hole -- the only problem is that it becomes singular at the 
horizon and thus observers freely falling into the black hole would notice a 
difference there. 
However, if that is acceptable, there would be no need nor reason for Hawking 
radiation. 

As a somewhat related issue, it should also be mentioned here that Hawking's 
derivation \cite{Hawking:1974rv,Hawking:1975vcx} is not without any problems. 
An important point is the so-called trans-Planckian problem 
\cite{Jacobson:1991gr,Unruh:1994je}
which can be illustrated in the following way:
If we consider an outgoing particle of the Hawking radiation far away from the
black hole and trace it back in time, we have to undo the gravitational red-shift.
Thus, its wave packet is squeezed against the horizon (when going back in time)
and the wavelengths become shorter and shorter.
Since the gravitational red-shift is exponential in time $t$ 
(where the exponent is determined by the surface gravity of the black hole),
these wavelengths become smaller than the Planck length 
$\ell_{\rm Planck}=\sqrt{\hbar G_{\rm N}/c^3}\approx1.6\times10^{-35}~\rm m$ 
after a comparably short time. 
At this point, one would probably not trust the approach used by Hawking
based on the propagation of quantum fields on a given classical background 
space-time anymore. 

\subsubsection{Black hole analogues}

\begin{figure}[tbp]
\centering
\includegraphics[width=.7\textwidth]{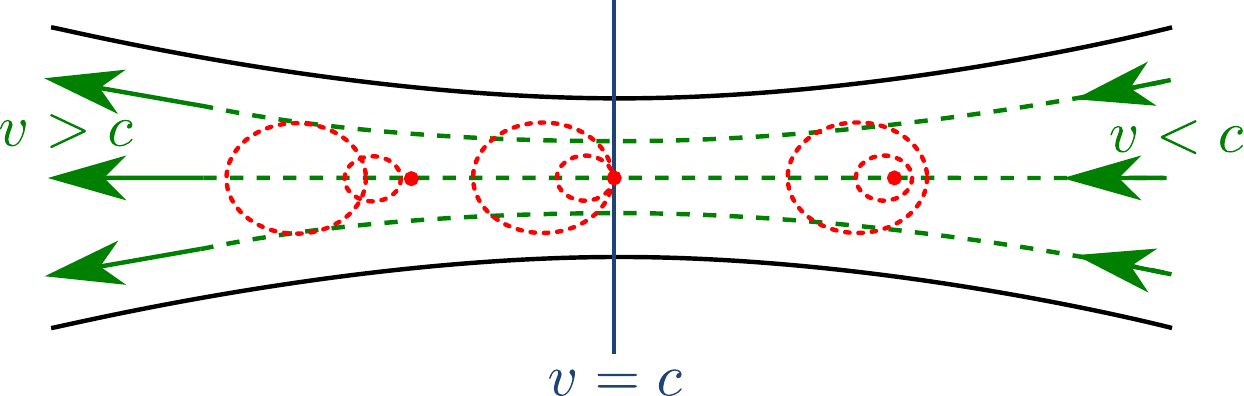}
\caption{De~Laval nozzle with streamlines (green) and sound waves (red). 
The border (blue) between sub-sonic (right) and the super-sonic (left) flow 
marks the analogue of the horizon.
In the super-sonic region (left), sound waves are swept away and form Mach cones.}
\label{fig:laval} 
\end{figure}

Actually, the trans-Planckian problem sketched above was one of the main motivations for Unruh 
\cite{Unruh:1980cg} to consider laboratory analogues for black holes -- 
or, more precisely, for Hawking radiation. 
His main idea is already explained in Sec.~\ref{Analogue gravity}
and can be based on the quantitative analogy between Eqs.~\eqref{sonic-metric} 
and~\eqref{minimally-coupled-scalar}.
To see how the analogue of a black hole and thus Hawking radiation can arise in such a 
laboratory set-up, let us consider a de~Laval nozzle, see Fig.~\ref{fig:laval}.
At the entrance of the nozzle, the flow is sub-sonic $v_{\rm c}<c_{\rm s}$ and thus 
sound waves (which represent the analogue of light rays) can propagate in all directions
(though dragged by the flow).
This region corresponds to the black-hole exterior $r>R_{\rm S}$. 
At the exit of the nozzle, the flow is super-sonic $v_{\rm c}>c_{\rm s}$ such that 
sound waves cannot propagate upstream anymore and are swept away. 
This region corresponds to the black-hole interior $r<R_{\rm S}$. 
At the narrowest point of the 
nozzle\footnote{\rs{These properties follow from the equations of fluid dynamics and do 
not require specific fine tuning. 
Actually, the somewhat counter-intuitive fact that the fluid flow,  
once it exceeds the speed of sound, speeds up further when the nozzle becomes wider again 
is used in rocket engines, for example.}}, 
the flow velocity equals the speed of sound 
$v_{\rm c}=c_{\rm s}$ and thus this point is the analogue of the horizon $r=R_{\rm S}$. 

\begin{figure}[tbp]
\centering
\includegraphics[width=.7\textwidth]{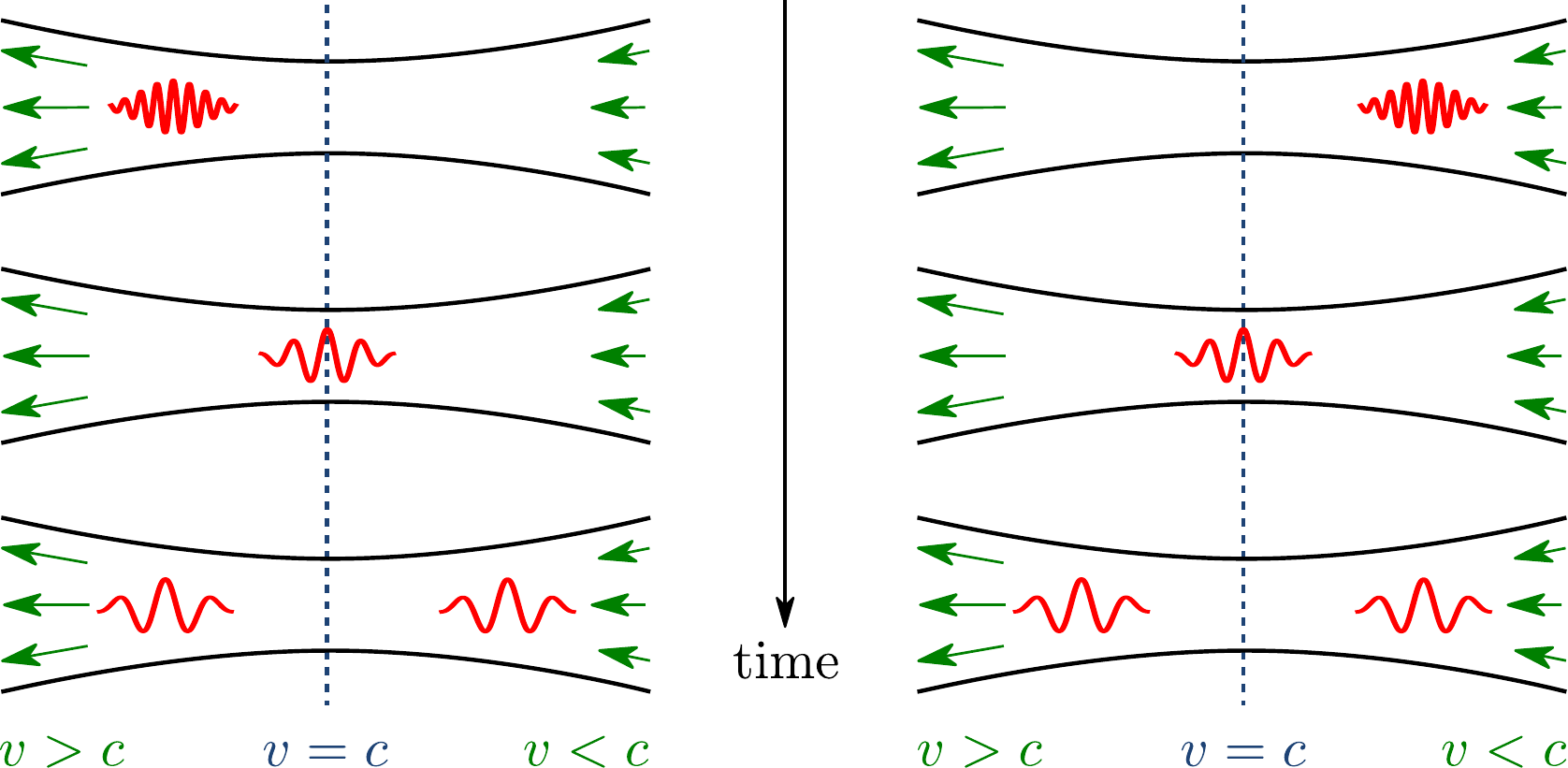}
\caption{Sketch (not to scale) of the evolution (tearing apart) of wave packets giving rise 
to Hawking radiation (and its in-falling partner particles) for a super-sonic (left) 
dispersion relation as in Eq.~\eqref{BEC-disperion} and for a sub-sonic dispersion relation
(right).}
\label{fig:laval-stages} 
\end{figure}

For Bose-Einstein condensates \cite{Garay:1999sk}, we know what happens at short wavelengths,
as explained in Sec.~\ref{Analogue gravity}, and thus we can address the trans-Planckian problem
by means of this analogue.
In this case, the healing length $\xi$ where the dispersion relation~\eqref{BEC-disperion} 
changes from linear to quadratic can be used as an analogue for the Planck 
length\footnote{\rs{For other scenarios, the dispersion relation and thus the analogue 
of the Planck length can be different. 
E.g., dipolar Bose-Einstein condensates can have a dispersion relation with an additional 
``roton'' dip, where the group velocity vanishes, see also \cite{Cha:2016esj,Ribeiro:2022gln}.}}. 
Then the trans-Planckian problem is resolved in the following way:
The initial wave packet has large wave number $k\gg1/\xi$ and is thus faster than $c_{\rm s}$
such that it can propagate upstream, i.e., it originates in the super-sonic region. 
While it propagates upstream, it is stretched (as the analogue of the gravitational red-shift) 
because the flow velocity on its right-hand side is smaller than on its left-hand side,
see Fig.~\ref{fig:laval-stages}. 
Around the time this wave packet reaches the horizon, its wave number is reduced to
$k=\ord(1/\xi)$ and thus its group velocity drops down to the sound speed. 
After that, the part of the wave packet which was already outside the horizon $r>R_{\rm S}$ 
can escape while the other part which was inside the horizon $r<R_{\rm S}$ is swept away. 
Hence, we find basically the same tearing apart process as discussed above. 
Actually, detailed calculations 
(see, e.g., \cite{Corley:1996ar,Unruh:2004zk,Robertson:2012ku} and references therein) 
show that -- if we start with a wave packet in its 
initial ground state in the locally co-moving fluid frame (which is analogous to the 
freely falling frame in gravity) -- the state of the part of the wave packet which 
escapes the nozzle on the right-hand side (i.e., its entrance) is indeed a thermal 
state with the analogue Hawking temperature 
\bea
\label{Hawking-analogue}
T_{\rm Hawking}=\frac{\hbar}{2\pi k_{\rm B}}\,
\left|
\frac{d(v_{\rm c}-c_{\rm s})}{dr}
\right|_{r=R_{\rm S}}
\,,
\ea
provided that the flow profile is smooth enough and that no other excitation mechanisms
(such as friction at the walls of the nozzle) are present. 
As one would expect, this quantity~\eqref{Hawking-analogue} matches precisely the 
surface gravity calculated from the acoustic metric~\eqref{sonic-metric} for this 
effectively one-dimensional set-up. 
Inserting typical values for Bose-Einstein condensates such as length scales in the 
micrometer regime and velocities of order millimeter per second, we find an effective
surface gravity $\kappa$ of the black-hole analogue in the kilo-Hertz range, which 
corresponds to a temperature~\eqref{Hawking-analogue} of a few nano-Kelvin. 
Although this is still a rather low temperature, it is much larger than the real Hawking
temperature~\eqref{Hawking} for typical astronomical black holes and, even more importantly,
it is in the realm of detectability for Bose-Einstein condensates. 

Actually, black-hole analogues in Bose-Einstein condensates have already been 
realized experimentally at the Technion 
\cite{Lahav:2009wx,Steinhauer:2015ava,Steinhauer:2015saa,MunozdeNova:2018fxv,Kolobov:2019qfs} 
and it was even possible to detect signatures of Hawking radiation in those systems. 
Instead of a de~Laval nozzle as in Fig.~\ref{fig:laval}, the effective horizon has been created 
by a step-like potential (generated by a laser beam) which moved through the condensate 
and thereby induced a waterfall like flow profile. 
Now one may transform into the frame co-moving with the potential step 
(i.e., the ``waterfall'') and adjust its sweeping speed such that the flow velocity 
before the potential step (i.e., upstream) is slower than the sound speed $v_{\rm c}<c_{\rm s}$ 
whereas after the potential step (i.e., downstream) it is faster $v_{\rm c}>c_{\rm s}$. 
Then, the resulting trans-sonic flow generates a sonic horizon in analogy to Fig.~\ref{fig:laval}. 
In order to detect the created (analogue) Hawking radiation, the Technion experiment 
\cite{Lahav:2009wx,Steinhauer:2015ava,Steinhauer:2015saa,MunozdeNova:2018fxv,Kolobov:2019qfs}
used the fact that the outgoing Hawking particles (in this case, phonons) and their 
in-falling partners are entangled (as explained above) and measured the density-density 
correlations induced by this entanglement.
To this end, the experiment was repeated many times and the density plots measured after 
each run where used to extract the correlations. 
Even though such a measurement of the density-density correlations alone does not facilitate 
a full quantum state tomography and thus does not allow us to demonstrate entanglement, 
arguments based a some additional assumptions indicate that the state was indeed entangled 
and hence that the Technion experiment indeed measured a quantum effect. 
However, at this point it is fair to say that the results and conclusions are not undisputed
in the community, see, e.g., \cite{Leonhardt:2016qdi}. 
Furthermore, our understanding of the interpretation of the results has been improved since
the first run of experiments. 
%
%
Nevertheless, I think that there is a good chance that the Technion experiment did indeed 
detect Hawking radiation for this black-hole analogue. 
In any case, I would also say that these experiments mark a decisive breakthrough 
in the field. 


\subsubsection{Hawking radiation as tunneling?}

To conclude this section, let us discuss another frequently used interpretation of 
Hawking radiation as a tunneling process, see, e.g., \cite{Parikh:1999mf}. 
Clearly, there are similarities: 
Both are quantum effects which generate a tail at finite spatial distances 
(the evanescent part of the wave function in tunneling and the correlation of the 
vacuum fluctuations of quantum fields) which does not exist classically.
However, when picturing Hawking radiation as tunneling of a particle from the 
interior of the black hole to the exterior \rs{in complete analogy to tunneling 
through a potential barrier in ordinary quantum mechanics,}
one should also be aware of some crucial differences.

Let us start by considering the case of relativistic quantum fields without modifications 
of the dispersion relation -- which is the case discussed in most of the papers devoted to 
the tunneling picture. 
First, tunneling in quantum mechanics usually refers the a case where, in classical mechanics,
there is no trajectory going from a point in front of the potential barrier to a point behind
the barrier if the initial kinetic energy is smaller than the barrier height. 
Yet, in quantum mechanics, the propagator is non-vanishing (though exponentially suppressed). 
In contrast, in relativistic quantum field theory, the causal propagator from any space-time 
point inside the black hole (i.e., behind the horizon) to any space-time point outside 
vanishes exactly\footnote{\rs{Note that the causal propagator is (already in flat space-time) 
different from the two-point (or Wightman) function. 
The latter can be non-zero for space-like separated points, which reflects the quantum vacuum 
correlations, but does not correspond to real propagation.}} 
due to causality 
(without any exponential tails which would correspond to tunneling in \rs{ordinary}
quantum mechanics). 
Second, as a related point, the WKB phase displays a branch-cut singularity 
at the classical turning points for tunneling -- but this singularity is resolved 
when calculating the full wave function (i.e., beyond the WKB approximation) 
in quantum mechanics.
The situation is different for Hawking radiation in relativistic quantum field theory 
since there the stationary mode functions at the horizon acquire a different kind of 
phase singularity which does not go away when considering the full wave 
equation.
%
%
Actually, this phase singularity is closely related to the trans-Planckian problem 
mentioned above. 
Since the tunneling picture of Hawking radiation makes heavy use of these singularities,
see, e.g., \cite{Parikh:1999mf}, this difference is also quite 
important\footnote{Probably one reason for the relative popularity of the tunneling 
picture for Hawking radiation is that it allows a ``short-cut derivation'' of the 
Hawking temperature without a full-fledged quantum-field theory calculation. 
However, in view of the complexity of the problem, it is important to remember that 
such a short cut can \rs{gloss} 
over some of the subtleties of the derivation -- which is \rs{one of the}
the main motivations for this subsection.}.
Third, for tunneling in \rs{ordinary} 
quantum mechanics, the state which comes out at the rear end of 
the barrier depends on the state (energy and amplitude etc.) which was incident on the front 
end (i.e., ''trying'' to tunnel through). 
The outgoing Hawking radiation, on the other hand, is independent of the quantum state inside 
the black hole (at least within relativistic quantum field theory).
If there was such a simple connection between the quantum state inside the black hole and the 
outgoing Hawking radiation as in ordinary tunneling in quantum mechanics, it would have 
profound consequences for our discussions of the black-hole information puzzle.
Fourth, for the Sauter-Schwinger effect discussed in 
Sec.~\ref{Sauter-Schwinger and Breit-Wheeler effect}, the Dirac equation in the presence 
of a constant electric field, for example, can be cast into a tunneling problem with 
a spatially oscillating wave function outside the potential barrier and an exponentially 
decaying wave function in between the classical turning points (i.e., inside the barrier).
%
%
Thus, in the case of the Sauter-Schwinger effect, the interpretation as tunneling 
(from the Dirac sea to the positive continuum) can be justified qualitatively as well as 
quantitatively in this way. 
For Hawking radiation, on the other hand, I am not aware of such a mapping and it is also 
not clear what the two classical turning points should be in this case. 
In view of the points above, such a direct mapping probably does not exist.

\rs{As a disclaimer, it should be stressed here that the above arguments apply to tunneling 
in {\rm real} space, i.e., from the interior to the exterior of the black hole. 
For tunneling in a different space -- such as in configuration space -- these arguments 
should be revisited. 
For example, it could well be that, after we understand Hawking radiation based on a full 
theory of quantum gravity, we shall find that it can be interpreted as tunneling from 
one configuration (of the coupled system space-time plus matter) to another one. 
In this case, the above arguments suggest that this tunneling process occurring in an 
abstract configuration space cannot be simply projected onto tunneling of matter 
from the interior to the exterior of the black hole, i.e., in real space.}

\rs{Also note} 
that, already in the original paper \rs{by Parikh and Wilczek} \cite{Parikh:1999mf}, 
it has been mentioned 
that an alternative interpretation could be to replace the tunneling of the Hawking 
particle from inside to outside by the tunneling of its partner particle into the black hole 
-- thereby liberating the Hawking particle to go off to infinity as a real particle. 
This alternative interpretation \rs{has already been described by Hawking in his seminal 
paper \cite{Hawking:1975vcx} and} evades some of the problems mentioned above.
\rs{It} 
also incorporates the important point of connecting the outgoing Hawking particles with 
their in-falling partners -- which was actually important for the Technion experiment 
mentioned above.
%

Taking into account modifications of the dispersion relation  such as in 
Eq.~\eqref{BEC-disperion}, some of the above arguments 
(e.g., the support of the propagator and the WKB approximation, see, e.g., 
\cite{Schutzhold:2013mba}) have to be modified as well. 
However, since these modifications strongly depend on the actual dispersion relation 
while the outgoing Hawking radiation does not \cite{Unruh:2004zk},
they do not resolve the problems mentioned above. 
Furthermore, as can be seen in Fig.~\ref{fig:laval-stages}, the semi-classical particle 
trajectories change (for early times) when modifying the dispersion relation and thus 
the tunneling picture must be adapted to this case \cite{DelPorro:2024tuw}.



To view the problem form a different angle, one could imagine simulating the Dirac 
equation in a black-hole background via fermionic atoms in an optical lattice along 
the lines sketched in Sec.~\ref{Dirac Hamiltonian}.
Having such a physical realization in mind, one could say that all non-trivial quantum 
evolution is tunneling (between the lattice sites). 
However, these tunneling processes leading to Hawking radiation would be more {\em along} 
the horizon instead of {\em across} the horizon -- and, if one has hopping across the horizon, 
it would be mostly from the outside to the inside\footnote{A lattice typically induces a 
sinusoidal dispersion relation where modes with large $k$ are slower than modes with small $k$.
In such a case, Hawking radiation originates from wave packets outside the horizon, 
see the right panel of Fig.~\ref{fig:laval-stages}.}.

\subsection{Unruh effect}\label{Unruh radiation}

Via the principle of equivalence, the Unruh effect is closely related to Hawking radiation. 
Actually, 
\rs{Unruh discovered it \cite{Unruh:1976db} basically simultaneously with 
the prediction of black-hole evaporation 
\cite{Hawking:1974rv,Hawking:1975vcx} and then realized the relation between the 
two phenomena.}
%
%
However, in order to explain \rs{the Unruh} 
effect, let us first go to flat space-time and consider the usual Minkowski vacuum.
Then, imagine placing a photon detector at rest in this vacuum state. 
Obviously (neglecting possible dark counts) the detector does not click.
If we now imagine moving this detector with a constant 
velocity\footnote{Which is of course smaller than the speed of light, 
see Sec.~\ref{Ginzburg effect}.} then it still does not click because the 
Minkowski vacuum is Lorentz invariant. 
However, the Unruh effect predicts that the situation changes if we accelerate the 
detector. 
In the case of uniform acceleration $a$, the detector responds as if it was immersed
in a thermal bath with the Unruh temperature 
\bea
\label{Unruh-temperature}
T_{\rm Unruh}=\frac{\hbar a}{2\pi k_{\rm B} c} 
\,.
\ea
One way to demonstrate where the thermal response comes from is to start with the 
usual two-point (Wightman) function of a mass-less scalar field in the 
3+1 dimensional Minkoswki vacuum (outside the light cone) 
\bea
\label{Minkowski-vacuum}
\langle\hat\phi(t,\f{r})\hat\phi(t',\f{r}')\rangle
=-\frac{1}{4\pi^2}\,\frac{1}{(t-t')^2-(\f{r}-\f{r}')^2}
\,.
\ea
Now we may evaluate this two-point function along a uniformly accelerated trajectory 
by inserting $t=\sinh(a\tau)/a$ and $z=\cosh(a\tau)/a$ where $\tau$ is the proper time 
of the detector and we have assumed acceleration in $z$-direction with $x=y=0$.
Note that this transformation is related to the Rindler coordinates. 
Insertion into Eq.~\eqref{Minkowski-vacuum} yields 
\bea
\label{accelerated}
\langle\hat\phi(\tau)\hat\phi(\tau')\rangle
=-\frac{a^2}{8\pi^2}\,\frac{1}{\cosh(a[\tau-\tau'])-1}
\,.
\ea
This expression is periodic in imaginary proper time with a period of $2\pi/a$. 
According to the Kubo-Martin-Schwinger (KMS) condition, this periodicity in 
imaginary time translates to thermal behaviour with the temperature being set 
by the period $2\pi/a$ which gives Eq.~\eqref{Unruh-temperature}. 

Now let us come back to the relation to Hawking radiation via the principle of 
equivalence as mentioned above. 
To this end, let us consider a series of static observers at fixed distances to the 
black-hole horizon and compare their findings with that of a freely falling observer 
which starts with small velocity far away from the black hole and then falls in. 
At large distances, the static and the freely falling observers 
\rs{do not differ much and thus} both see basically 
the same Hawking radiation emitted by the black hole. 
However, this agreement changes when the freely falling observer approaches the 
black hole. 
On the one hand -- as we have already discussed above -- the freely falling
observer does not notice anything special at the horizon but locally sees a 
vacuum state. 
On the other hand, in view of the gravitational red-shift of the photons 
propagating away from the black hole, static observers closer to the horizon 
must find a higher temperature than static observers further away for their 
observations to be consistent (in a stationary quantum state).
Thus, the closer \rs{the static observers} 
are to the horizon (from which the gravitational red-shift
becomes infinitely strong) the higher their observed temperatures must be. 
This explains the connection to the Unruh effect since these static observers 
feel a strong gravitational pull and thus must be strongly accelerated 
(in view of the principle of equivalence) in order to retain their fixed 
distance to the horizon. 
In contrast, the freely falling observer corresponds locally to the inertial 
Minkowski observer.
The latter does not detect any photons while the former sees a thermal bath. 

\subsubsection{Analogue of Unruh effect}\label{Unruh analogue}

Due to the factor $\hbar/c$ in Eq.~\eqref{Unruh-temperature}, everyday accelerations
of order $\rm m/s^2$ correspond to extremely low temperatures of order $10^{-21}~\rm K$,
i.e., zepto-Kelvin. 
One way to increase this temperature would be to consider much higher accelerations,
e.g., with electrons in ultra-strong laser fields, see, e.g.,
\cite{Chen:1998kp,Schutzhold:2006gj,Schutzhold:2008zza}. 
For ultra-cold atoms, one usually follows a different route and effectively replaces 
the speed of light $c$ by the speed of sound $c_{\rm s}$ which improves the situation 
by more than eleven orders of magnitude, such that we are approaching the nano-Kelvin 
regime which is more accessible experimentally. 
Thus, the idea would be to consider the phonon field in a approximately homogeneous 
condensate at rest. 
As explained in Sec.~\eqref{Analogue gravity}, this field simulates a mass-less scalar 
quantum field with the speed of light being replaced by the speed of sound. 

Now, another important ingredient for realizing an analogue of the Unruh effect is 
the detector. 
Instead of photons, this detector should detect phonons, i.e., density and phase 
fluctuations of the Bose-Einstein condensate. 
One promising option could be an atomic quantum dot, i.e., a very tight optical potential 
(like optical tweezers) which may hold one or a small number of atoms, see, 
e.g., \cite{Fedichev:2003id,Fedichev:2003dj,Recati:2005,Jaksch:2005}. 
The internal states of this detector could then correspond to the internal states of 
the trapped atoms or their motional states or even to the number of trapped atoms inside 
the atomic quantum dot.
The atomic quantum dot could be coupled to remaining Bose-Einstein condensate via tunneling 
or optical transitions, possibly in connection with Rydberg couplings or dipolar couplings.  

In order to simulate the accelerated detector, one could imagine moving the 
atomic quantum dot through the condensate on a non-inertial trajectory, see, e.g., 
\cite{Retzker:2007vql}.
This would model the coupling of the detector to the field along such a non-inertial 
trajectory.  
However, another important ingredient for the Unruh effect is the relativistic time 
dilatation of the detector (i.e, the difference between the Minkowski time $t$ and 
the proper time $\tau$ of the detector), which is not represented by the 
atomic quantum dot {\em a priori}.
To take this effect into account, one could externally tune the internal eigen-energies 
of the atomic quantum dot such that they represent the proper time $\tau$.
Alternatively, one could replace the uniformly accelerated trajectory by a circular 
motion of the detector, e.g., $x(t)=R\cos(\Omega t)$, $y(t)=R\sin(\Omega t)$ and $z=0$.
In this case, the absolute value of the acceleration would be constant but its direction 
would rotate constantly. 
The two-point function along the detector trajectory can be obtained by inserting 
$x(t)=R\cos(\Omega t)$, $y(t)=R\sin(\Omega t)$ and $z=0$ into Eq.~\eqref{Minkowski-vacuum}. 
The result would be stationary, i.e., only depend on $t-t'$, 
\rs{but this dependence is no longer periodic in imaginary time.}
%
As a consequence, while the detector on a circular trajectory would also experience 
particles, their spectrum would not be exactly thermal anymore. 
On the other hand, one would not have to tune the internal eigen-energies of the atomic 
quantum since the Lorentz factor determining the relativistic time dilatation of the detector 
is constant for the circular motion. 
In principle, this non-thermal excitation spectrum could be measured by moving the 
atomic quantum dot through the condensate on a circular trajectory, see, e.g., 
\cite{Retzker:2007vql}.

However, in order to detect the analogue of the Unruh effect as a quantum phenomenon,
one should be careful to avoid creating classical excitations by stirring the condensate 
with the moving atomic quantum dot -- which then could also excite it. 
This is often a problem when trying to observe quantum effects, as competing classical  
background effects can easily be larger than the usually small quantum phenomena one
is interested in. 
As an idea to avoid these problems, it has been suggested \cite{Gooding:2020scc} to 
replace the atomic quantum dots by laser beams which measure the local density of the 
condensate and move through it on circular orbits, see also 
\cite{Biermann:2020bjh,Bunney:2023vyj}.
By analysing the density fluctuations measured by those laser beams 
(in an interferometric set-up), one could then reconstruct the detector response. 
Using two laser beams, one red and one blue detuned from the resonance frequency of the 
condensate atoms, one can suppress the classical back-reaction of those laser beams onto 
the condensate, i.e., avoid unwanted excitations. 
However, one should still be careful because every measurement perturbs the quantum state 
of the system -- which might in turn react back on the density fluctuations.
This back-reaction effect could be minimized by using weak instead of strong measurements 
of the density fluctuations. 

There have also been proposals which focus on other aspects of the Unruh effect instead of 
the non-inertial trajectories. 
For example, in \cite{Rodriguez-Laguna:2016kri,Kosior:2018vgx}
an optical lattice simulator for the Rindler geometry 
(i.e., the Minkowski metric in terms of the Rindler coordinates) 
has been proposed.
In order to observe particle creation, the idea is to start in the ground state of the 
Minkowski metric and then to suddenly switch (i.e., quench) to the Rindler geometry, 
\rs{see also \cite{Louko:2018pij}.} 
This scenario displays some similarities to the (sudden) collapse to a black hole, 
as depicted in Fig.~\ref{fig:hawking}, and also to the phenomenon of cosmological 
particle creation as discussed in the next section. 
Other proposals focus on the squeezing mechanism because the Minkoswki vacuum is a 
squeezed state in terms of the Rindler creation and annihilation operators
(associated to the uniformly accelerated observers).
However, since squeezing is important for many phenomena (including Hawking radiation 
as discussed in the previous section as well as cosmological particle creation as 
discussed in the next section), this feature alone is probably not sufficient for 
claiming to have a simulator for the Unruh effect. 
As in all of these simulators, it is important to carefully determine which aspects 
of the phenomenon to be simulated are reproduced and which not -- or, equivalently, 
how many and which steps lie between the original phenomenon and the simulator. 




\subsection{Cosmological particle creation}\label{Cosmological particle creation}

Already in the early days of quantum mechanics (long before Hawking and Unruh etc.),  
Schr\"odinger \cite{Schrodinger:1939}
understood that the expansion of the Universe can create particles out of the vacuum, 
see also \cite{Parker:1968mv}. 
In principle, this phenomenon can also be traced back to the tearing apart process as 
already discussed in Sec.~\ref{Hawking radiation} but now this tearing apart is caused 
by the time-dependent expansion of the Universe instead of the stationary gravitational
field of the black hole \rs{(long after the gravitational collapse)}.
However, one may also understand the basic mechanism by means of a single harmonic 
oscillator. 
To show this, let us consider the spatially flat Friedmann-Lemaitre-Robertson-Walker 
metric\footnote{In case of spatial curvature, the spatial line element 
(which reads $d\f{r}^2=dx^2+dy^2+dz^2$ in 3+1 dimensions) would acquire 
additional position dependent metric coefficients, as discussed below.} 
\bea
\label{FLRW}
ds^2=a^2(\eta)\left[d\eta^2-d\f{r}^2\right]=d\tau^2-a^2(\tau)d\f{r}^2
\,,
\ea
where the time-dependent scale factor $a$ corresponds to the cosmic expansion.
Here, we used the representation in terms of the conformal time $\eta$ 
and the proper time $\tau$ which are related by $d\tau=ad\eta$. 
When simulating such an expanding Universe via an analogue system such as a Bose-Einstein
condensate, another important time is the laboratory time $t$ and its relation to 
$\eta$ and $\tau$ depends on the concrete physical realization, which we shall discuss below.

\begin{figure}[tbp]
\centering
\includegraphics[width=.7\textwidth]{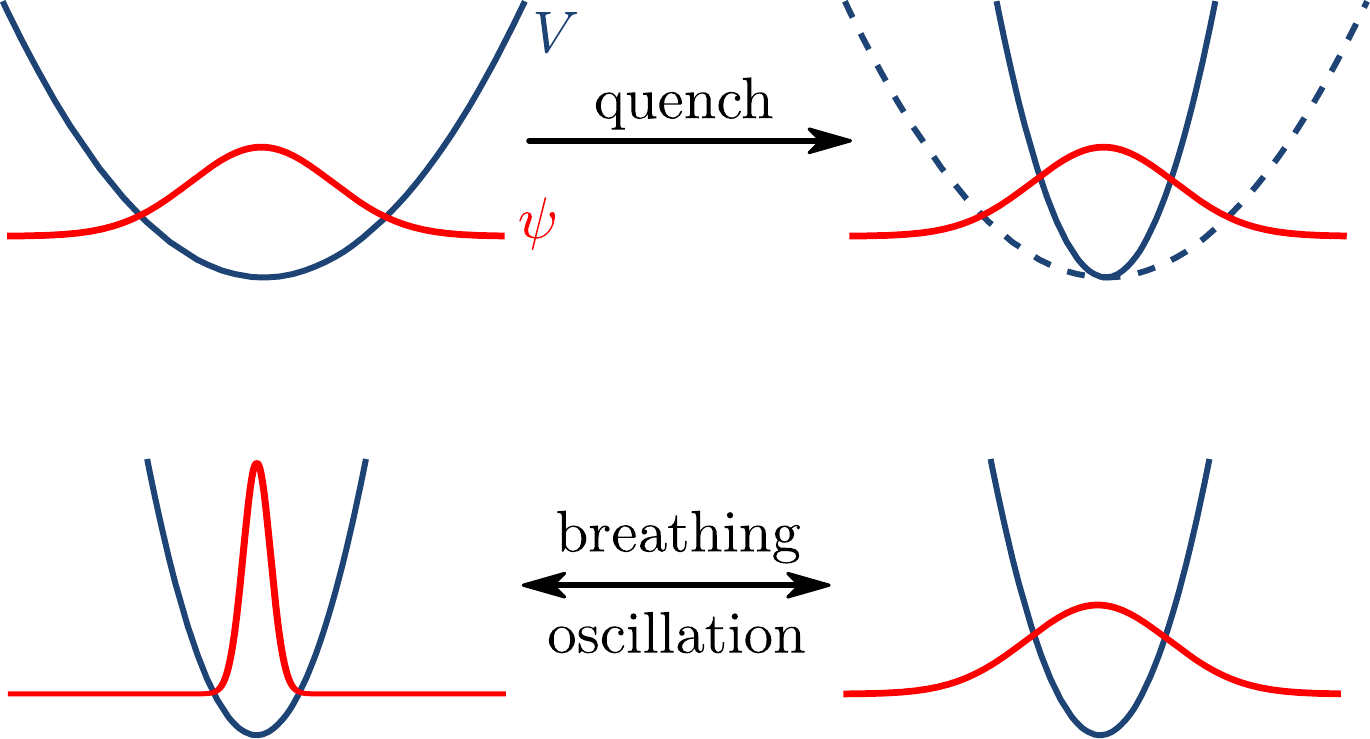}
\caption{Sketch of the main mechanism for cosmological particle creation: 
When the harmonic oscillator potential $V$ (blue curve) is rapidly (e.g., suddenly) switched 
to a new value (top right), the wave function $\psi$ (red curve) has no time to adapt and thus 
becomes a squeezed state. 
After that, the wave function $\psi$ undergoes an oscillatory breathing motion (bottom).}
\label{fig:quench} 
\end{figure}

For later convenience, let us introduce yet another time coordinate defined by 
$dT=a^{-D}d\tau=a^{1-D}d\eta$ depending on the number $D$ of spatial dimensions, 
which we shall refer to as internal oscillator time. 
In terms of this time coordinate, the metric~\eqref{FLRW} reads 
$ds^2=a^{2D}dT^2-a^2d\f{r}^2$. 
As a result, we have $\sqrt{-g}=a^{2D}$ and thus the 
Klein-Fock-Gordon equation~\eqref{minimally-coupled-scalar} 
of a mass-less and minimally coupled scalar field in such a 
curved space-time simplifies to 
\bea
\label{harmonic-oscillator}
\rs{
\left(\frac{d^2}{dT^2}
+
a^{2(D-1)}\f{k}^2\right)\phi_{\fk{k}}=0
} 
\,,
\ea
after a spatial Fourier expansion of the scalar field 
(exploiting the spatial homogeneity). 
Apart from the special case of one spatial dimension $D=1$ 
(where the field is conformally invariant, see also Sec.~\ref{Gibbons-Hawking effect})
this is the equation of a harmonic oscillator with a time-dependent frequency 
$\Omega^2_{\fk{k}}(T)=a^{2(D-1)}(T)\f{k}^2$, see Fig.~\ref{fig:quench}. 

Now, in analogy to the line of arguments in Sec.~\ref{Hawking radiation}, 
if we start in the ground state of this harmonic oscillator and change its 
frequency $\Omega_{\fk{k}}(T)$ slowly enough (i.e., 
$\dot\Omega_{\fk{k}}\ll\Omega^2_{\fk{k}}$ and $\ddot\Omega_{\fk{k}}\ll\Omega^3_{\fk{k}}$ 
etc.) the quantum evolution will be adiabatic, i.e., the quantum state will stay close to 
the instantaneous ground state in view of the adiabatic theorem.
However, if the time-dependence of $\Omega_{\fk{k}}(T)$ is too fast, there will be 
deviations, i.e., the final state will differ from the ground state -- 
it will be a squeezed state.  
As an extreme example, let us consider a sudden change of $\Omega_{\fk{k}}(T)$ which is 
usually referred to as a quench. 
For such a sudden switch, the wave function has no time to change, i.e., 
immediately after the quench it will have the same form as before, see Fig.~\ref{fig:quench}. 
For the initial value $\Omega_{\fk{k}}^{\rm in}$ this wave function represents the 
ground state and thus does not change with time. 
After the quench to $\Omega_{\fk{k}}^{\rm out}\neq\Omega_{\fk{k}}^{\rm in}$, 
however, this wave function has the ``wrong'' width and thus does not represent the 
ground state for the new value of $\Omega_{\fk{k}}^{\rm out}$. 
Instead, it is an excited state 
-- \rs{or, more precisely, a superposition state of the ground state and (even) excited states} 
-- in this case, a squeezed state containing pairs of excitation quanta. 
As a result, it is also no longer a stationary state, but it starts to ``breathe''
by changing its width in an oscillatory manner, see Fig.~\ref{fig:quench}. 

Going back from this mode picture to the particle picture, one finds that the expanding
Universe creates pairs of particles moving in opposite directions (i.e., with opposite
momenta $\f{k}$ and $-\f{k}$), as already expected from momentum conservation. 
Thus, we may also visualize this effect as the tearing apart of the initial quantum
vacuum fluctuations by the cosmic expansion -- in analogy to the picture based on the 
coupled harmonic oscillators in Fig.~\ref{fig:hawking}.
As a result, these particles do also generate a specific imprint in the space-time 
dependent correlation functions (see also \cite{Lang:2018tpj}), quite in analogy to the 
black-hole case, which the Hawking particles and their in-falling partners move away from 
the horizon in opposite directions.

\subsubsection{Analogue Universe in condensate at rest}
\label{Analogue Universe in condensate at rest}

In order to simulate an expanding Universe, let us first consider an effectively two-dimensional
homogeneous condensate at rest with $\vau_{\rm c}^0=0$ and $\rho_{\rm c}^0=\rm const$.
In this case ($D=2$), the effective line element~\eqref{sonic-metric} reads 
\bea
ds^2_{\rm eff}
=
\left(\frac{\rho_{\rm c}^0}{c_{\rm s}(t)}\right)^{2/(D-1)}
\left[c_{\rm s}^2(t)dt^2-d\f{r}^2
\right] 
=
\rho_{\rm c}^0
\left[dt^2-\frac{d\f{r}^2}{c_{\rm s}^2(t)}\right] 
\,.
\ea
We see that (up to the irrelevant constant pre-factor $\rho_{\rm c}^0$) 
the laboratory time $t$ equals the proper time $\tau$ in this case. 
Furthermore, changing the coupling strength $g_s$ and thereby the speed of sound 
$c_{\rm s}$ in a time-dependent manner (e.g., via Feshbach resonances \cite{Chin:2010crf})
models a dynamical scale factor $a(t)$ in Eq.~\eqref{FLRW}.
More specifically, decreasing the speed of sound $c_{\rm s}$ increases the sound 
distance (i.e., acoustic run-time $\Delta t=|\f{r}_1-\f{r}_2|/c_{\rm s}$) between 
two positions $\f{r}_1$ and $\f{r}_2$ and thus corresponds to the expansion of the 
Universe\footnote{More involved schemes open up additional possibilities.
For example, a two-component condensate supports two species of quasi-particles:
In addition to the mode associated with fluctuations of the total density, one has the
mode corresponding to the density differences, which is typically much softer and thus
easier to manipulate and to excite, see, e.g., 
\cite{Fischer:2004bf,Syu:2022cws,Syu:2022ajm,Syu:2024txa}.}.

Such an expanding Universe has been successfully simulated in the Heidelberg experiment
\cite{Viermann:2022wgw}, see also \cite{Chatrchyan:2020cxs,Tolosa-Simeon:2022umw}.
As in the Technion experiment 
\cite{Lahav:2009wx,Steinhauer:2015ava,Steinhauer:2015saa,MunozdeNova:2018fxv,Kolobov:2019qfs} 
aiming at Hawking radiation, the phenomenon of cosmological particle creation has been
detected by measuring the density-density correlations.
In this way, it was also possible to observe how the phonons move away from each other 
(into opposite directions) with the speed of sound after their creation. 
Translating from the particle picture in real space to the mode picture in momentum space 
facilitated the observation of Sakharov oscillations (in analogy to those observed in the
cosmic microwave background), which are associated to the oscillatory
breathing motion sketched in Fig.~\ref{fig:quench} (bottom), see also \cite{Hung:2012nc}. 

Besides the spatially flat metric~\eqref{FLRW} corresponding to an approximately constant 
condensate density, it was also possible to simulate Universes with spatial curvature in the
Heidelberg experiment \cite{Viermann:2022wgw}.
For example, in a parabolic confinement potential, the condensate density decreases when 
going from the centre to the edge of the condensate. 
This implies that the speed of sound shrinks such that the acoustic distance grows -- 
which corresponds to a spatially hyperbolic geometry. 
In a nut-shell, the spatial dependence of the sound speed represents the spatial curvature
of the Universe while the time dependence of the sound speed models the cosmic expansion.

Instead of a monotonically expanding Universe, an oscillating scale factor has been simulated
in the Chicago experiment \cite{Hu:2018psq} by modulating the external magnetic field near a
Feshbach resonance \cite{Chin:2010crf}.
By measuring the number of excited atoms {\em in situ}, it was possible to observe the 
thermal character of the radiation emitted into a certain direction. 
In terms of the coupled harmonic oscillator picture sketched in Fig.~\ref{fig:hawking}, 
the reduced density matrix of one oscillator is thermal, i.e., a mixed state, 
even though the total quantum state of \rs{the two} 
oscillators is a pure (entangled) state. 
Indeed, the coherence associated to this two-mode squeezed state could also be tested in 
the Chicago experiment \cite{Hu:2018psq}.
Note that the title of the publication~\cite{Hu:2018psq} refers to Unruh radiation,
which can probably be explained by the fact that squeezing underlies both 
phenomena, cosmological particle creation and the Unruh effect 
(as well as many others), but the analogy to cosmological particle creation
as considered in this section is apparently closer and more quantitative.

The creation of pairs of particles with opposite momenta has also been observed in the 
Palaiseau experiment \cite{Jaskula:2012ab}.
Instead of varying the interaction strength $g_s$ as in the Heidelberg experiment 
\cite{Viermann:2022wgw} or the Chicago experiment \cite{Hu:2018psq},
here the trap stiffness was modulated -- which changes the 
condensate density in time and thereby the speed of sound, see also the next section.
Besides a ramp of the trap stiffness, it was also modulated in an oscillatory manner in the 
Palaiseau experiment \cite{Jaskula:2012ab}, which would correspond to a Universe with an 
oscillating scale factor (as in the Chicago experiment \cite{Hu:2018psq}). 
%
%
Via parametric resonance, such a modulation predominantly creates specific phononic 
modes whose energy is half that of the trap modulation frequency. 
The phonon pairs created in this way were detected via time-of-flight measurements 
(i.e., in momentum space, see Sec.~\ref{Measurement schemes}) and the correlations
between the two phonons in a pair have also been observed in the 
Palaiseau experiment \cite{Jaskula:2012ab}. 
%
%
Although the title of the publication~\cite{Jaskula:2012ab} refers to the dynamical Casimir 
effect, it can also be viewed as an analogue to cosmological particle creation, see also 
Sec.~\ref{Dynamical Casimir effect}. 

In this context, one should also mention the Boulder experiment \cite{Roberts:2001,Donley:2001} 
where a ``Bose nova'' was created by switching the coupling $g_s$ from a positive value
(corresponding to repulsion between the condensate atoms) to a negative value 
(corresponding to attraction). 
In terms of the speed of sound, this would translate to replacing $c_{\rm s}^2$ by
$-c_{\rm s}^2$.
As a result, modes with sufficiently long wavelengths (above the healing length $\xi$), 
such that the term $c_{\rm s}^2\f{k}^2$ in the dispersion relation~\eqref{BEC-disperion}
dominates in comparison to the term $\f{k}^4/(2m)$, become unstable and start to grow 
exponentially. 
In terms of the effective metric~\eqref{sonic-metric} for $D=2$, for example, 
this would correspond to changing the signature form $(+,-,-)$ to $(+,+,+)$,
i.e., going from a hyperbolic to an elliptic case. 
Fortunately, this probably does not happen in our Universe. 












\subsubsection{Expanding condensates}
\label{Expanding condensates}

Apart from a homogeneous condensate at rest with a time-dependent speed of sound, there are 
also other options for simulating an expanding 
Universe\footnote{\rs{Besides bosons, one could also employ fermionic atoms, 
see also \cite{Fedichev:2003bv}.}}.
As one might already intuitively expect 
(see also the Palaiseau experiment \cite{Jaskula:2012ab}), 
letting the condensate expand has an analogous effect. 
To see this, let us go back to the metric~\eqref{FLRW} and transform it to  
co-moving coordinates $\f{R}=a(\tau)\f{r}$ such that 
$d\f{R}=a(\tau)d\f{r}+\dot a(\tau)\f{r}d\tau$.  
Then, insertion into the metric~\eqref{FLRW} yields 
\bea
\label{co-moving}
ds^2
=
d\tau^2-\left(d\f{R}-\f{R}\,\frac{\dot a}{a}\,d\tau\right)^2
=
\left(1-\f{R}^2\frac{\dot a^2}{a^2}\right)d\tau^2
+2\,\frac{\dot a}{a}\,\f{R}\cdot d\f{R}\,d\tau
-d\f{R}^2
\,.
\ea
For these co-moving coordinates $\f{R}$, the scale factor in front of the spatial 
line element $d\f{R}^2$ disappears at the expense of additional terms in the 
temporal and mixed components containing the Hubble velocity $\f{R}\dot a/a$. 
Thus, apart from the pre-factor, we find a form quite reminiscent of the acoustic
metric~\eqref{sonic-metric} where the condensate velocity $\vau_{\rm c}$ represents 
the Hubble velocity $\f{R}\dot a/a$. 
Note, however, that the pre-factor must be considered as well because the expansion 
of the condensate naturally leads to a decreasing density -- which, in turn, reduces 
the speed of sound (unless this is counterbalanced by a corresponding growth of 
the coupling $g_s$).
On the other hand, as explained above, a shrinking speed of sound does also model
the expansion of the Universe such that one can consider both effects together,
see also \cite{Uhlmann:2005hf,Schutzhold:2008zzb}.

In summary, different laboratory scenarios -- condensates at rest with time-dependent 
speed of sound or expanding condensates -- can represent the same expanding Universe.
One of the main differences between the two scenarios is the role of the laboratory time $t$
and how it compares to the conformal time $\eta$, the proper time $\tau$ or the oscillator
time $T$.
As another important difference, the frequency or wave number scale where the dispersion
relation~\eqref{BEC-disperion} changes from linear to quadratic (i.e., the analogue of the 
Planck scale) behaves differently in the two scenarios.
In view of those differences, one scenario might be better suited for observing a specific 
phenomenon of an expanding Universe than the other one -- while it could be the other way 
around for another specific phenomenon. 

An expanding ring-shaped Bose-Einstein condensate has been realized in the 
Maryland experiment \cite{Eckel:2017uqx,Banik:2021xjn}.
Since the wave-length of the phonons expands with the condensate, such a set-up is well 
suited for observing the red-shift of the field modes within an expanding Universe. 
In the Maryland experiment \cite{Eckel:2017uqx,Banik:2021xjn}, this has been achieved by 
imprinting a sound wave (i.e., a classical excitation) in the initial state of the condensate. 
As a related but more subtle phenomenon, the Maryland experiment 
\cite{Eckel:2017uqx,Banik:2021xjn}
did also find evidence for Hubble friction, see also \cite{Eckel:2020qee}. 
To explain this phenomenon, let us consider the equation of 
motion~\eqref{minimally-coupled-scalar} for a mass-less scalar field in the 
Friedmann-Lemaitre-Robertson-Walker metric~\eqref{FLRW} in 3+1 dimensions 
in terms of the proper time $\tau$ 
\bea 
\label{Hubble-friction}
\left(
\frac{d^2}{d\tau^2}+\frac{3\dot a}{a}\,\frac{d}{d\tau}+\frac{\f{k}^2}{a^2}
\right) 
\phi_{\fk{k}}
=0\,,
\ea
again after a spatial Fourier transformation.
If we now insert the scale factor $a(\tau)$ of the de~Sitter metric 
$a(\tau)=\exp\{H\tau\}$ with the Hubble constant $H$, we see that the 
above equation looks like that of a harmonic oscillator with a constant 
damping term $\propto H$ and a potential term 
(i.e., spring strength) which is exponentially decaying in time
$\propto\exp\{-2H\tau\}$.
This exponential decay is a consequence of the red-shift of the modes due to 
the cosmic expansion. 
As a result of Eq.~\eqref{Hubble-friction}, all modes go through the same 
sequence (though at different times):
Initially $\f{k}^2/a^2\gg H^2$, they are in the under-damped regime and 
oscillate. 
Once $a$ has grown sufficiently, however, they reach the over-damped regime 
and freeze in to an asymptotically constant (and typically non-zero) value.
This transition from oscillation to freezing is an important mechanism for 
amplifying the initial quantum vacuum fluctuations of the inflaton field during 
cosmic inflation, which are supposed to the responsible for the seeds of 
structure formation in our Universe.

After the initial period of linear evolution, the Maryland experiment 
\cite{Eckel:2017uqx,Banik:2021xjn} was also able to observe the subsequent non-linear
dynamics, which are qualitatively analogous to the pre-heating and re-heating 
periods in the early Universe after the end of inflation, 
\rs{see also \cite{Wang:2023wld}.} 









\subsection{Gibbons-Hawking effect}\label{Gibbons-Hawking effect}

The Gibbons-Hawking effect \cite{Gibbons:1977mu}
is sometimes mixed up with the phenomenon of cosmological 
particle creation, but it is actually a different mechanism.
Even though both can occur in an expanding Universe, 
cosmological particle creation refers to the generation of particle pairs as lasting 
excitations -- whereas the Gibbons-Hawking effect is associated to the response of a 
detector in such a curved space-time (analogous to the Unruh effect in 
Sec.~\ref{Unruh radiation}). 
In order to clearly separate the two concepts, let us study a scenario in which there 
is no cosmological particle creation.  

For simplicity -- and because it will be relevant for the experimental proposal 
discussed below -- we consider the mass-less and minimally coupled scalar 
field~\eqref{minimally-coupled-scalar} within a spatially flat 
Friedmann-Lemaitre-Robertson-Walker metric~\eqref{FLRW} in 1+1 dimensions. 
In this case, the internal oscillator time $T$ coincides with the conformal time $\eta$ 
and thus the evolution equation~\eqref{harmonic-oscillator} becomes independent of scale 
factor -- which reflects the conformal invariance of the field in this case. 
As a result, the solutions of the wave equation~\eqref{minimally-coupled-scalar} 
have exactly the same form $\exp\{-i\omega\eta+ikx\}$ as in flat space-time 
when expressed in terms of $\eta$. 
Thus, the same applies to the two-point function in the conformal vacuum state 
which reads 
\bea
\label{conformal-vacuum}
\langle\hat\phi(\eta,x)\hat\phi(\eta',x')\rangle
=
-\frac{1}{4\pi}\,\ln\left[(\eta-\eta')^2-(x-x')^2\right]
\,,
\ea
up to an 
undetermined additive term which is spatially constant and 
related to the infra-red divergence of the mass-less and minimally coupled scalar 
field in 1+1 dimensions. 
Because the time-dependent scale factor $a(\eta)$ does neither show up in this expression 
nor in the evolution equation~\eqref{harmonic-oscillator}, there is no cosmological 
particle creation in this case, which becomes also evident by a direct comparison 
to Sec.~\ref{Cosmological particle creation}. 

However, that fact that the two-point function~\eqref{conformal-vacuum} has the same form 
as in the usual Minkoswki vacuum does {\em not} mean that a particle detector would not click. 
The two-point function~\eqref{conformal-vacuum} is expressed in conformal time $\eta$ 
whereas the internal dynamics of a detector at rest would be determined by the proper 
time $\tau$, and they are different in an expanding Universe. 
To simplify the analysis, let us consider the de~Sitter space-time with 
$a(\tau)=\exp\{H\tau\}$ where $H$ is the Hubble 
constant\footnote{In contrast to the general Hubble parameter $H(\tau)=\dot a/a$ 
it is really a constant here.}. 
In this case, the conformal time is given by $\eta=-\exp\{-H\tau\}/H$ 
and runs from $-\infty$ to zero. 
Inserting this transformation in the two-point function~\eqref{conformal-vacuum}
we find that it acquires a more involved dependence on $\tau$ and does no longer 
coincide with the vacuum behaviour which would go as 
$\ln[(\tau-\tau')^2-(x-x')^2]$. 
Instead, we again find periodicity along the imaginary time axis, which is related to 
thermal behaviour via the Kubo-Martin-Schwinger (KMS) condition 
already discussed
in Sec.~\ref{Unruh radiation} and allows us to read off the Gibbons-Hawking temperature 
\bea
\label{Gibbons-Hawking-temperature}
T_{\rm GH}=\frac{\hbar H}{2\pi k_{\rm B}}
\,.
\ea
In analogy to the Unruh effect, the detector experiences the conformal vacuum 
state~\eqref{conformal-vacuum} as a thermal state with the Gibbons-Hawking 
temperature~\eqref{Gibbons-Hawking-temperature}. 
Note that all three temperatures, the Hawking temperature~\eqref{Hawking}, 
the Unruh temperature~\eqref{Unruh-temperature}, as well as the 
Gibbons-Hawking temperature~\eqref{Gibbons-Hawking-temperature}, 
have exactly the same form -- one just has to replace the surface gravity $\kappa$ 
of the black hole by the acceleration $a$ or the Hubble constant $H$, respectively. 

Note that, strictly speaking, concluding thermal behaviour would also require showing 
that the detector response is stationary. 
Due to the aforementioned infra-red divergence, this is a bit more involved in this case.
On the other hand, in view of the time-dependent scale factor $a(\tau)$, one could wonder 
whether such an expanding space-time can actually lead to a stationary response at all. 
This objection can be resolved by transforming to the co-moving coordinates~\eqref{co-moving}
after which the metric becomes stationary for the de~Sitter Universe. 
By yet another coordinate transformation of the time variable, one can get rid of the 
mixed components and cast this stationary metric into a static form 
\bea
\label{static-de-Sitter} 
ds^2=\left(1-H^2r^2\right)dt^2-\frac{dr^2}{1-H^2r^2}
\,,
\ea
at the expense of introducing coordinate singularities at the de~Sitter horizon 
\rs{$r=\pm1/H$.} 
In 3+1 dimensions, we would find the same form with an additional $r^2d\Omega^2$ 
for the angular part. 

\subsubsection{Analogue of Gibbons-Hawking effect in expanding condensate}
\label{Gibbons-Hawking analogue}

In view of the coordinate singularities at the de~Sitter horizon $r=1/H$, 
the static form~\eqref{static-de-Sitter} is not very suitable for the experimental
realization as an analogue model, so let us go back to the original form~\eqref{co-moving}
corresponding to an expanding condensate. 
Indeed, it has been proposed to observe the analogue of the Gibbons-Hawking effect in an 
expanding cigar-shaped (i.e., effectively 1+1 dimensional) Bose-Einstein condensate
\cite{Fedichev:2003id}. 
Similar to Sec.~\ref{Unruh analogue}, the idea is to realize the detector by an atomic 
quantum dot at the centre of the condensate. 
Assuming that we have two species of atoms (e.g., different hyper-fine levels), 
the condensate is supposed to be formed by one species $a$ while the tightly confining 
atomic quantum dot potential traps atoms of the other species $b$. 
Then, laser-driven transitions between the two states $a$ and $b$ with adjustable Rabi 
frequency and detuning can ensure that the internal time of the detector indeed 
corresponds to the proper time $\tau$ and that it is coupled to the phonon field 
of the expanding condensate \cite{Fedichev:2003id}. 

%

\subsection{Ginzburg effect}\label{Ginzburg effect}

In Sec.~\ref{Unruh radiation}, we stated that a detector moving through the Minkowski vacuum 
with a constant velocity does not click because it can be Lorentz transformed to a detector 
at rest. 
Strictly speaking, this is only true for sub-luminal detector velocities. 
Trajectories with super-luminal velocities always stay super-luminal, i.e., space-like,  
under (proper) Lorentz transformations. 
Indeed, if a detector would move through the Minkowski vacuum with a constant velocity 
that is larger than the speed of light, it would experience the quantum vacuum fluctuations 
as real excitations and thus it would click. 
In contrast to the Unruh effect, the spectrum would not be thermal -- but still the 
detector response would be non-zero.

This phenomenon is referred to as the Ginzburg effect 
\cite{Ginzburg:1986,Ginzburg:1996zz}
or Ginzburg radiation or the 
anomalous Doppler effect, even though the latter is also used to describing frequency 
inversion on the purely classical level. 
Note that -- in contrast to Cherenkov radiation for a charge moving with
super-luminal  velocity --
the Ginzburg effect is a purely quantum phenomenon which is also related to 
quantum friction. 
Obviously, the laws of special relativity prohibit observing this effect with 
real detectors in the real Minkowski vacuum, but one could try to realize an 
analogue with ultra-fast atoms moving through a dielectric medium with a reduced 
speed of light, for example, see, e.g., \cite{Lang:2021rez}. 

On the other hand, considering the phonon field in Bose-Einstein condensates, 
the comparably slow speed of sound (of order mm/s) allows us to realize an analogue 
of the Ginzburg effect at much smaller detector velocities, 
as proposed in \cite{Marino:2016nof}. 
In analogy to our discussion of the Unruh effect in Sec.~\ref{Unruh analogue} 
and the Gibbons-Hawking effect in the previous section, the detector could be realized
by an atomic quantum dot, but now moving through 
the condensate with a constant velocity above the sound speed. 
In the proposal \cite{Marino:2016nof}, the idea is to have one ``impurity'' atom 
inside the atomic quantum dot which has two internal levels which couple differently 
to the density of the surrounding condensate. 
In contrast to the simulator for the Gibbons-Hawking effect in the previous section,
there is no tunnel coupling between the atom in the condensate and the ``impurity'' 
atom inside the atomic quantum dot, but only a collisional coupling.  
The coupling of this ``impurity'' atom to the phonon modes of the condensate is then 
analogous to the usual dipole coupling of an atom to the electromagnetic field -- 
which allows us to simulate phenomena such as the Casimir effect
(see also Sec.~\ref{Dynamical Casimir effect}) or the Ginzburg effect
\cite{Marino:2016nof}.

Nevertheless, one should also keep in mind that basically the same problems arise as 
in the Unruh effect since one should also take into account the disturbance caused by 
the  ``impurity'' atom on the classical level. 
Actually, these problems might even become more pronounced here, since its 
super-sonic velocity opens up more classical excitation channels
(think of the Landau criterion etc.).





\subsection{Super-radiance and Penrose process}\label{Super-radiance and Penrose process}

In order to avoid confusion, this section should start with a disclaimer:
The term ``super-radiance'' is used in different contexts in physics and 
thus has no unique meaning. 
One important example is Dicke super-radiance \cite{Dicke:1954zz}
which refers to the amplification of 
spontaneous emission from an ensemble of atoms due to the constructive interference 
of their emission amplitudes, see also \cite{tenBrinke:2015,tenBrinke:2016}, 
That is not the subject of this section.
To explain what is meant by ``super-radiance'' here, let us write down the effective 
Lagrangian ${\cal L}_{\rm eff}$ for the phonon field $\delta\phi_{\rm c}$ in a given 
background flow determined by $\vau_{\rm c}^0$ and $\rho_{\rm c}^0$ such that the 
Euler-Lagrange equation yields the wave equation~\eqref{wave-equation-sound}
\bea
\label{effective-Lagrangian-sound}  
{\cal L}_{\rm eff}
=
\frac{1}{2}
\left[
\frac{m}{g_s}
\left(\partial_t\delta\phi_{\rm c}+\vau_{\rm c}^0\cdot\na\delta\phi_{\rm c}\right)^2
-\rho_{\rm c}^0\left(\na\delta\phi_{\rm c}\right)^2
\right] 
\,.
\ea
This allows us to derive the canonical field momentum density 
$\Pi=\partial{\cal L}_{\rm eff}/\partial\delta\dot\phi_{\rm c}$
and the effective Hamiltonian density\footnote{Note that, for deriving the Hamilton 
equations, one should write down the Hamiltonian density as a function of 
$\na\delta\phi_{\rm c}$ and $\Pi$ instead of $\partial_t\delta\phi_{\rm c}$. 
However, for our purposes, the form~\eqref{effective-Hamiltonian-sound} is sufficient.} 
\bea
\label{effective-Hamiltonian-sound}
{\cal H}_{\rm eff}
=
\Pi\delta\dot\phi_{\rm c}-{\cal L}_{\rm eff}
=
\frac{1}{2}
\left[
\frac{m}{g_s}
\left(\partial_t\delta\phi_{\rm c}\right)^2
+\rho_{\rm c}^0\left(\na\delta\phi_{\rm c}\right)^2
-\frac{m}{g_s}
\left(\vau_{\rm c}^0\cdot\na\delta\phi_{\rm c}\right)^2
\right] 
\,.
\ea
For stationary background flows $\vau_{\rm c}^0=\vau_{\rm c}^0(\f{r})$
and $\rho_{\rm c}^0=\rho_{\rm c}^0(\f{r})$, the associated effective energy 
$E_{\rm eff}=\int d^3r\,{\cal H}_{\rm eff}$ is conserved for all solutions 
$\delta\phi_{\rm c}(t,\f{r})$ of the wave equation~\eqref{wave-equation-sound}.
Now the major point here is the following:
Comparison of the last two terms in the square brackets in
Eq.~\eqref{effective-Hamiltonian-sound} shows that the effective energy density
could become negative for supersonic flow
$|\vau_{\rm c}^0|>c_{\rm s}=\sqrt{\rho_{\rm c}^0g_s/m}$. 
Actually, this is related to the energy balance in Hawking radiation because 
the in-falling partner particles carry negative effective energy $E_{\rm eff}$ 
(from the point of view of an outside observer).

For one-dimensional $\vau_{\rm c}^0=v_{\rm c}^0(x)\f{e}_x$ or radial flow profiles 
$\vau_{\rm c}^0=v_{\rm c}^0(r)\f{e}_r$, the point where the flow becomes 
supersonic is the horizon of the black-hole analogue. 
For more complex flow profiles, however, this coincidence is no longer granted. 
As a simple example, let us consider the draining vortex or draining ``bathtub'' flow 
(see, e.g., \cite{Schutzhold:2002rf})
in 2+1 dimensions 
\bea
\label{draining vortex flow}
\vau_{\rm c}^0=-\frac{C}{r}\,\f{e}_r+\frac{L}{r}\,\f{e}_\varphi
\,,
\ea 
where $C$ and $L$ are constant. 
Apart from the singularity at $r=0$, this flow is locally irrotational
$\na\times\vau_{\rm c}^0=0$. 
The effective horizon (i.e., point of no return) is then the radius at which the 
radial flow velocity $|\vau_{\rm c}^0\cdot\f{e}_r|=C/r$ equals the speed of sound.
However, the radius at which the total flow velocity $|\vau_{\rm c}^0|$ equals
the speed of sound,
i.e., where the effective energy can become negative, is larger. 
In analogy to rotating black holes, this radius marks the beginning of the ergo-region. 
In this region, sound waves cannot propagate into all directions anymore, 
but they could still escape (unless they are already beyond the horizon). 

Now one could imagine the following situation:
A wave packet $\delta\phi_{\rm c}(t,\f{r})$ approaches the draining vortex 
flow~\eqref{draining vortex flow} from the outside and enters the ergo-region. 
There, it is torn apart by the strong inhomogeneous flow 
profile~\eqref{draining vortex flow} such that part of it falls into the horizon 
and the other part escapes to the outside again.
Then, for certain parameters (azimuthal number and frequency) of the initial
wave packet $\delta\phi_{\rm c}(t,\f{r})$, it is possible that the part 
falling into the horizon has a negative effective energy $E_{\rm eff}$. 
Since the effective energy $E_{\rm eff}$ of the total solution 
$\delta\phi_{\rm c}(t,\f{r})$ is conserved, this means that the 
effective energy $E_{\rm eff}$ of the outgoing wave packet is higher 
than that of the initial wave packet. 
This amplification process is also called super-radiance -- 
and that is the denomination we use here.

So far, the phenomenon of super-radiance discussed above refers to purely classical 
wave dynamics. 
Actually, it even has an analogue in terms of classical particles, which is the 
Penrose process \cite{Penrose:1969pc} and allows us to extract energy from a rotating 
black hole.
Furthermore, the phenomenon of super-radiance can be generalized to other scenarios
such as a rotating body \cite{Zeldovich:1971}.  
Generally speaking, super-radiance requires an effective energy which can become 
negative, some sort of scattering or tearing apart process and some mechanism to 
absorb the parts with negative energy. 
For surface waves \cite{Schutzhold:2002rf}, this phenomenon has actually been observed 
for such a vortex flow~\eqref{draining vortex flow} in the Nottingham experiment 
\cite{Torres:2016iee}. 
Replacing the water as used in the Nottingham experiment \cite{Torres:2016iee} 
by a ``photon super-fluid'', an analogous effect has been observed in the  
Glasgow experiment \cite{Braidotti:2021nhw}, see also 
\cite{Vocke:2017tif,Marino:2019flp,Hod:2021qem,Ciszak:2021xlw} and 
\cite{Braidotti:2024} as well as \cite{Boulier:2020}. 

On the other hand, as one should already expect from our discussion of Hawking radiation,
this classical wave phenomenon has also a quantum counterpart. 
When the draining vortex flow~\eqref{draining vortex flow} takes an initial
(classical) wave packet and returns it with a higher energy, i.e., amplitude,
it acts as an amplifier.
Since the underlying evolution equation~\eqref{wave-equation-sound} is linear,
we can transfer the results to the quantum regime, where this amplification process 
translates into a squeezing operation. 
Thus, such a quantum amplifier takes the initial quantum vacuum state and returns 
a squeezed, i.e., excited state, where the emitted particles are again entangled with 
their in-falling partners, see also \cite{Unruh:2011vm}. 

\subsubsection{Quantum vortices in Bose-Einstein condensates}
\label{Vortices in Bose-Einstein condensates}

Actually, flow profiles as in Eq.~\eqref{draining vortex flow}, but with $C=0$,  
are naturally realized in Bose-Einstein condensates (and other super-fluids) 
in the form of vortices. 
In this case, the Bose-Einstein condensate or super-fluid ``drills a hole in itself'' 
and changes its phase by an integer multiple (which is called the winding number) 
of $2\pi$ when going around this hole (the vortex core) because it must end up with the 
same wave function. 
As a result, the circulation is quantized 
\bea
\oint d\f{r}\cdot\vau_{\rm c}\in\frac{2\pi\hbar}{m}\,{\mathbb Z}
\,,
\ea
such that we have $L\in{\mathbb N}\hbar/m$ in Eq.~\eqref{draining vortex flow}. 
In this case, we do not have a horizon (because $C=0$), but we find an ergo-region -- 
where one would expect potentially negative effective energies. 
However, if we consider a singly quantized vortex with $L=\pm\hbar/m$ and calculate 
the radius of the ergo-region (i.e., the value of $r$ where 
$|\vau_{\rm c}^0|=c_{\rm s}=\sqrt{\rho_{\rm c}^0g_s/m}$) we find that it coincides 
with the healing length $\xi$ as discussed in Sec.~\ref{Analogue gravity} after 
Eq.~\eqref{BEC-disperion}, i.e., the length scale below which the effective 
description~\eqref{effective-Lagrangian-sound} breaks down. 
This is perhaps no too surprising since we know that singly quantized vortices 
are stable to linear perturbations. 

If we consider higher winding numbers, on the other hand, the ergo-region becomes 
larger and thus eventually overlaps with the region where the effective 
description~\eqref{effective-Lagrangian-sound} provides a good approximation. 
Hence, one may ask the question of whether super-radiance phenomena could be 
found in such systems, see also \cite{Federici:2006}. 
This is a valid question since we know that multiply quantized vortices 
are unstable in the absence of additional stabilization mechanisms 
(such as a repulsive potential at the vortex core), 
see also \cite{Fischer:2003zz,Svancara:2023yrf,Patrick:2023}. 
As an important difference to the case studied above, there is no horizon here 
and thus one should wonder what happens to the negative-energy parts moving inwards. 
When approaching the vortex core, the condensate density $\rho_{\rm c}^0$
drops and at some point, the linearization 
$\phi_{\rm c}=\phi_{\rm c}^0+\delta\phi_{\rm c}$ and 
$\rho_{\rm c}=\rho_{\rm c}^0+\delta\rho_{\rm c}$ breaks down. 
Thus, these parts may trigger non-linear processes such as the splitting of the 
multiply quantized vortex into several single quantized vortices -- which is 
how the multiply quantized vortex decays. 
This process is somewhat analogous to the absorption of the in-falling negative-energy 
parts by the horizon or their absorption by some sort of dissipation. 
In this sense, one could relate the decay of a multiply quantized vortex to 
super-radiance \cite{Giacomelli:2019tvr,Patrick:2021oqk,Patrick:2021dxw}. 

Of course, it would be nice to study such super-radiance phenomena in the quantum regime 
-- in analogy to the Technion experiment (see Sec.~\ref{Hawking radiation}) 
devoted to Hawking radiation, for example. 
Unfortunately, such an experiment seems to be even harder than the Technion experiment. 
On the one hand, the background flow containing a multiply quantized vortex is 
intrinsically unstable -- which complicates the experiment.
On the other hand, the in-falling partners are scrambled by the non-linearity at 
the vortex core such that the detection method based on the correlations does also 
require special care.
In view of these obstacles, it might be advantageous to add a radial flow component
$C\neq0$ as in Eq.~\eqref{draining vortex flow} which could stabilize the flow and 
might help us to extract the in-falling partners. 
However, this would require a non-trivial (e.g., funnel like) geometry or removing 
condensate atoms from the core by some mechanism. 

Apart from the single-component Bose-Einstein condensate considered above, 
one can also study super-radiance phenomena in more complex scenarios 
(which typically offer more options for manipulation and control) such as 
two-component Bose-Einstein condensates (see, e.g., \cite{Berti:2022btt}) 
or additional synthetic gauge fields 
(see, e.g., \cite{Giacomelli:2020evu,Giacomelli:2021xfn}).
The latter scenario displays interesting connections to the Sauter-Schwinger effect 
discussed in Sec.~\ref{Sauter-Schwinger effect}. 

As a further outlook, the evolution (motion, re-connection etc.) of vortices in 
Bose-Einstein condensates is also closely related to the subject of 
quantum turbulence, see, e.g., 
\cite{Volovik:2003rb,Nowak:2010tm,Tsubota:2017ncj,Madeira:2020ill,Barenghi:2023mcj,Barenghi-book}.
However, a more detailed discussion of this topic is outside the scope of the present 
article. 

\subsubsection{Quasi-normal modes}\label{Quasi-normal modes}

To conclude this section, let us briefly discuss a somewhat related topic, 
the quasi-normal modes of black holes (and their analogues), 
\rs{see, e.g., \cite{Kokkotas:1999bd,Berti:2009kk,Konoplya:2011qq}.}
In the discussion above, we explained super-radiance via the scattering picture 
where an incoming wave packet is scattered at the inhomogeneous flow 
profile~\eqref{draining vortex flow} and thus splits (or is torn apart) 
into an in-falling and an outgoing part. 
Of course, using such a scattering picture, one is led to the question of whether 
there are bound states -- which would then affect the dynamics. 
Even though there are typically no real bound states in such black-hole space times, 
there can be quasi-bound states or quasi-normal modes.

In order to sketch the main idea by means of a simple picture, let us consider the 
one-dimensional Schr\"odinger equation in the presence of a rectangular (i.e., box) 
potential barrier such as $V(x)=V_0\Theta(L^2-x^2)$.  
In classical mechanics, a particle with an energy $E>V_0$ starting at $x=0$ with 
$\dot x>0$ would just fall off the potential edge at $x=L$ and continue its propagation. 
In quantum mechanics, however, we have the phenomenon of quantum reflection and the 
particle can actually be reflected at the potential edge.  
Thus, we may approximate the two potential edges at $x=\pm L$ by (imperfectly) reflecting 
boundary conditions -- where we find approximate bound state at quasi-discrete energy levels. 
However, since there is also a non-vanishing transmission probability at the edges, 
these are not exact bound state because the wave function would slowly leak out.  
They are quasi-bound states or quasi-normal modes or resonances, characterized by a
discrete set of frequencies with real and imaginary parts. 
These frequencies allow us to infer properties of the potential.
Roughly speaking, the spacing between the real parts tells us the potential
length $L$ while the lowest real part and the imaginary parts are related to 
the potential height $V_0$. 

Very analogous phenomena occur in other wave equations, such as that for a 
mass-less and minimally coupled scalar field~\eqref{minimally-coupled-scalar} 
in the Schwarzschild metric, for example. 
In terms of the Regge-Wheeler tortoise coordinate $r_*$ and after the usual 
separation ansatz $\phi(t,r_*,\vartheta,\varphi)=
e^{-i\omega t}\phi_{\omega, \ell}(r_*)Y_{\ell,m}(\vartheta,\varphi)$, 
the remaining equation for $\phi_{\omega,\ell}(r_*)$ can be cast into a 
Schr\"odinger-like 
form\footnote{\rs{Note that one should be a bit careful when relating solutions 
(even for fixed $\omega$)
of the Klein-Fock-Gordon equation~\eqref{minimally-coupled-scalar}, 
which is relativistic, to those of the Schr\"odinger equation, which is 
non-relativistic, especially if they are not exactly stationary since 
$\omega$ is not purely real. 
However, the main ideas do still apply.}}
with an effective potential $V_{\rm eff}(r_*)$.
In the case of gravitational waves $h_{\mu\nu}$, the underlying wave equation 
is a bit more complicated, but it also supports quasi-normal modes. 
Actually, they are 
related\footnote{\rs{In analogy to the spectroscopy of atoms and molecules, 
this relation is exploited in the field of ``black-hole spectroscopy'', see,
e.g., \cite{Dreyer:2003bv,Cabero:2019zyt}.}} 
to the ring-down signal at late times which has been 
observed at LIGO for the black-hole merging event \cite{LIGOScientific:2016aoc}. 

Even though quasi-normal modes can already be discussed on the level of purely 
classical wave equations, it has been speculated that they might be related to black-hole
quantization, see, e.g., \cite{Hod:1998vk}. 
For this specific relation, black-hole analogues can probably not be used as a 
fully quantitative analogue, because they do not reproduce the Einstein equations
(which are presumably important for black-hole quantization).
Nevertheless, black-hole analogues can serve as a qualitative analogue for 
quasi-normal modes themselves, see, e.g., 
\cite{Berti:2004ju,Barcelo:2007ru,Torres:2020tzs,Jacquet:2021scv,Vieira:2023ylz,Burgess:2023pny}. 

As a related phenomenon, one can also study modes which propagate {\em along} the vortex. 
For free vortices in super-fluids such as Bose-Einstein condensates, one important example 
are Kelvin waves which correspond to oscillating displacements of the vortex core. 
If the vortex is pinned (e.g., by a laser potential in a Bose-Einstein condensate or a 
wire in super-fluid Helium), one can have other phenomena such as an analogue of 
whispering-gallery modes which are trapped in the vicinity of the vortex and thus 
can only propagate along it, see, e.g., \cite{Marecki:2013}. 


\subsection{Electromagnetic fields: Sauter-Schwinger and Breit-Wheeler effect}
\label{Sauter-Schwinger and Breit-Wheeler effect}

\subsubsection{Sauter-Schwinger effect}
\label{Sauter-Schwinger effect}

While in Hawking radiation and cosmological particle creation, the quantum vacuum fluctuations 
are torn apart by the strong gravitational field of the black hole or the cosmic expansion,
respectively, this tearing apart is caused by a strong electric field in the 
Sauter-Schwinger effect \cite{Sauter:1931zz,Schwinger:1951nm}. 
In the latter, the picture based on the particles and anti-particles 
(here electrons and positrons) works a bit better than in the former two 
-- but the tunneling picture provides a much more adequate description
(in contrast to the case of Hawking radiation). 
To understand why, let us first consider a massive charged (i.e., complex) 
scalar field $\psi$ minimally coupled to the electrostatic potential $\phi$ via 
\bea
\label{charged-scalar-field}
\left[\left(\partial_t+iq\phi\right)^2-\na^2+m^2\right]\psi=0
\,.
\ea
For simplicity, let us assume that the electrostatic potential depends on $x$ only 
$\phi=\phi(x)$, for example $\phi(x)=-qEx$ for a constant electric field $E$.  
This allows us to insert the ansatz 
$\psi(t,x,y,z)=e^{-i\omega t+ik_yy+ik_zz}\psi_\omega(x)$ into Eq.~\eqref{charged-scalar-field}
which then simplifies to 
$[-(\omega-q\phi)^2-\partial_x^2+\f{k}_\perp^2+m^2]\psi_\omega=0$ where the transversal 
wave number $\f{k}_\perp^2=k_y^2+k_z^2$ can be absorbed into an effective
mass $m^2_{\rm eff}=\f{k}_\perp^2+m^2$.
We see that this differential equation is formally equivalent to a stationary 
Schr\"odinger equation in ordinary (non-relativistic) quantum mechanics where 
$-(\omega-q\phi)^2$ plays the role of the potential. 
E.g., for a constant electric field $\phi(x)=-qEx$, it would just be an inverted parabola. 
Up to a minus sign, the effective mass $m^2_{\rm eff}=\f{k}_\perp^2+m^2$ determines the 
effective energy, which is negative and thus lies below the top of the potential barrier.
As a result, we have the analogue of a Schr\"odinger tunneling scenario where we can apply 
the same tools and methods as in ordinary (non-relativistic) quantum mechanics.

\begin{figure}[tbp]
\centering
\includegraphics[width=.7\textwidth]{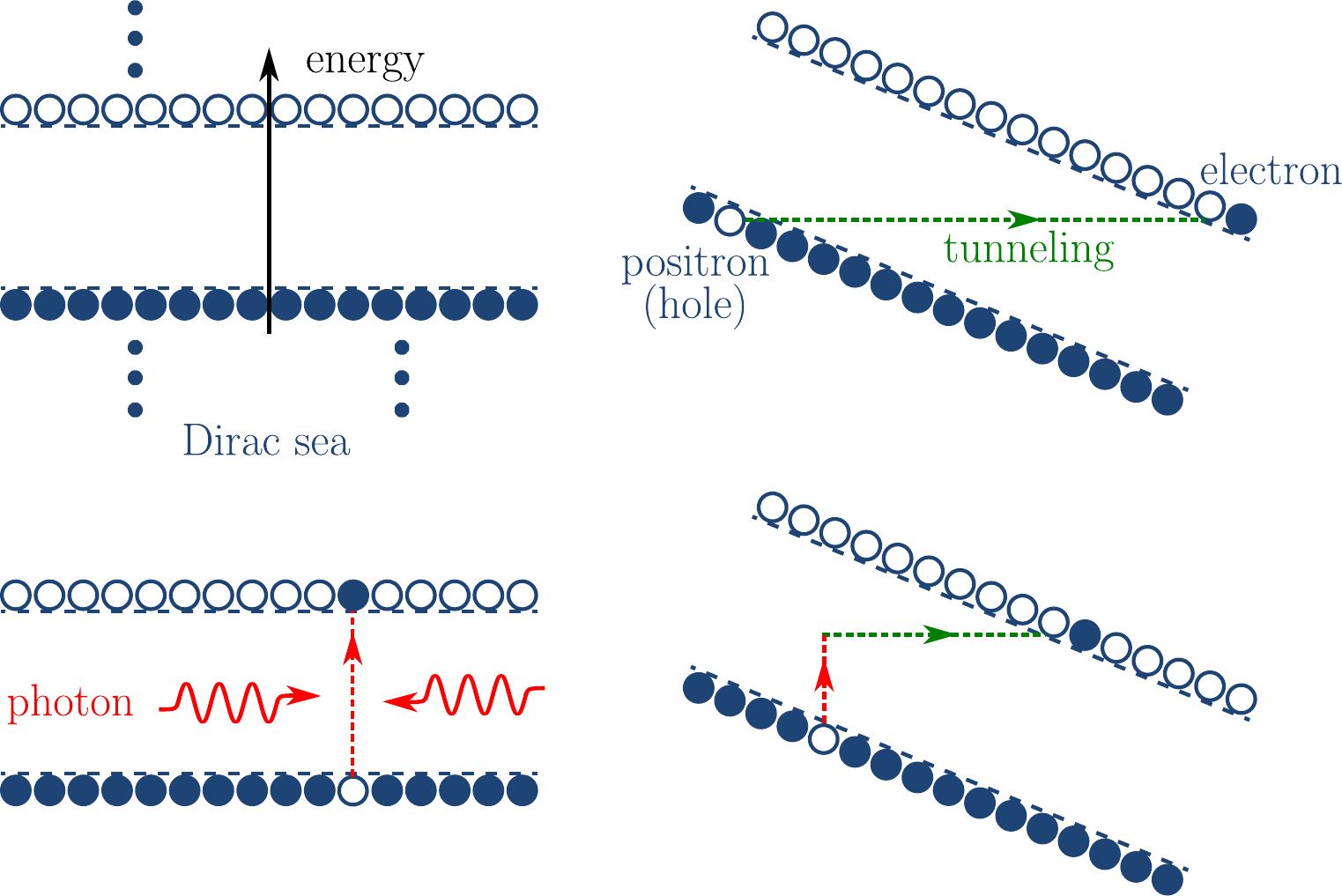}
\caption{Sketch of the Dirac sea (top left) where the filled blue dots depict occupied 
electron states while the empty blue dots denote unoccupied states. 
The Sauter-Schwinger effect (top right) can then be visualized by tunneling through 
the classically forbidden region. 
In the Breit-Wheeler effect (bottom left), on the other hand, an electron is directly 
lifted up from the Dirac sea to the positive continuum, where the required energy is 
provided by photons. 
Finally, the dynamically assisted Sauter-Schwinger effect (or the tunnel-assisted 
Breit-Wheeler effect) can be understood as a combination of the two effects (bottom right).}
\label{fig:sauter}
\end{figure}

However, building a quantum simulator based on ultra-cold atoms for the dynamics in 
Eq.~\eqref{charged-scalar-field} may be non-trivial\footnote{
\rs{For example, the non-relativistic (many-body) Schr\"odinger equation for the atoms 
contains a first-order time derivative while the relativistic Eq.~\eqref{charged-scalar-field} 
is of second order in time (and space).} 
}.
Thus, let us instead focus on the Dirac equation in 1+1 dimensions from 
Eq.~\eqref{Dirac-Hamiltonian-1+1} as already discussed in Sec.~\ref{Dirac Hamiltonian}
\bea
\label{Dirac+phi}
i\partial_t\Psi
=
\left(-i\alpha\partial_x+m\beta+q\phi\right)\Psi 
=
\left(i\sigma_y\partial_x+m\sigma_z+q\phi\right)\Psi 
\,,
\ea
again minimally coupled to the electrostatic potential $\phi(x)$. 
%
%
In analogy to the discussion above, this stationary problem facilitates the 
ansatz $\Psi(t,x)=e^{-i\omega t}\Psi_\omega(x)$ after which Eq.~\eqref{Dirac+phi} 
simplifies to 
%
%
\bea
\partial_x\Psi_\omega
=
\left(
\begin{array}{cc}
0 & q\phi-\omega-m \\
\omega-q\phi-m & 0 
\end{array}
\right)\cdot\Psi_\omega
=
\f{M}_\omega\cdot\Psi_\omega
\,.
\ea
The eigen-values of the matrix $\f{M}_\omega$ read 
$\lambda_\omega=\pm\sqrt{m^2-(\omega-q\phi)^2}$. 
In complete analogy to the effective Schr\"odinger tunneling problem derived above
from Eq.~\eqref{charged-scalar-field}, they vanish at the classical turning points 
where $\omega-q\phi=\pm m$. 
In between these turning points (i.e., inside the potential barrier), 
these eigen-values are real, which corresponds to the exponential decaying (or growing) 
wave function in the classically forbidden region. 
Outside these turning points, the eigen-values are imaginary, which represents the 
spatially oscillating wave function in the classically allowed region.  

However, as a major distinction to ordinary tunneling in non-relativistic 
quantum mechanics, here the tunneling occurs between the negative continuum 
$\omega-q\phi<-m$ (i.e., the Dirac sea) and the positive continuum 
$\omega-q\phi>+m$, as in Klein tunneling. 
Thus, we arrive at the following 
picture\footnote{
\rs{It should be stressed here that the picture of the Dirac sea is a means to 
visualize quantum vacuum fluctuations, but one should be careful when taking 
it too seriously. 
For example, besides an infinite vacuum energy and mass, it would also correspond to 
an infinite vacuum charge. 
Furthermore, in view of the $CPT$ symmetry, one could also say that the Dirac 
sea is full of positrons and the electrons are the holes.}}: 
Even in the vacuum state, the Dirac sea is filled with electrons and due to the 
electric field $E$, it can happen that one of these electrons tunnels through the 
classically forbidden region (i.e., the mass gap between $-m$ and $+m$) into the 
positive continuum, where it emerges as a real particle.
Since the total number of fermions is conserved, it leaves behind a hole in the 
Dirac sea -- which is then a positron. 
Thus, the electric field $E$ should be able to spontaneously create electron-positron 
pairs out of the vacuum, as depicted in Fig.~\ref{fig:sauter}. 

\begin{figure}[tbp]
\centering
\includegraphics[width=.7\textwidth]{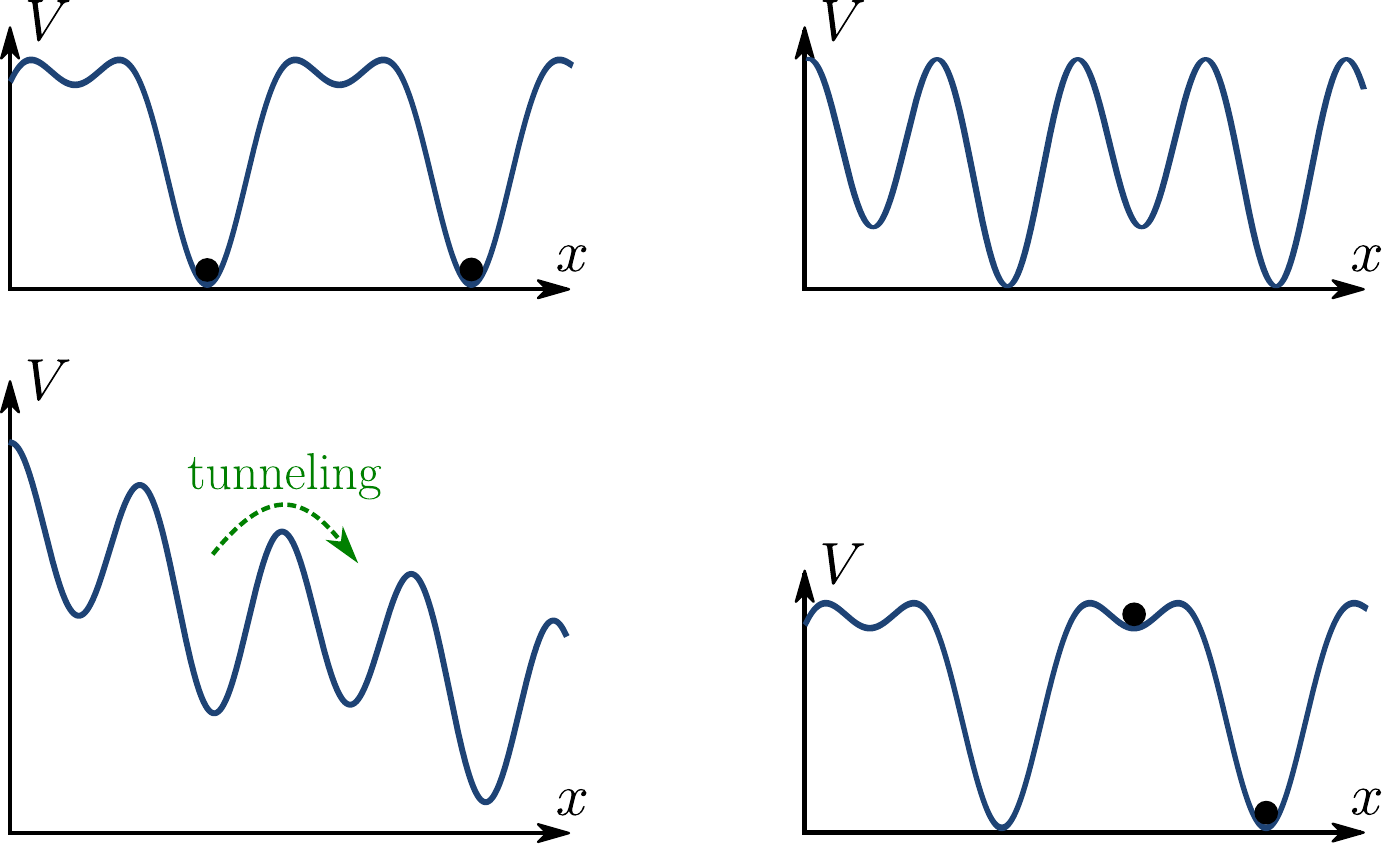}
\caption{Possible sequence for an analogue of the Sauter-Schwinger effect in optical lattices, 
see also Fig.~\ref{fig:dirac-dispersion}.
Starting with a strongly bi-chromatic optical lattice (top left), one could place atoms in 
the energetically low even lattice sites. 
Then, by slowly (i.e., adiabatically) lowering the energy gap to the desired value (to right),
which determines the effective mass $m_{\rm eff}$ in the Dirac equation~\eqref{Dirac+phi}, 
one can ensure that the upper (conduction) band in Fig.~\ref{fig:dirac-dispersion} remains
empty while the lower (valence) band stays filled (and represents the Dirac sea).
Then, one can apply the lattice deformation or tilt (bottom left) which simulates the
external electric field $E$ and allows the atoms to tunnel from the lower to the upper band. 
Finally, after reversing this sequence, atoms on the high-lying (odd) lattice sites would 
be the hallmark of the Sauter-Schwinger effect (bottom right).}
\label{fig:sauter-analogue}
\end{figure}

In order to estimate the probability for such as process, we may employ the 
WKB approximation and integrate $\lambda_\omega$ over $x$ from one turning point to 
the other which then yields the tunneling exponent. 
For a constant electric field $E$, we find the tunneling and thus 
electron-positron pair creation probability 
\bea
\label{Schwinger-scaling}
P_{e^+e^-}
\sim
\exp\left\{-\pi\,\frac{m^2}{qE}\right\}
=
\exp\left\{-\pi\,\frac{E_{\rm crit}}{E}\right\}
\,,
\ea
where $E_{\rm crit}=m^2c^3/(q\hbar)\approx1.3\times10^{18}~\rm V/m$ 
is usually referred to as the Schwinger limit or the Schwinger critical field. 
Reaching such a strong field in large enough space-time volumes (such that it can be 
considered as approximately constant) is extremely 
hard\footnote{Such extremely high field strengths are reached very close to highly 
charged nuclei, but the spatial extent of these high field strengths is too small to 
induce the Sauter-Schwinger effect. 
Otherwise, these nuclei would not be stable. 
Note that this phenomenon is related to the behaviour of the Dirac equation 
for point-like nuclei, were the energy eigen-values become complex for atomic 
numbers $Z$ larger than $1/\alpha_{\rm QED}\approx137$.} 
which explains why a conclusive experimental verification of this prediction is still lacking. 
%
%
On the other hand, for a quantum simulator, the effective speed of light $c_{\rm eff}$ 
as well as the effective mass gap $m_{\rm eff}c_{\rm eff}^2$ are typically many 
orders of magnitude smaller -- which helps us to realize a suitable analogue of this 
effect, see Fig.~\ref{fig:sauter-analogue}. 

Note that the applicability of the WKB approximation requires that the eigen-values 
do not change too fast, which translates to $qE\ll m^2$, i.e., $E\ll E_{\rm crit}$. 
Furthermore, the WKB approximation breaks down at the classical turning points 
(as usual in such tunneling scenarios). 
In addition, if the electric field is not constant but changes with space and/or time, 
the associated rate of change must be taken into account as well. 
Especially a time-dependence of the electric field may also lead to electron-positron 
pair creation -- in addition to the tunneling phenomenon explained above. 
This brings us to the Breit-Wheeler effect \cite{Breit:1934zz}. 

\subsubsection{Breit-Wheeler effect}
\label{Breit-Wheeler effect}

To explain this phenomenon, let us consider a spatially homogeneous but time-dependent 
electric field $\f{E}(t)$.  
In this case, it is advantageous to perform a gauge 
transformation\footnote{For lattice systems, such a gauge transformation is usually 
referred to as Peierls transformation or Peierls substitution. 
The basic idea is the same as already discussed in Secs.~\ref{Dirac Hamiltonian} 
and~\ref{Gauge fields}.} 
such that the electric field $\f{E}(t)$ is represented by the vector potential $\f{A}(t)$. 
For simplicity, we again start with the complex scalar field 
\bea
\label{charged-scalar-field+A}
\left[\partial_t^2-\left(\na-iq\f{A}\right)^2+m^2\right]\psi=0
\,.
\ea
In analogy to Eq.~\eqref{harmonic-oscillator} in Sec.~\ref{Cosmological particle creation}, 
exploiting the spatial homogeneity by an expansion into spatial plane waves 
$\exp\{i\f{k}\cdot\f{r}\}$, this equation becomes equivalent to that of a 
harmonic oscillator with a frequency $\Omega^2_{\fk{k}}(t)=[\f{k}-q\f{A}(t)]^2+m^2$. 
Using again the analogy to the Schr\"odinger equation in ordinary (non-relativistic) 
quantum mechanics, this now corresponds to a negative effective potential -- such that 
we have scattering above the potential (e.g., quantum reflection)
instead of tunneling through the potential.
The reflection and transmission coefficients then determine the Bogoliubov coefficients 
for this problem, which in turn yield the pair-creation probability $P_{e^+e^-}$.

However, before going into more details, let us switch back to the Dirac equation in 
1+1 dimensions
\bea
\label{Dirac+A}
i\partial_t\Psi
=
\left(-i\alpha\partial_x-\alpha qA+m\beta\right)\Psi 
=
\left(i\sigma_y\partial_x+\sigma_yqA+m\sigma_z\right)\Psi 
\,,
\ea
which is more suitable for a quantum simulator. 
After an expansion into spatial plane waves $e^{ikx}$, it can be written as 
$i\partial_t\Psi_k=\f{M}_k\cdot\Psi_k$ where the eigen-values of the matrix 
$\f{M}_k$ are the eigen-frequencies $\pm\Omega_k(t)=\pm\sqrt{[k-qA(t)]^2+m^2}$. 
The corresponding eigen-vectors $u_k(t)$ and $v_k(t)$ represent the particle and 
anti-particle components and thus an expansion of the wave function into those 
eigen-vectors $\Psi_k=\alpha_k u_k + \beta_k v_k$ can be used to define the 
Bogoliubov coefficients $\alpha_k$ and $\beta_k$.
Inserting this expansion into Eq.~\eqref{Dirac+A} and considering the ratio 
$R_k=\beta_k/\alpha_k$ we arrive at the Riccati 
equation\footnote{Here we have used the convention that the phases $e^{\pm2iS_k(t)}$ 
are absorbed into the eigen-vectors $u_k(t)$ and $v_k(t)$.}
\bea
\label{Riccati}
\dot R_k(t)=\frac{mqE(t)}{\Omega_k^2(t)}
\left(e^{2iS_k(t)}+R_k^2(t)e^{-2iS_k(t)}\right) 
\,,
\ea
with the phase (eikonal) function $\dot S_k(t)=\Omega_k(t)$. 
For slowly varying fields $E(t)$, the solution to this equation reproduces the 
Sauter-Schwinger result~\eqref{Schwinger-scaling}. 
Rapidly varying fields, on the other hand, can open up additional channels for 
electron-positron pair creation.
For example, if the electric field is oscillating $E(t)\propto\cos(\omega_Et)$ 
with a frequency $\omega_E$ in resonance with the $e^{2iS_k(t)}$ term, 
which requires $\omega_E\geq2m$, we would obtain a resonant growth of $R_k(t)$ 
and thus $\beta_k$.
In this case, we have a quadratic scaling of the pair-creation
probability $P_{e^+e^-}\sim E^2$ instead of the exponential 
scaling~\eqref{Schwinger-scaling}. 

As an intuitive picture, the frequency $\omega_E$ of the electric field can directly 
transfer enough energy $\hbar\omega_E$ to an electron in the Dirac sea to lift it up 
to the positive continuum -- depicted as a vertical line in Fig.~\ref{fig:sauter}
(bottom left). 
In contrast, the tunneling in the Sauter-Schwinger effect for a constant 
electric field occurs at {\em constant} energy and is thus depicted as a 
horizontal line in Fig.~\ref{fig:sauter} (top right). 
One can also have a mixture of these two effects, which is usually referred to as  
the dynamically assisted Sauter-Schwinger effect, see, e.g., \cite{Schutzhold:2008pz}. 
Again using the above intuitive picture, a frequency $\omega_E<2m$ can lift up an 
electron from the Dirac sea a little bit into the classical forbidden region
(but ot all the way up to the positive continuum)
such that the remaining way to tunnel is shorter and thus the tunneling 
exponent~\eqref{Schwinger-scaling} is reduced, see Fig.~\ref{fig:sauter} (bottom right).

One might object that the case of a purely time-dependent electric field considered above 
is un-physical (e.g., because it is not a vacuum solution of the Maxwell equations). 
While this is a valid point, it does not spoil the analysis of the 
Sauter-Schwinger or Breit-Wheeler effect, which are a consequence of the 
Dirac~\eqref{Dirac+A} or Klein-Fock-Gordon~\eqref{charged-scalar-field+A} 
equations and {\em per se} independent of the Maxwell equations.
E.g., if $\f{A}$ would instead be a Proca field, the solutions of the
Dirac~\eqref{Dirac+A} or Klein-Fock-Gordon~\eqref{charged-scalar-field+A} 
equations (for the same $\f{A}$) would not change. 
This situation is a bit similar to the case of Hawking radiation discussed in 
Sec.~\ref{Hawking radiation}, which can be derived 
from the Klein-Fock-Gordon equation~\eqref{minimally-coupled-scalar} in a given background 
metric, but does not require this metric to be a vacuum solution of the Einstein equations. 

Coming back to the Sauter-Schwinger and Breit-Wheeler effect, another way to resolve the 
objection above is to consider the collision of two plane-wave laser pulses 
(which are vacuum solutions of the Maxwell equations). 
Then it can be shown \cite{Kohlfurst:2021skr} that electron-positron pair creation is 
dominated by the field at the collision point and that the local pair-creation probability 
(density) can be well approximated by the scenario of a purely time-dependent field.  


\subsubsection{Quantum simulator for Sauter-Schwinger and Breit-Wheeler effect}
\label{Quantum simulator for Sauter-Schwinger and Breit-Wheeler effect}

As already explained in Sec.~\ref{Dirac Hamiltonian}, the 1+1 Dirac 
Hamiltonian~\eqref{Dirac-Hamiltonian-1+1} can modeled by fermionic atoms in a bi-chromatic 
optical lattice, see Fig.~\ref{fig:dirac-dispersion}. 
Quite intuitively, an electric field can then be simulated by applying a force to the 
atoms in the lattice -- such that we arrive at a quantum simulator as sketched in 
Fig.~\ref{fig:sauter-analogue}, see also \cite{Szpak:2011pra,Szpak:2011jj}.
The bosonic\footnote{Due to the Pauli principle, bosonic atoms can be cooled down much better 
than fermionic atoms.} 
analogue of such a simulator has been realized in another experiment \cite{Pineiro:2019uzb} 
at Maryland (not to be confused with the expanding Universe simulator 
\cite{Eckel:2017uqx,Banik:2021xjn}).
Since the bosonic (Rubidium) atoms used in this experiment \cite{Pineiro:2019uzb}
do not obey the Pauli principle, one does not have the precise analogue of the Dirac sea
as sketched above.
Therefore, the experiment \cite{Pineiro:2019uzb} instead created an initial state where all 
the states in the lower band (which would correspond to the Dirac sea) were occupied by 
some bosons (with a certain probability) while the upper band was still empty. 
To ensure that the upper band was not excited in this initialization process, 
they started with a large gap, as in Fig.~\ref{fig:sauter-analogue} (top left), 
and later adiabatically reduced it to the desired value, as in Fig.~\ref{fig:sauter-analogue} 
(top right), thereby enabling tunneling transitions from the lower band to the upper band,
i.e., the analogue of the Sauter-Schwinger effect\footnote{As one might expect, 
tunneling from the valence band to the conduction band as an analogue of the Sauter-Schwinger 
effect is not restricted to ultra-cold atoms, similar effects can also occur in other systems, 
such as semi-conductors, see, e.g., \cite{Linder:2015fba} and references therein.}.

So far, in the considerations in this section, the gauge field has been treated as an external 
field, i.e., the back-reaction of the created electron-positron pairs onto the electric field 
was not included, see also the discussion in Sec.~\ref{Gauge fields}.
A very first step into this direction could be the combined quantum simulator sketched in
Fig.~\ref{fig:dirac+scalar} which corresponds to the coupled field theory in 
Eq.~\eqref{dirac+scalar-field}.
More elaborated schemes have been discussed as well, see, e.g., 
\cite{Kasper:2015cca,Kasper:2016mzj,Zache:2018jbt,Aidelsburger:2021mia}
as well as the references mentioned in Sec.~\ref{Gauge fields}.
However, they are already beyond the realm of linear fields as discussed in this chapter,
but belong to the class of non-linear fields considered in the next 
chapter~\ref{Non-linear fields}. 
In this context, it might also be worth mentioning here that the mass gap $m_{\rm eff}$ 
can also be generated (instead of using a bi-chromatic lattice) by the interaction $U$ 
between the atoms in Eqs.~\eqref{Bose-Hubbard Hamiltonian} and \eqref{Fermi-Hubbard},
see also \cite{Queisser:2012,Queisser:2023aee}
and references therein. 

\subsection{Dynamical Casimir effect}\label{Dynamical Casimir effect}

To conclude this chapter, let us briefly discuss the dynamical Casimir effect,
see, e.g., \cite{Dodonov:2020eto} for a review of recent developments.
While the original Casimir effect \cite{Casimir:1948dh} refers to the attraction of two 
parallel mirrors in vacuum and related phenomena, i.e., static situations, 
the dynamical Casimir effect typically denotes generalizations to the time-dependent case,
see, e.g., \cite{Moore:1970tmc}. 
Even though there are also alternative derivations, the static (i.e., original) Casimir
effect can be explained by the modification of the quantum vacuum fluctuations by the
presence of the two mirrors (or analogous external conditions, such a dielectric media).
This modification changes (reduces) the vacuum energy which then results in an attraction
between the mirrors.

In order to visualize this attractive force, it is sometimes argued that there are less
modes allowed in between the mirrors than outside -- and that, therefore, the vacuum energy
must be smaller.
While it is correct that there are \rs{fewer} 
allowed modes in between the mirrors
(perpendicular to the mirrors, we have a discrete spectrum inside but a continuous
spectrum outside), this picture is oversimplified:
Even though there are \rs{fewer} 
modes, they have a higher weight and thus the calculation of the
vacuum energy requires a subtle subtraction procedure (where the infinite vacuum energy cancels).
The failure of the above oversimplified arguments can be illustrated by comparing different
boundary conditions at the mirrors.
While two mirrors with the same boundary conditions (e.g., both Dirichlet or both Neumann
\rs{for the scalar field case}) 
indeed attract each other, two different mirrors, one with Dirichlet and one with  Neumann
boundary conditions, repel each other, 
\rs{see. e.g., \cite{Asorey:2013wca}.}
Similarly, two dielectric or two paramagnetic media typically attract each other while a
dielectric and a paramagnetic medium tend to have a repulsive Casimir force,
\rs{cf.~\cite{Kenneth:2002ij}}. 

After this detour to the static Casimir effect, let us turn to the dynamical Casimir effect.
Using the example of the two parallel mirrors, one could now imagine moving one or both
of them.
In this way, one should be able to create pairs of particles (in this case, photons)
out of the initial vacuum state.
There are two main contributions to this phenomenon.
First, the squeezing effect stemming from the change of the distance between the
mirrors (or, more precisely, the variation of the eigen-frequencies), which results in
single-mode squeezing -- analogous to the parametrically driven harmonic oscillator
as in Eq.~\eqref{harmonic-oscillator}.
Second, the acceleration effect is induced by the motion of the mirrors and
results in multi-mode squeezing, see, e.g., \cite{Schutzhold:1997yh}. 
Even if we consider a single moving mirror, we could still have the second effect
(but not the first one) such that we could also emit pairs of particles.

If we consider a cavity with a finite volume instead, we may enhance the effect by
harmonically oscillating one (or more) of the mirrors where the driving frequency matches
twice one of the discrete eigen-frequencies of the cavity such that we achieve parametric
resonance in this mode according to Eq.~\eqref{harmonic-oscillator}.
In this way, it may be possible to resonantly select one mode in which the pairs
are predominantly created via the squeezing effect while the other modes and the acceleration
effect can be neglected since they are off-resonant.

As a generalization, one could also imagine a cavity filled with some dielectric medium
whose index of refraction is modulated.
Even if there is no real motion of the mirrors encapsulating the cavity, the effective
distance ``felt'' by the electromagnetic field (i.e., the light travel time) changes and
thus it should not be surprising that one can also create pairs of particles in this case.
Actually, even without any mirrors, a temporal $n(t)$ or spatio-temporal $n(t,\f{r})$
change of the refractive index can create pairs of particles.
In order to accommodate all these scenarios, one could associate the dynamical Casimir effect
to the creation of pairs of photons (or other particles) out of the vacuum (or ground state)
induced by time-dependent external conditions, which could be mirrors or media with an
index of refraction.
However, it is fair to say that probably not everybody in the community would fully agree
to this definition.

As a result, although the static Casimir effect has been verified experimentally 
(see, e.g., \cite{Lamoreaux:1996wh}),
the situation for the dynamical Casimir effect is more complicated.
The analogue of the dynamical Casimir effect has been observed in the Gothenburg
\cite{Wilson:2011rsw} and Helsinki \cite{Lahteenmaki:2011cwo} experiments 
in time-dependent electromagnetic 
wave guides\footnote{Electromagnetic wave guides with suitable properties
could also be used to generate the analogue of Hawking radiation, 
see also \cite{Schutzhold:2004tv}.}, but a conclusive verification of photon pair creation 
due to the real motion of a mirror in vacuum remains yet to be observed. 
As another point, 
one could argue that the well-known phenomenon of parametric down conversion
in non-linear optical media\footnote{\rs{Note that the mapping of such non-linear 
media to refractive indices $n$ is very simplified: 
Depending on the symmetries of the medium, one should distinguish between $\chi^{(2)}$
and $\chi^{(3)}$ media, for example, where the material properties are typically encoded 
in tensorial quantities such as $\chi^{(2)}_{ijk}$ and $\chi^{(3)}_{ijkl}$ etc.}} 
can be interpreted as the creation of photon pairs by a
spatio-temporal change of the refractive index $n(t,\f{r})$.
%

Depending on which aspects are deemed to be more important and which less, the 
dynamical Casimir effect can be related to many of the phenomena discussed above. 
Thus, it is not so easy to draw a clear and unique separation line between the
dynamical Casimir effect (or its analogues) and the signatures of the Unruh effect,
Hawking radiation or cosmological particle creation (and their analogues). 
For example, a purely time-dependent refractive index $n(t)$ is very analogous to 
cosmological particle creation discussed in Sec.~\ref{Cosmological particle creation}.
In this case, photons are also created in pairs with opposite momenta and the underlying 
evolution equation for a single mode is formally equivalent to 
Eq.~\eqref{harmonic-oscillator}.
In view of this formal equivalence, it is perhaps not surprising that the paper 
reporting on the Palaiseau experiment \cite{Jaskula:2012ab} was referring to the 
dynamical Casimir effect instead of cosmological particle creation.

As a more involved scenario, one could also consider a space-time dependent 
refractive index $n(t,\f{r})=n(\f{r}-\vau t)$ moving with a constant velocity
$\vau$.
If this velocity $\vau$ is smaller than the speed of light inside the entire medium, 
no photons would be created. 
If it is faster (e.g., generated by a flying laser focus in a non-linear Kerr medium),
one could create signatures of the Ginzburg effect discussed in Sec.~\ref{Ginzburg effect}.
If this velocity $\vau$ lies in between the medium speed of light of the regions  
with large $n$ and small $n$, one could obtain an optical or dielectric black-hole 
analogue and create the analogue of Hawking radiation discussed in Sec.~\ref{Hawking radiation},
see, e.g., \cite{Schutzhold:2004tv,Leonhardt:2000fd,Visser:2000pk,Leonhardt:2000hf,
Schutzhold:2001fw,Philbin:2007ji}. 
A refractive index $n(t,\f{r})=n(\f{r}-\f{r}_{\rm D}[t])$ moving along an accelerated
trajectory $\f{r}_{\rm D}[t]$, on the other hand, could create 
(in addition to the effects mentioned above) 
signatures of the Unruh effect discussed in Sec.~\ref{Unruh radiation}.

In summary, we see that quantum electrodynamics in media with space-time dependent refractive 
indices $n(t,\f{r})$, moving dielectric bodies or flowing dielectric fluids is a very rich 
field and offers many fascinating phenomena -- where it is not always easy to draw a clear 
line between the different effects, see also \cite{Schutzhold:2011ze}. 
As in Bose-Einstein condensates~\eqref{BEC-disperion}, it is also important to take into 
account the dispersion relation, see, e.g., \cite{Linder:2015vmk} and references therein. 
Furthermore, remembering the Kramers-Kronig relations, one should also investigate the 
impact of dissipation on those phenomena, see also \cite{Lang:2019avs}. 
Note that such dissipative effects are also present in Bose-Einstein condensates, e.g., 
in the form of three-body losses (see, e.g., \cite{Jack:2002,Raetzel:2021}) and should 
there be investigated as well. 


\newpage
\section{Non-linear fields}\label{Non-linear fields}

After having discussed several effects of linear fields (propagating on non-trivial backgrounds),
let us turn our attention to fields which obey inherently non-linear evolution equations. 
Of course, for all the phenomena discussed in the previous chapter, one can ask the question 
of how they change for non-linear fields, i.e., in the presence of interactions. 
For example, the Coulomb attraction between electrons and positrons should affect their creation
via the Sauter-Schwinger or Breit-Wheeler effect, 
see Sec.~\ref{Sauter-Schwinger and Breit-Wheeler effect}.
As another example, one would expect that small black holes with a Hawking temperature far above 
the QCD scale $\Lambda_{\rm QCD}=\ord(100~\rm MeV)$ emit quasi-free quarks and gluons, which only 
later (outside the black hole) combine to hardons. 
For larger black holes with a Hawking temperature below the QCD scale 
$\Lambda_{\rm QCD}=\ord(100~\rm MeV)$, on the other hand, this behaviour should change 
drastically due to confinement. 
Confirming this expectation by solid calculations, however, is a highly non-trivial task. 

Even though these are very interesting and important questions -- 
which could, at least qualitatively, be addressed with quantum simulators, see also 
chapter~\ref{Summary and outlook} -- the next sections instead focus on new phenomena 
caused by non-linearities of the field equations. 

\subsection{Sine-Gordon model}\label{Sine-Gordon model}

An important example for a non-linear field theory is the sine-Gordon model as described by the
Lagrangian density
\bea
\label{sine-gordon}
{\cal L}
=
\frac{1}{2g^2_\phi}\,
\left[(\partial_t\phi)^2-(\partial_x\phi)^2+2\cos\phi\right]
\,.
\ea
Although the pre-factor $g_\phi$ is not relevant for the classical equation of motion $\Box\phi+\sin\phi=0$,
it does play a role for the associated quantum field theory or thermal field theory since it is
then combined with the Planck constant $\hbar$ or the temperature, respectively.
Note that $g_\phi$ can be identified with a coupling strength.
To see that, let us consider a field re-definition $\phi\to g_\phi\phi$ after which the pre-factor
disappears and the potential terms is modified to $\cos(g_\phi\phi)$.
Taylor expanding this term for small $g_\phi$, we see that the theory approaches a free field in
the limit $g_\phi\to0$.

Let us first discuss the classical field theory.
The sine-Gordon model~\eqref{sine-gordon} combines at least three important features.
First, it is relativistic, i.e., invariant under Lorentz transformations.
Second, it is integrable (in 1+1 dimensions).
Third, it is invariant under the symmetry $\phi\to\phi+2\pi$ and displays topological properties.
This combination entails several interesting consequences.
For example, the sine-Gordon model~\eqref{sine-gordon} supports soliton solutions in the
form of kinks and anti-kinks.
These are topological defects which connect the asymptotic field values $\phi(x\to-\infty)=0$
and $\phi(x\to+\infty)=2\pi$ for kinks or
$\phi(x\to-\infty)=0$ and $\phi(x\to+\infty)=-2\pi$ for anti-kinks -- both modulo global shifts
$\phi\to\phi+2\pi$ due to the aforementioned symmetry.
Note that the difference between these asymptotic values $\phi(x\to-\infty)$
and $\phi(x\to+\infty)$ corresponds to the topological charge or winding number
multiplied by $2\pi$.
Due to the Lorentz invariance of the sine-Gordon model~\eqref{sine-gordon}, these kinks and
anti-kinks obey the relativistic energy-momentum relation $E^2=m^2+p^2$ and thus can be
viewed as toy models for relativistic particles.
Moreover, the sine-Gordon model~\eqref{sine-gordon} facilitates simple analytic expressions
(in terms of the inverse tangent and exponential functions) for single or even multiple solitons.
These could be scattering solutions where the solitons just go through each other and even
bound-state solutions (breathers).

Upon quantization, the sine-Gordon model~\eqref{sine-gordon} becomes more complicated.
Apart from the non-trivial structure of the vacuum state, the spectrum contains
single-particle states as described by the same relativistic energy-momentum relation
$E^2=m^2+p^2$ as in the classical case (with suitably renormalized values).
The states with two or more particles are more involved due to their interaction --
which is the usual case in interacting quantum field theories.
Instead of the eigen-states of the Hamiltonian, let us consider the $n$-point functions
$\langle\hat\phi(t_1,x_1)\dots\hat\phi(t_n,x_n)\rangle$ in the vacuum state.
They allow us to calculate the $S$-matrix or the generating functional, for example.
For the case of a free (linear) field, the vacuum state is a Gaussian state and hence
the Wick theorem allows us to express all $n$-point functions as a sum of products of 
two-point functions.
%
%
For interacting quantum field theories, on the other hand, this is no longer true and
the deviations from a Gaussian state depend on the parameters of the model, such as $g_\phi$.
Turning this argument around, these deviations from Gaussianity, i.e., the  ``error''
in the Wick expansion, can be used to infer those parameters.

To conclude this brief introduction into the sine-Gordon model, it should be mentioned here 
that it is equivalent (in a certain region of parameter space, see \cite{Coleman:1974bu}) 
to the massive Thirring model \cite{Thirring:1958in} describing fermions in 1+1 dimensions 
with a four-fermion interaction term 
\bea
\label{Thrirring}
{\cal L}=\bar\Psi\left(i\gamma^\mu\partial_\mu-m\right)\Psi   
-\frac{g_\Psi}{2}\left(\bar\Psi\gamma^\mu\Psi\right) \left(\bar\Psi\gamma_\mu\Psi\right) 
\,.
\ea
Generalizing this ansatz to more than one fermion species $\Psi_a$ and replacing the 
interaction term by $(\bar\Psi_a\Psi^a)(\bar\Psi_b\Psi^b)$, 
one obtains the Gross-Neveu model \cite{Gross:1974jv}. 
As usual in such bosonization approaches (think of the Jordan-Wigner transformation 
of the quantum Ising model, for example), the mapping between the fermionic degrees 
of freedom in the Thirring model~\eqref{Thrirring} and the bosonic degrees of freedom 
in the sine-Gordon model~\eqref{sine-gordon} is non-local in space and works in this 
way in one spatial dimension only. 

\subsubsection{Analogues for the sine-Gordon model}\label{Sine-Gordon analogue}

In view of the equivalence between the sine-Gordon model~\eqref{sine-gordon} and the 
Thirring model~\eqref{Thrirring}, let us start with ultra-cold atom simulators for the 
latter. 
As already explained in Sec.~\ref{Dirac Hamiltonian}, the bi-linear part of the Dirac 
Hamiltonian in Eq.~\eqref{Thrirring}, i.e., the free-field contribution,  
can be modeled with fermionic atoms in bi-chromatic optical lattices. 
In this representation, the remaining interaction terms  
$(\bar\Psi\gamma^\mu\Psi)(\bar\Psi\gamma_\mu\Psi)$ in the Thirring model or that of the 
Gross-Neveu model $(\bar\Psi_a\Psi^a)(\bar\Psi_b\Psi^b)$ can be modeled by suitable 
on-site interaction terms between the different species such as the $U$ term in the 
Fermi-Hubbard model~\eqref{Fermi-Hubbard}, see also \cite{IgnacioCirac:2010us}.
As another possibility, a Boson-Fermion interaction as sketched in Fig.~\ref{fig:dirac+scalar}
could generate, after integrating out the bosonic degrees of freedom, to an effective 
four-Fermion interaction, as explained after Eq.~\eqref{dirac+scalar-field}. 
By arranging the lattice representation is a suitable way, one could then generate the 
desired structure of the interaction term. 

However, since bosonic atoms can be cooled down much easier than fermionic atoms 
(due to the Pauli principle), let us now turn to the original 
sine-Gordon model~\eqref{sine-gordon}.
One way of simulating this model could be a lattice of Rydberg atoms as sketched in 
Fig.~\ref{fig:scalar} with a harmonic on-site potential $V$ which can be obtained by 
a standing wave laser field in $z$ direction (i.e., along the $q_I$). 
Another option has been realized in the Vienna experiment 
\cite{Schweigler:2017,Zache:2019xkx,Schweigler:2021,Tajik:2022ycs}
where two effectively one-dimensional Bose-Einstein condensates were generated 
parallel to each other at a small distance such that the atoms could still tunnel 
from one condensate to the other.
Then, the dynamics of the phase difference between the two condensates can be 
approximately described by the sine-Gordon model~\eqref{sine-gordon}.
As an intuitive picture, the atoms can most efficiently reduce their energy by 
tunneling between the two condensates if they have the same phase (modulo $2\pi$)
which explains the $\cos(\phi)$ potential in Eq.~\eqref{sine-gordon}.
In order to measure the relative phase (as a function of $x$, i.e., the coordinate 
along the effectively one-dimensional Bose-Einstein condensates), one can release 
the condensates (in analogue to the time-of-flight technique sketched in 
Sec.~\ref{Measurement schemes}) and let them interfere. 
Then, taking a picture of the resulting pattern, the measured interference fringes 
allow us to reconstruct the relative phase, 
cf.~\cite{Schweigler:2017,Zache:2019xkx,Schweigler:2021,Tajik:2022ycs}.

Having realized such a simulator for the sine-Gordon model~\eqref{sine-gordon},
one could first study its classical solutions such as kinks and anti-kinks and 
how they go through each other 
(as in Newton's cradle, see also \cite{Kinoshita:2006xby}).
Going to the quantum regime, one can also study $n$-point functions
$\langle\hat\phi(t_1,x_1)\dots\hat\phi(t_n,x_n)\rangle$, how they deviate 
from a Gaussian state (i.e., the ``error'' in the Wick expansion), and how this deviation is 
related to the effective Lagrangian~\eqref{sine-gordon}, 
cf.~\cite{Schweigler:2017,Zache:2019xkx,Schweigler:2021}.
In addition to the correlation  between different space-time points 
$(t_1,x_1)\dots(t_n,x_n)$, one can also study the correlation between two (or more)
regions $A$ and $B$ by considering the von~Neumann entropy of the reduced density matrix 
of these regions $S_A$, $S_B$ and $S_{A\cup B}$, see also \cite{Tajik:2022ycs}. 




Note that the schemes for experimental realization mentioned above are not the only ones. 
As another possibility, the Innsbruck experiment \cite{Haller:2010} realized the 
sine-Gordon model~\eqref{sine-gordon} in a suitable limiting case of an optical lattice 
realizing the Bose-Hubbard Hamiltonian~\eqref{Bose-Hubbard Hamiltonian} in one 
spatial dimension. 

\subsection{Kibble-Zurek mechanism}\label{Kibble-Zurek mechanism}

The Kibble-Zurek mechanism \cite{Kibble:1976sj,Kibble:1980mv,Zurek:1985qw,Zurek:1996sj}
refers to the creation of topological defects in symmetry-breaking phase transitions.
Before turning to their creation mechanism, let us briefly discuss the concept of topological defects,
where we focus on classical fields for the moment.
Examples for topological defects are the kinks and anti-kinks discussed in the previous section,
as well as vortices in Bose-Einstein condensates discussed in
Sec.~\ref{Vortices in Bose-Einstein condensates}.
Generally speaking, topological defects can occur in a phase of broken symmetry where the
ground state is not unique but we have a set or manifold of ground states.
In such a case, the vacuum would be a state with the same position in the ground-state
manifold everywhere -- often called the order parameter.
A topological defect is then an inhomogeneous deviation from this homogeneous vacuum state
where different spatial points assume distinct order parameters such that smooth deformations
of the order parameter cannot transform the defect away (i.e., it is topologically protected).

As a rule of thumb\footnote{A more rigorous approach can be based on homotopy groups or homotopy classes.},
topological defects can be characterized by the dimension $D_{\rm gs}$ of the ground-state manifold in
the broken-symmetry phase and the space-time dimension $D_{\rm st}$, which then determine the
dimensionality $D_{\rm td}$ of possible topological defects via $D_{\rm td}=D_{\rm st}-D_{\rm gs}-2$.
Kinks and anti-kinks in the sine-Gordon model~\eqref{sine-gordon} correspond to a breaking of the
discrete $\mathbb Z$ symmetry $\phi\to\phi+2\pi\mathbb Z$ such that ground-state manifold is
discrete, i.e., zero-dimensional $D_{\rm gs}=0$.
Thus, in 1+1 dimensions $D_{\rm st}=2$, they are point-like defects (i.e., also zero-dimensional)
$D_{\rm td}=0$.
In 2+1 dimensions $D_{\rm st}=3$, they would be line-like defects (i.e., one-dimensional)
$D_{\rm td}=1$, and so on.
As our second example, vortices in Bose-Einstein condensates correspond to a breaking of the
$U(1)$ symmetry of the original Hamiltonian~\eqref{Hamiltonian-general} by the phase of the 
condensate wave function $\psi_{\rm c}$.
Thus, the ground-state manifold is one-dimensional, i.e., a circle, $D_{\rm gs}=1$.
As already discussed in Sec.~\ref{Vortices in Bose-Einstein condensates}, a vortex is a point-like
defect $D_{\rm td}=0$ in 2+1 dimensions $D_{\rm st}=3$ and a line-like defect $D_{\rm td}=1$
in 3+1 dimensions $D_{\rm st}=4$.
As an intuitive picture, if we encircle a (singly quantized) vortex in real space, the phase
(i.e., order parameter) describes a circle in order-parameter space.
For multiply quantized vortices, the phase would wind around the circle in internal space
several times, where the number of revolutions corresponds to the winding number,
i.e., the topological charge.

When investigating possible symmetry-breaking phase transitions in the early Universe,
Kibble \cite{Kibble:1976sj,Kibble:1980mv} found that they could entail the creation of 
such topological defects --
which might (due to their topological protection) still be present today, for example
in the form of cosmic strings.
In spite of extensive search for such objects, their existence has not been
conclusively confirmed (yet).
Later on Zurek \cite{Zurek:1985qw,Zurek:1996sj}
realized that an analogous mechanism should also be possible in
symmetry-breaking phase transitions in condensed matter, e.g., in Helium.
As an important point, the response time of the system diverges when approaching
the critical point, which facilitates the observation of non-equilibrium phenomena
such as the creation of topological defects.
Note that the condensed-matter systems do typically not obey precisely the same equations of
motion as the fields in the early Universe, such that we do not have an exact quantitative
analogy -- instead the qualitative features such as symmetry-breaking phase transitions
with diverging response times (where the scaling depends on the universality class)
are reproduced.

\begin{figure}[tbp]
\centering
\includegraphics[width=.7\textwidth]{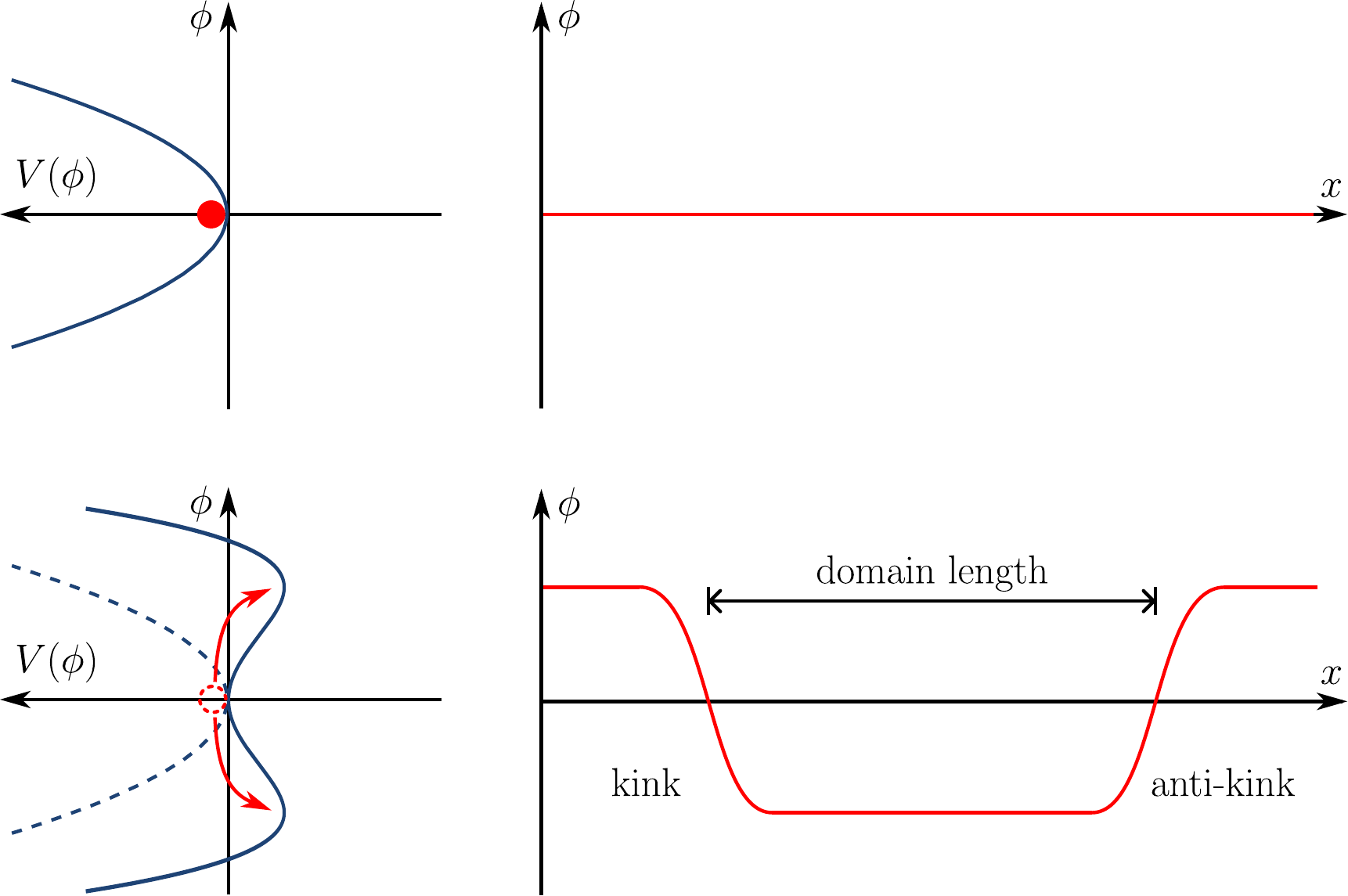}
\caption{Schematic of the Kibble-Zurek mechanism for the example of a 1+1 dimensional real 
scalar field~\eqref{double-well}.}
\label{fig:kibble-zurek} 
\end{figure}

In order to explain the main idea of the Kibble-Zurek mechanism, let us consider the
simple example of a real scalar field $\phi$ in 1+1 dimensions with a double-well potential,
see also Fig.~\ref{fig:kibble-zurek} 
\bea
\label{double-well}
{\cal L}
=
\frac{1}{2}\,
\left[(\partial_t\phi)^2-(\partial_x\phi)^2+c_2\phi^2-c_4\phi^4\right]
\,.
\ea
Symmetry-breaking phase transitions are often realized via decreasing the temperature.
However, according to the theme ``quantum simulators'', let us consider zero-temperature phase
transitions or quantum phase transitions instead, where an external parameter is varied
rather than the temperature\footnote{In this way, we also avoid problems due to the
Mermin-Wagner theorem \cite{Mermin:1966fe}.}.
In this case~\eqref{double-well}, changing $c_2$ with time brings us from the symmetric phase
$c_2<0$ with one potential minimum at $\phi=0$ to the broken-symmetry phase $c_2>0$ with
two potential minima at $\phi=\pm\sqrt{2c_4/c_2}$ (assuming $c_4>0$).
Now, if we start in the symmetric phase $c_2<0$ with $\phi=0$ and imagine increasing $c_2$
infinitely slowly, the system would stay in the ground state and thus has to ``choose''
one of the two minima (either left $\phi=-\sqrt{2c_4/c_2}$ or right $\phi=+\sqrt{2c_4/c_2}$)
when passing the critical point at $c_2=0$.
However, when traversing the critical point with a finite velocity, the system has not enough time
to ``communicate'' its choice to all other spatial positions and thus it could be that it chooses
the left minimum at one point but the right minimum at some other point.
As a result, there must be at least one topological defect in the form of a kink in between
these two points.
If there happens to be an anti-kink close to a kink, the two can annihilate each other in the
subsequent course of time\footnote{Note that this is not possible for the
sine-Gordon model~\eqref{sine-gordon} since this model is integrable and thus kinks and anti-kinks
would just go through each other or form stable bound states.
This shows that topological defects and solitons are not always the same -- even though these
terms are sometimes mixed up.}
but topological defects well separated from others would be long lived.

Since the Kibble-Zurek mechanism itself already refers to laboratory analogues of 
cosmological phenomena, it is probably redundant to speak of an analogue. 
There have been various experimental realizations, especially in the classical regime 
(e.g., in thermal phase transitions).
Besides Helium (see, e.g., \cite{Baeuerle:1996,Ruutu:1996}) as the system originally 
envisaged by Zurek \cite{Zurek:1985qw}, defect creation has been observed in several 
scenarios, including liquid crystals, non-linear optical systems, and superconductors 
etc., see also  \cite{Kibble:2007zz}.
%
%
Consistently with our topic and in order to avoid spreading out too much,
let us focus on ultra-cold atoms.
As in the previous Section, one option would be to use a lattice simulator as sketched in 
Fig.~\ref{fig:scalar}, which directly models the Lagrangian~\eqref{double-well} and
would allow us to study the formation of kinks and anti-kinks, 
see Eq.~\eqref{scalar-field+potential} in Sec.~\ref{Scalar field}. 
On the other hand, vortices in Bose-Einstein condensates are also topological defects 
(as discussed above) and vortex formation in Bose-Einstein condensates consistent with 
the Kibble-Zurek mechanism has already been observed \cite{Weiler:2008,Beugnon:2016mko}
in different geometries, e.g., harmonic (i.e., simply connected) and toriodal (i.e., ring). 
Going from the case of higher dimensions to very elongated, i.e., effectively one-dimensional, 
condensates, one would have kinks instead of vortices and their creation has also been observed 
\cite{Lamporesi:2013}.
If we have more than one species of atoms, e.g., spinor or multi-component condensates, 
the effective spin can play the role of the order parameter (instead of the $U(1)$ breaking 
phase of the wave function) and the formation of the associated topological defects in such 
systems has also been observed \cite{Sadler:2006cok,Anquez:2016}. 
As a completely different scenario, one could replace the bosonic atoms by fermionic ones. 
Of course, they cannot directly form a Bose-Einstein condensate, but two Fermions can 
combine to a Cooper pair, which can then condense to the analogue of a super-conducting 
state. 
This super-conducting state can also support vortices -- which facilitate the Kibble-Zurek 
mechanism \cite{Ko:2019ntc,Lee:2024kvo}. 

A completely different kind of phase transition has been realized \cite{Keesling:2018ish} 
with Rydberg atoms where two effects compete with each other. 
On the one hand, an external optical laser induces (detuned) Rabi oscillations between 
the ground state level and the highly excited Rydberg level of these atoms. 
In terms of the quantum Ising model, this would correspond to an on-site $\sigma^x_I$ term.
On the other hand, two atoms $I$ and $J$ in the highly excited Rydberg state experience
the van-der-Waals interaction, which translates to a $\sigma^z_I\sigma^z_J$ contribution
in terms of the quantum Ising model.
(The remaining on-site $\sigma^z_I$ terms can be canceled by suitably choosing the detuning.)
The competition between these two contributions, similar to a local magnetic field in 
$x$ direction $\sigma^x_I$ versus the (anti) ferromagnetic interaction $\sigma^z_I\sigma^z_J$
induces a symmetry-breaking quantum phase transition. 



Having realized such a (quantum) simulator for the Kibble-Zurek scenario, 
one can study several interesting questions. 
As one example, one can investigate the number density of the created topological defects 
(e.g., the inverse domain length in Fig.~\ref{fig:kibble-zurek}) depending on the transition 
velocity, such as $\dot c_2$ in the case discussed above.
For sufficiently slow transitions (which are still non-adiabatic because the response time 
diverges at the critical point), one would usually expect the Kibble-Zurek scaling 
$\propto\dot c_2^{\nu/(1+z\nu)}$ which is based on the idea to employ the equilibrium 
critical exponents (i.e., the critical exponent $\nu$ for the correlation length and the 
dynamical exponent $z$ for the response time) to infer non-equilibrium properties such as 
the created defect density.
For faster transitions, however, there could be deviations from this behaviour 
(see, e.g., \cite{Schaller:2023} and references therein). 
Apart from their pure number (density), the correlations between the created defects
is also in interesting question.
For the simple ${\mathbb Z}_2$ breaking case of the double-well potential in 
Eq.~\eqref{double-well} and Fig.~\ref{fig:kibble-zurek}, it is obvious that one can only 
have a kink and an anti-kink beside each other.
For other systems, however, such as the ${\mathbb Z}$ breaking case of the $\cos\phi$ 
potential in the sine-Gordon model~\eqref{sine-gordon}, one can have two kinks or two 
anti-kinks beside each other -- and thus the question of their correlations becomes 
more interesting. 
Going to higher dimensions, one can ask an analogous question regarding the correlation 
between (anti) vortices in Bose-Einstein condensates (created by the Kibble-Zurek mechanism), 
which correspond to a breaking of the $U(1)$ symmetry.
As one possible measure, these correlations would determine how the expectation value of the 
total winding number squared\footnote{The expectation value of the total winding number
itself must vanish due to symmetry.}, after going around a circle 
(which may encompass several vortices and anti-vortices),
scales with the radius of this circle.
These question can be generalized to other broken symmetries and higher dimensions, 
where it is also interesting to investigate the difference in scaling between topological
defects and other quasi-particles (e.g., phonons in Bose-Einstein condensates) 
created by traversing the critical point, see, e.g.,
\cite{Uhlmann:winding,Uhlmann:2010gg,Uhlmann:2010zz} and references therein.










\subsection{False-vacuum decay}\label{False-vacuum decay}

\begin{figure}[tbp]
\centering
\includegraphics[width=.7\textwidth]{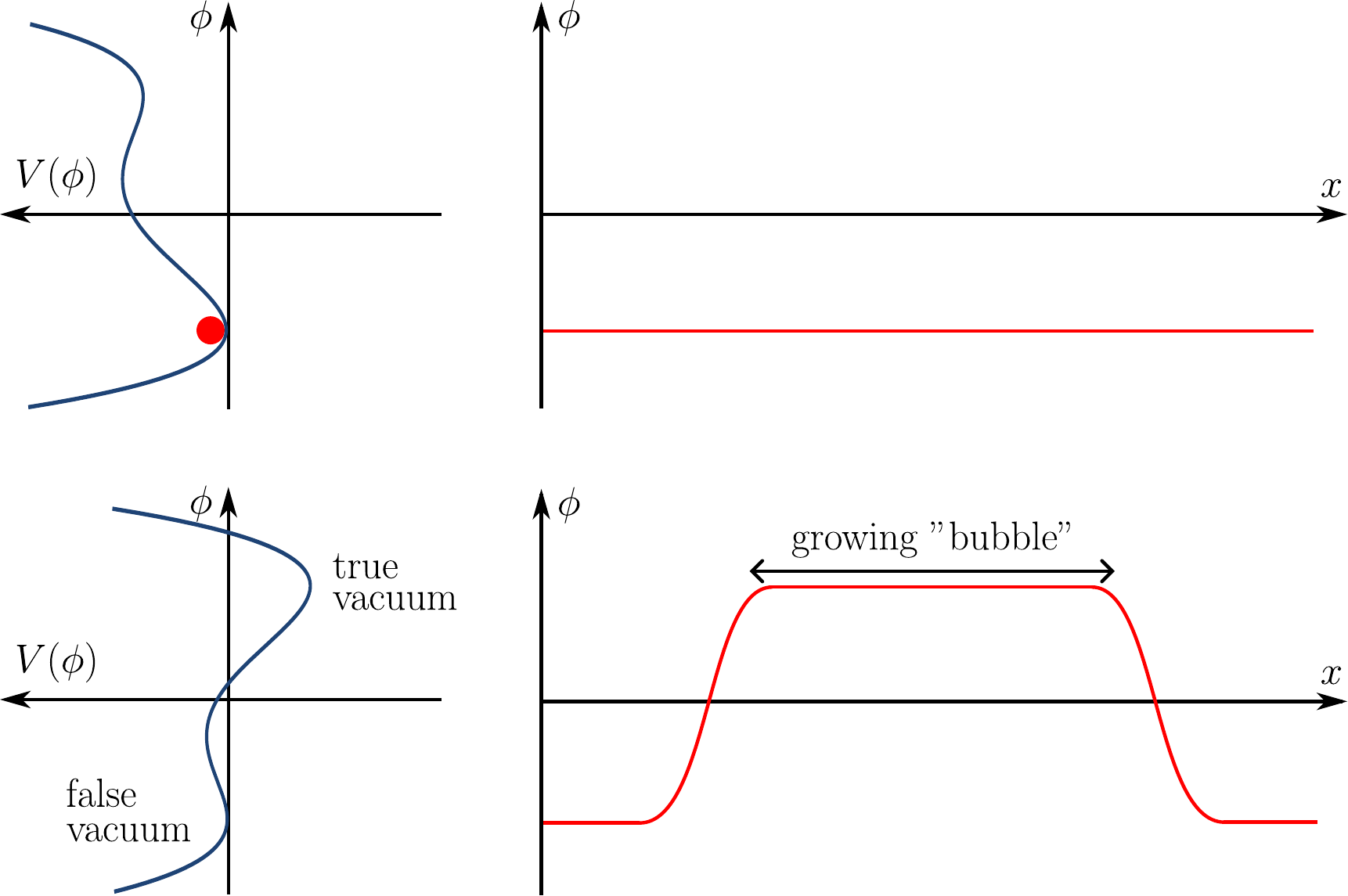}
\caption{Sketch of false-vacuum decay for a 1+1 dimensional scalar field~\eqref{false-vacuum}.
Initially (top), the field $\phi$ sits in the true ground state, i.e., 
the global minimum of the potential $V(\phi)$.
After this potential has been deformed (bottom), however, this field value is no longer the 
global minimum, but only a local minimum -- i.e., the false vacuum. 
In order to go to the true minimum, the field forms a ``bubble'' which must be large enough 
such that the energy gained inside the bubble compensates the energy cost of the kink and 
the anti-kink at its boundary.}
\label{fig:false-vacuum} 
\end{figure}

Another interesting phenomenon where we can apply many of the concepts discussed above is
false-vacuum decay \cite{Coleman:1977py,Callan:1977pt}. 
Again, let us explain the main idea by means of a very simple model, the real scalar field
in 1+1 dimensions
\bea
\label{false-vacuum}
{\cal L}
=
\frac{1}{2}\,
\left[(\partial_t\phi)^2-(\partial_x\phi)^2\right]-V(\phi)
\,,
\ea
but now with a slightly different potential $V(\phi)$.
As in the double-well potential~\eqref{double-well}, let us assume that $V(\phi)$ possesses two
local minima -- which are, however, no longer degenerate but slightly differ in depth by $\Delta V$.
One possibility would be to add a small linear term $c_1\phi$ to the double-well
potential~\eqref{double-well}.
The global (i.e., lower) minimum is then referred to as the true vacuum while the other
(local) minimum is called the false vacuum, see Fig.~\ref{fig:false-vacuum}. 

Now let us imagine that we initialize the system in the false vacuum at all positions $x$.
Similar to the Kibble-Zurek mechanism discussed in the previous section, this could happen
via a change of the potential landscape $V(\phi)$ in the early Universe
(then in 3+1 dimensions, of course) after which one could end up in a local minimum instead
the global one, i.e., the true vacuum.
However, in contrast to the scenario of the Kibble-Zurek mechanism which is analogous to a
second-order (symmetry-breaking) phase transition, this change of the potential landscape
$V(\phi)$ would correspond to a first-order phase transition\footnote{Note that this
important distinction bears similarities to the case of adiabatic quantum algorithms
discussed in Sec.~\ref{Quantum ground-state algorithms}.}.
Even though the false vacuum is not the lowest-energy state, it can be a metastable state --
in analogy to supercooled or undercooled water which can stay liquid after cooling it down
carefully below the freezing temperature.
Taking the analogy even further, the transition or decay to the real ground state
(such as the freezing of the water)
does typically not occur homogeneously, but in the form of expanding domains which require
a nucleation seed.
For the model~\eqref{false-vacuum}, this seed would be a kink and an anti-kink at a
sufficient distance $L$ such that energy gain $L\Delta V$ in the region given by 
this distance is 
large enough to compensate the energy cost of creating the kink and anti-kink.
In the absence of external perturbations
(such as sticking a spoon into the supercooled water)
this kink--anti-kink pair could be created by tunneling trough the potential barrier or
by quantum vacuum (or thermal) fluctuations which happen to conspire such that the field
can climb over the potential barrier on a sufficiently large length $L$.
Comparing this kink--anti-kink pair creation to the Sauter-Schwinger effect discussed in
Sec.~\ref{Sauter-Schwinger effect} we find several similarities.
After the kink--anti-kink pair has been created, it can mediate or catalyze the transition
to the true vacuum by expanding, i.e., increasing the distance
(similar to the electron-positron pairs in the Sauter-Schwinger effect 
which move way from each other after their creation). 

In 2+1 (or 3+1) dimensions, this nucleation seed would be an expanding bubble of the true
vacuum state surrounded by a domain of the false vacuum.
In this case, the energy gain given by $\Delta V$ times the bubble area (or the bubble volume)
should be large enough to compensate the energy cost of the domain boundary which scales with
the bubble radius (or the bubble surface area), see Fig.~\ref{fig:false-vacuum}. 

As we have seen in the undercooled water example discussed above, there are everyday 
analogues to false-vacuum decay -- at least in the classical regime.
Therefore, let us focus on ultra-cold atoms.
Similar to the previous sections, the correspondence between Eq.~\eqref{false-vacuum} above 
and Eq.~\eqref{scalar-field+potential} in Sec.~\ref{Scalar field} shows that one way to 
model false-vacuum decay could be the lattice simulator sketched in Fig.~\ref{fig:scalar}.
Going from the lattice to the continuum case, there have been several proposals for 
quantum simulators based on multi-component or spinor Bose-Einstein condensates 
under the influence of an external drive (e.g., microwave radiation), see, e.g., 
\cite{Fialko:2014xba,Braden:2017add,Braden:2019vsw,Billam:2020xna,Billam:2021psh,
Ng:2020pxk,Billam:2021nbc,Billam:2022ykl,Jenkins:2023eez,Jenkins:2023npg}.
Such an external drive can generate new minima of effective potential and thereby generate 
a potential landscape as in Fig.~\ref{fig:false-vacuum}. 
The emergence of a new potential minimum in driven systems can be visualized by an example 
from classical mechanics, the Kapitza pendulum.
If its pivot point oscillates vertically sufficiently fast and with large enough amplitude,
then the stable configuration would be that the pendulum points up instead of down 
(this phenomenon is also referred to as "dynamical stabilization"). 
In systems with more degrees of freedom such as Bose-Einstein condensates, 
one has to make sure that the external drive does not generate additional instabilities, 
e.g., acts as a seed for the decay of the false vacuum (similar to sticking a spoon into 
undercooled water). 
On the other hand, it has also been suggested to introduce such a seed in a controlled way, 
e.g., in the form of a vortex in the Bose-Einstein condensate \cite{Billam:2018pvp}. 
Similar to the Kibble-Zurek mechanism discussed in the previous section, a different 
realization could be based on Rydberg atoms, see, e.g., \cite{Darbha:2024srr}. 

An important step into the direction of simulating false-vacuum decay in the 
laboratory is the Trento experiment \cite{Zenesini:2023afv}
where an elongated (cigar shaped) spinor Bose-Einstein condensate has been realized. 
In analogy to the Vienna experiment 
\cite{Schweigler:2017,Zache:2019xkx,Schweigler:2021,Tajik:2022ycs}, 
one may split up the degrees of freedom into the total density 
$\rho_+=\rho_\uparrow+\rho_\downarrow$ (and phase) components plus density (and phase) 
differences $\rho_-=\rho_\uparrow-\rho_\downarrow$, which represent the pure spin degrees 
of freedom. 
For the latter, we get an effective double-well like potential as in Fig.~\ref{fig:false-vacuum}
where the energy of the two minima can be changed by varying the detuning of the external 
microwave radiation which induces Rabi oscillations between the spin states $\uparrow$ 
and $\downarrow$, similar to the Kibble-Zurek experiment \cite{Keesling:2018ish} 
using Rydberg atoms discussed in the previous section. 
Due to the strong elongation of the Bose-Einstein condensate, the system is effectively
1+1 dimensional, as in Eq.~\eqref{false-vacuum}.
%
%
Indeed, bubble formation in analogy to false-vacuum decay has been observed in the 
Trento experiment \cite{Zenesini:2023afv}, but one should be aware of two major deviations 
from original idea.
First, in contrast to the homogeneous case as in Eq.~\eqref{false-vacuum}, 
the Bose-Einstein condensate is inhomogeneous even in $x$ direction and hence 
bubbles are formed predominantly at or near the centre of cloud.
This is in contrast to spontaneous bubble creation mechanism which can occur anywhere 
with equal probability\footnote{In this regard, the centre of cloud acts a bit 
like the nucleation seeds (e.g., spoon) for under-cooled water discussed above.}. 
Second, the experimental results and parameters indicate that the false-vacuum decay 
in the Trento experiment \cite{Zenesini:2023afv} is not caused by quantum fluctuations.  
%
Probably it is also not induced by purely thermal (equilibrium) fluctuations, 
but instead by classical noise generated by the preparation steps. 
As a result, it is not fully clear (yet) to which extent the quantitative analogy 
to false-vacuum decay -- such as caused by relativistic tunneling in the 
system~\eqref{false-vacuum} -- can be applied. 
Nevertheless, the experimental results, such as the observed exponential dependence of 
the bubble formation probability on experimental parameters over more than one order of 
magnitude, are a real impressive achievement. 


As an alternative realization, one should also mention the Cambridge experiment 
\cite{Song:2021pyy}.
In contrast to the scenarios above which are based on weakly interacting condensates, 
the Cambridge experiment \cite{Song:2021pyy} employs strongly interacting 
(i.e., correlated) bosonic atoms in an optical lattice\footnote{Thus, the Cambridge 
experiment \cite{Song:2021pyy} simulating false-vacuum decay displays
some similarities to the Sauter-Schwinger or Breit-Wheeler effect in 
Sec.~\ref{Sauter-Schwinger and Breit-Wheeler effect} which can also be modeled
with strongly interacting fermionic or bosinic atoms in optical lattices, 
see, e.g., \cite{Queisser:2012,Queisser:2023aee} and references therein.}.
They can be described by an extended Bose-Hubbard Hamiltonian as in 
Eq.~\eqref{Bose-Hubbard Hamiltonian}, but generalized to two atomic species 
in the presence of an external drive.  
By resonantly shaking the optical lattice, one can destabilize the Mott insulator state
such that it is analogous to the meta-stable false vacuum state, while the true vacuum 
state (of the driven system) would be a super-fluid state with a staggered order
parameter\footnote{Such a spatially modulated super-fluid order parameter is often
also referred to as super-solid and the associated instability can be visualized
via the dispersion relation, where an effective ``roton'' dip touches the $k$ axis,
see, e.g., \cite{Schutzhold:supersolid} and references therein.}.
The meta-stability of the Mott insulator state and its decay to the true ground state have 
been observed in the Cambridge experiment \cite{Song:2021pyy} as a function of the experimental
parameters such as the shaking amplitude. 
Similarly to the Trento experiment \cite{Zenesini:2023afv}, the  
Cambridge experiment \cite{Song:2021pyy} is clearly a qualitative analogy to false-vacuum 
decay (and thus a very important step), but the extent of the quantitative analogy 
requires further studies. 

Having realized a quantum simulator for false-vacuum decay would allow us to address 
several interesting questions. 
As a non-perturbative phenomenon of an interacting quantum field theory, 
false-vacuum decay is a highly non-trivial problem and it is probably fair to 
say that our understanding is far from complete, see also 
\cite{Pirvu:2021roq,Braden:2022odm} and references therein. 
As already indicated above, open questions exist regarding the microscopic mechanism, 
including the role of quantum and thermal fluctuations, their space-time conspiracy 
versus a simple one-dimensional tunneling picture etc. 
Similar to the Kibble-Zurek mechanism discussed in the previous section, the correlations 
between different bubbles is also an interacting (and related) question. 
The same applies to the symmetry of bubbles: usually, one starts with an ansatz of a highly 
symmetric (e.g., spherical) bubble, but the probabilities for symmetric and asymmetric bubbles 
should be determined and compared. 
As another point, it would also be very interesting to generalize the static 
Lagrangian~\eqref{false-vacuum} to an explicitly time-dependent background 
(such as an expanding Universe). 
The additional time dependence could dynamically assist (or even suppress) 
false-vacuum decay via tunneling, see also \cite{Kohlfurst:2021dfk,Ryndyk:2023hfm}
and references therein. 
As a final remark, such a quantum simulator would also allow us to investigate the 
impact of the (omnipresent) environment. 
For example, it could stabilize the false vacuum via the quantum Zeno effect \cite{Misra:1976by}. 












\subsection{Back-reaction and cosmological constant}
\label{Back-reaction and cosmological constant}

A less obvious but very important point where the non-linearity plays a role is the back-reaction 
of the quantum effects discussed in chapter~\ref{Linear fields} onto the (approximately classical) 
background. 
As an example, let us take the case of Hawking radiation where the in-falling partner particles 
carry negative energy (from the point of view of a static observer far away). 
They should decrease the mass of the black hole in order to compensate for the energy emitted 
via Hawking radiation.
However, how this should work in detail is still a matter of 
debate\footnote{This debate is also related to the black-hole information puzzle.
On can try to gain some insight using simplified toy models (see, e.g.,
\cite{Maia:2007vy,Plunien:2004bq,Schutzhold:2008zzc}),
but there are many open question.
\rs{For the back-reaction in optical systems, see, e.g., \cite{Marino:2019flp} and 
\cite{Baak:2024ajn}.}}.

In this respect, one can hope to learn something from the black-hole analogues,  
e.g., in Bose-Einstein condensates, because there the microscopic physics is better 
understood.
However, one should also be aware of some important differences between the two cases. 
First, the fluid analogue such as the de~Laval nozzle in Fig.~\ref{fig:laval}
is not a closed system 
-- there is a constant in- and out-flux of matter and energy which changes the concept 
of energy balance in comparison to the real black-hole case, 
\rs{see also \cite{Ribeiro:2021fpk}.}
Second, as already discussed above, the black-hole analogues in Bose-Einstein condensates
do not reproduce the Einstein equations, which are crucial for the back reaction in real 
black holes. 

Taking into account the above differences, there have been several theoretical 
investigations regarding the back-reaction for black-hole analogues 
\rs{and other scenarios} in Bose-Einstein condensates,
see, e.g., \cite{Balbinot:2004da,Balbinot:2004dc,Schutzhold:2005ex,Balbinot:2005zkf,
Fagnocchi:2006cz,Schutzhold:2007fg,Liberati:2020mdr,Baak:2022hum,Pal:2024qno}. 
It should also be mentioned here that there were experimental studies 
\cite{Patrick:2019kis} of the back-reaction in black-hole analogues -- 
but instead of phonons in Bose-Einstein condensates, 
surface waves in flowing liquids \cite{Schutzhold:2002rf} 
where used to realize a black-hole analogue 
(which then typically recovers classical aspects instead of quantum phenomena,
at least for the experiments so far). 

\subsubsection{Cosmological constant problem}
\label{Cosmological constant problem}

In order to circumvent some of the difficulties of the black-hole analogues mentioned above, 
let us consider the much simpler case of a homogeneous condensate at rest. 
Still, the quantum fluctuations of the phonon field (in its ground state) should have some 
impact -- which might be interpreted as a zero-point pressure contribution.
This zero-point pressure can be considered as a toy model for studying the 
cosmological constant problem, see also \cite{Jannes:2011em,Finazzi:2011zw}. 

When trying to measure how the expansion of our present Universe slows down,
it has been found that it does not, but is actually accelerating.
This striking finding has been awarded with the Nobel prize in physics 2011. 
To understand why this came as a surprise to many, let us consider the 
Friedmann equation (with vanishing spatial curvature)
\bea 
\label{Friedmann}
\frac{\ddot a}{a}
=
-\frac{4\pi G_{\rm N}}{3}\left(\rho+3p\right) 
\,,
\ea
which governs the time-dependence of the scale factor $a$ representing the 
expansion of our Universe. 
Since all matter we know of has positive energy or mass density $\rho>0$ 
and positive (or negligible) pressure $p\geq0$, one would expect $\ddot a<0$,
i.e., a deceleration of the expansion. 
Intuitively speaking, gravity is attractive for all kinds of ordinary matter 
-- which should slow down the expansion. 
Thus, the observed acceleration of the expansion indicates some sort of 
repulsive effect of gravity.
These observations suggest adding a cosmological constant $\Lambda$ to the 
Einstein equations or, equivalently, a term $\propto\Lambda g_{\mu\nu}$ to 
the energy-momentum tensor $T_{\mu\nu}$. 
Such a cosmological constant $\Lambda$ yields an effective pressure 
$p_\Lambda=-\rho_\Lambda$ 
which equals the {\em negative} density $\rho_\Lambda$.  
In analogy to dark matter, such a contribution is often referred to as  
``dark energy''.
However, this is a bit misleading since the energy density 
$\rho_\Lambda=\Lambda>0$ is not the dominant term in the 
Friedmann equation~\eqref{Friedmann}, i.e., it does not change the 
sign of the right-hand side -- it is the negative pressure 
$p_\Lambda=-\rho_\Lambda$ which is important here. 
Thus, ``dark energy'' should actually better be called ``dark pressure''. 

A prominent idea to explain this contribution $\propto\Lambda g_{\mu\nu}$
to the energy-momentum tensor $T_{\mu\nu}$ is the energy and pressure of the 
vacuum itself. 
However, this requires calculating that quantity. 
A naive approach would be to consider the zero-point energy $\hbar\omega_{\fk{k}}/2$
of each mode $\f{k}$ and then to sum up all these contributions. 
Of course, this yields an ultra-violet divergence scaling with $\f{k}^4$
in 3+1 dimensions.
Now, one could argue that this $d^3k$-integration should be cut off at the Planck 
scale $k_{\rm Planck}\sim1/\ell_{\rm Planck}=\sqrt{c^3/\hbar G_{\rm N}}$ 
already discussed in Sec.~\ref{Hawking radiation} because the quantum field theory 
we are using is probably only valid for $\omega$ and $k$ scales below $k_{\rm Planck}$.
Using $k_{\rm Planck}$ as a sharp cut-off would give us 
$\Lambda=\ord(k_{\rm Planck}^4)$ which is a huge energy or mass density. 

Comparing the value $\Lambda=\ord(k_{\rm Planck}^4)$ estimated in this way with 
the value for $\Lambda$ obtained from the observed acceleration of our Universe 
$\Lambda=\ord(10^{-26}~\rm kg/m^3)$, one finds that they differ by 120~orders of magnitude. 
This is sometimes called the ``biggest discrepancy'' or the ``worst prediction'' in physics. 
However, calling the above estimate giving $\Lambda=\ord(k_{\rm Planck}^4)$ a prediction is 
an exaggeration. 
There are at least two major problems. 
First, the dominant contribution to the value $\Lambda=\ord(k_{\rm Planck}^4)$ stems from 
fluctuations at the Planck scale -- where we do not know what happens. 
For the above estimate, one implicitly assumes that the low-energy behaviour we know 
is correct up to the Planck scale $k<k_{\rm Planck}$ and that afterwards $k>k_{\rm Planck}$
no further contributions arise. 
This is a very strong assumption which is not really justified.
For example, it could well be that modifications at the Planck scale lead to a completely 
different behaviour, possibly even with negative contributions -- which might cancel the 
low-energy part, see also \cite{Volovik:2010vx}. 
Second, a proper calculation should start with the full underlying interacting theory 
in terms of the fundamental degrees of freedom and their correct operator ordering 
(which unfortunately we do not have). 
Then, one should split this up (if it is possible at all) into an approximately classical 
background space-time plus quantum fluctuations in top of it. 
For example, there might be some sort of symmetry principle which enforces the effective 
cosmological constant $\Lambda$ derived in this way to vanish in flat 
space-times\footnote{Such a symmetry principle could perhaps also explain why the 
QCD trace anomaly -- which would also lead to an effective cosmological constant
which is far too large -- does not contribute, at least not in flat space-times, 
see also \cite{Schutzhold:2002sg}.}. 

\subsubsection{Back-reaction in fluids and Bose-Einstein condensates}
\label{Back-reaction in fluids and Bose-Einstein condensates}

When faced with a difficult problem such as the one described above, it can be useful to consider 
a much simpler scenario where an analogous problem occurs. 
Thus, let us study a perfect fluid described by the Bernoulli equation 
$\dot\phi+V/m+g_s\rho/m+(\na\phi)^2/2=0$ as derived in Sec.~\ref{Analogue gravity}. 
Now we take this classical equation and assume that it is valid for operator-valued fields 
$\hat\phi$ and $\hat\rho$ (see also the next subsection).
Then we split these quantum fields up into their classical background values
$\phi_{\rm class}=\langle\hat\phi\rangle$ and $\rho_{\rm class}=\langle\hat\rho\rangle$
plus quantum fluctuations 
$\hat\phi=\phi_{\rm class}+\delta\hat\phi$ and $\hat\rho=\rho_{\rm class}+\delta\hat\rho$. 
Inserting this split into the (quantum) Bernoulli equation and taking the expectation value 
yields 
\bea
\label{Bernoulli+back-reaction-1}
\dot\phi_{\rm class}+\frac{V}{m}+\frac{g_s\rho_{\rm class}}{m}
+\frac12\left(\na\phi_{\rm class}\right)^2
=
-\frac12\left\langle\left(\na\delta\hat\phi\right)^2\right\rangle 
\,,
\ea
where the quantum correction on the right-hand side describes the back-reaction of the 
quantum fluctuations of the phonon field onto the classical background. 
Qualitatively, this would be the analogue to the cosmological constant discussed above. 
Even quantitatively, we find the same dependence $k^4_{\rm cut}$ on the cut-off scale
since the phonon field $\delta\hat\phi$ behaves as a mass-less scalar field at small $k$, 
see Sec.~\ref{Analogue gravity}. 

For Bose-Einstein condensates, the first natural cut-off scale would be the healing 
length $\xi$ where the dispersion relations changes, as discussed in 
Sec.~\ref{Analogue gravity}. 
At these length scales, we also have to include the ``quantum pressure'' term in 
Eq.~\eqref{Hamilton-Jacobi}. 
Taking the ``quantum pressure'' term into account, we may now apply the same procedure 
as for Eq.~\eqref{Bernoulli+back-reaction-1}. 
Assuming that the quantum fluctuations $\delta\hat\rho$ are much smaller than the 
classical background $\rho_{\rm class}$ we may Taylor expand in terms of $\delta\hat\rho$
and arrive at an additional back-reaction term \cite{Schutzhold:2007fg} 
\bea
\label{Bernoulli+back-reaction-2}
\dot\phi_{\rm class}+\frac{V}{m}+\frac{g_s\rho_{\rm class}}{m}
+\frac12\left(\na\phi_{\rm class}\right)^2
=
-\frac12\left\langle\left(\na\delta\hat\phi\right)^2\right\rangle 
-\frac{\hbar^2}{8m^2\rho_{\rm class}^2}
\left\langle\delta\hat\rho\na^2\delta\hat\rho\right\rangle 
+\ord(\delta\hat\rho^4) 
\,.
\ea
After an integration by parts, we see that the additional term containing the density 
fluctuations $\delta\hat\rho$ and the previous term~\eqref{Bernoulli+back-reaction-1}
from the phase fluctuations $\delta\hat\phi$ have the opposite sign and thus partly 
cancel each other. 
Moreover, if we take the naive analogy to the scalar field and forget about the 
modified dispersion relation, this additional term would have an even stronger 
ultra-violet divergence $k^6_{\rm cut}$ because $\delta\hat\rho$ represents the 
field momentum density $\propto\partial_t\delta\hat\phi$. 

When taking into account the full dispersion relation, however, the leading divergences 
cancel each other -- including the $k^4_{\rm cut}$ divergence which would be the analogue 
to the estimate in Sec.~\ref{Cosmological constant problem}. 
What remains is a $k^3_{\rm cut}$ divergence, but now the cut-off is no longer determined 
by the healing length since the associated change of the dispersion relation is already 
included \cite{Schutzhold:2007fg}.
Instead, this cut-off stems from the approximation of a contact interaction 
$W(\f{r}-\f{r}')\approx g_s\delta^3(\f{r}-\f{r}')$. 
For a smooth and regular function $W(\f{r}-\f{r}')$ such as a Gaussian, this 
divergence would disappear. 
Still, it would be premature to associate this result $k^3_{\rm cut}$ to the 
back-reaction of the quantum fluctuations. 
As we shall see below, this $k^3_{\rm cut}$ contribution is just an artifact of 
operator ordering. 

For Bose-Einstein condensates, we actually know what the correct result should be
(up to the level of accuracy under consideration). 
Instead of the effective Bernoulli equation, one should start from the evolution 
equation~\eqref{pre-gp} in terms of the original many-body field operators $\hat\psi$
and insert the split $\hat\psi=\psi_{\rm c}+\hat\chi$ (or its particle-number conserving version). 
Then we find the modified Gross-Pitaevskii equation including the back-reaction corrections  
\bea
\label{Gross-Pitaevskii+back-reaction} 
i\hbar\,\frac{\partial\psi_{\rm c}}{\partial t}
=
\left(-\frac{\hbar^2}{2m}\,\na^2+V+g_s|\psi_{\rm c}|^2+
2g_s\langle\hat\chi^\dagger\hat\chi\rangle\right)\psi_{\rm c}
+g_s\langle\hat\chi^2\rangle\psi_{\rm c}^*
\,.
\ea
The term $\langle\hat\chi^\dagger\hat\chi\rangle$ is referred to as the quantum depletion 
and corresponds to the (small) density of atoms which are pushed out of the condensate 
state $\psi_{\rm c}$ by the finite interaction $g_s$. 
Even using the contact interaction 
$W(\f{r}-\f{r}')\approx g_s\delta^3(\f{r}-\f{r}')$,
this term $\langle\hat\chi^\dagger\hat\chi\rangle$ is finite. 
The so-called anomalous term $\langle\hat\chi^2\rangle$ displays weak ultra-violet 
divergence $\sim k_{\rm cut}$ which disappears when more realistic interaction potentials  
$W(\f{r}-\f{r}')$ are considered, see, e.g., \cite{Mean-field-expansion}. 

Even though many important details are left out here, this analogy shows that the naive  
estimate $k^4_{\rm cut}$ sketched in the previous Sec.~\ref{Cosmological constant problem}
should not be trusted too much. 
Of course, apart from these theoretical investigations, one could ask for experimental 
verification, e.g., how to actually measure the quantum depletion 
$\langle\hat\chi^\dagger\hat\chi\rangle$. 
One way of doing so are time-of-flight measurements as discussed in Sec.~\ref{Measurement schemes} 
where one can also increase the signal by imposing an optical lattice potential which effectively
enhances the interaction strength, see, e.g., \cite{Xu:2006}.
Note that a non-adiabatic evolution is crucial for such a measurement of 
$\langle\hat\chi^\dagger\hat\chi\rangle$. 
To see that, let us imagine the following scenario:
We start with the many-body ground state of $N$ atoms in a finite volume and with a finite 
interaction strength $g_s$, which implies a finite $\langle\hat\chi^\dagger\hat\chi\rangle>0$. 
If we then increase the volume or decrease the interaction strength $g_s$ (or both) very slowly, 
i.e., adiabatically, we stay in the many-body ground state.
Thus, we would end up in a final ground state of negligible density or interaction strength $g_s$ 
(or both) which then also has a vanishing quantum depletion 
$\langle\hat\chi^\dagger\hat\chi\rangle=0$.
In contrast, if we suddenly (or rapidly) switch off the interaction strength $g_s$ or let the 
condensate expand quickly, i.e., non-adiabatically, not all the atoms can ``find their way back''  
to the ground state and thus we are left with a finite number of atoms in excited states -- 
which are a smoking gun for the initial quantum depletion 
$\langle\hat\chi^\dagger\hat\chi\rangle$.
Thus, by changing the interaction strength $g_s$ or by letting the condensate expand
-- both depending on how fast those variations occur -- we may influence the quantum depletion 
$\langle\hat\chi^\dagger\hat\chi\rangle$, which can also be measured {\em in situ} by 
momentum-selective two-photon Bragg scattering, see, e.g., \cite{Lopes:2017}.
As a result, we find close analogies to the analogue of cosmological particle creation,
as discussed in Sec.~\ref{Cosmological particle creation}. 

\subsubsection{Quantum fluid dynamics}
\label{Quantum fluid dynamics}

To conclude this section, let us sketch the operator-ordering problems mentioned above. 
We start from the full microscopic description in terms of the many-body field operators 
$\hat\psi$ and $\hat\psi^\dagger$ and their Hamiltonian (in normal ordering)
\bea
\label{Hamiltonian-back}
\hat H=\int d^3r
\left(
\frac{(\na\hat\psi^\dagger)\cdot(\na\hat\psi)}{2m}
+
V\hat\psi^\dagger\hat\psi
+
\frac{g_s}{2}\,
(\hat\psi^\dagger)^2\hat\psi^2
\right) 
\,.
\ea
The next step would be to generalize the Madelung split~\eqref{Madelung} 
to the quantum case where the many-body field operators are represented by 
self-adjoint density and phase operators 
(more precisely, operator-valued distributions)
\bea
\label{Madelung-quantum}
\hat\psi=\exp\left\{i\hat S\right\}\sqrt{\hat\rho} 
\,.
\ea
Note that the commutation relation 
$[\hat\psi^\dagger(\f{r}),\hat\psi(\f{r}')]=\delta^3(\f{r}-\f{r}')$
implies that $\hat S$ and $\hat\rho$ do not commute.
As a result, operator ordering is important here and other forms such 
as $\hat\psi=\sqrt{\hat\rho}\exp\{i\hat S\}$ are inconsistent with 
$\hat\rho=\hat\psi^\dagger\hat\psi$.
Apart from the difficulties stemming from the character of $\hat S$ and $\hat\rho$ as 
operator-valued distributions, the phase operator $\hat S$ should be treated with special 
care\footnote{Some of the problems can already arise for the simpler case of a  
harmonic oscillator if we set $\hat a^\dagger=\sqrt{\hat n}\exp\{-i\hat S\}$
and $\hat a=\exp\{i\hat S\}\sqrt{\hat n}$.}.

At this point, we can already identify the operator-ordering problem mention in the 
previous Sec.~\ref{Back-reaction in fluids and Bose-Einstein condensates}. 
If we translate the interaction term $(\hat\psi^\dagger)^2\hat\psi^2$ in the 
Hamiltonian~\eqref{Hamiltonian-back} into density operators, we find 
$(\hat\psi^\dagger)^2\hat\psi^2=\hat\rho^2-\hat\rho\delta^3(0)$ in view of the 
commutation relation 
$[\hat\psi^\dagger(\f{r}),\hat\psi(\f{r}')]=\delta^3(\f{r}-\f{r}')$. 
Then, if we try to derive the quantum version of the Bernoulli equation and calculate 
$\partial_t\hat S$ via the commutator $[\hat S,\hat H]$, we see that it acquires a 
singular contribution $\propto\delta^3(0)$ which explains the ultra-violet singularity
$k^3_{\rm cut}$ found above. 

In order to avoid such problems, let us start from the evolution equation
$\partial_t\hat\psi$ of the original field operator and insert the quantum Madelung
representation~\eqref{Madelung-quantum}. 
As expected, we find a quantum version of the equation of continuity 
\bea
\label{continuity-quantum}
\partial_t\hat\rho+\na\cdot\hat{\f{j}}=0
\qquad 
{\rm with}
\qquad 
\hat{\f{j}}=\frac{\hat\psi^\dagger\na\hat\psi-{\rm h.c.}}{2mi}
=\frac{\sqrt{\hat\rho}\,(\na\hat S)\sqrt{\hat\rho}}{m}
\,,
\ea
which is exact. 
Note that the symmetric form $\sqrt{\hat\rho}\,(\na\hat S)\sqrt{\hat\rho}$ 
ensures that $\hat{\f{j}}$ is self-adjoint. 
In order to define a velocity operator $\hat\vau$ one should multiply $\hat{\f{j}}$
with $\hat\rho^{-1/2}$ from left and right -- which becomes problematic if the density 
can become zero (such as at a vortex core). 

Since the quantum version of the equation of continuity~\eqref{continuity-quantum} 
has exactly the same for as in the classical case, one could say that it does not 
acquire any quantum back-reaction corrections\footnote{Note that such a statement depends on
the definition of the classical background.
For example, if takes the condensate wave function $\psi_{\rm c}$ as the classical background,
the associated density $\rho_{\rm c}$ would acquire a quantum correction
$\langle\hat\rho\rangle=\rho_{\rm c}+\langle\hat\chi^\dagger\hat\chi\rangle$
due to the quantum depletion discussed after Eq.~\eqref{Gross-Pitaevskii+back-reaction}.
}.
However, the situation is more complicated for the quantum version of the Euler 
or Bernoulli equation 
\bea
i\partial_t\hat\psi
=
i\partial_t\left(e^{i\hat S}\sqrt{\hat\rho}
\right)
=
\left(
-\frac{i\na^2\hat S-(\na\hat S)^2}{2m}\,
+V+g_s\hat\rho
\right)
e^{i\hat S}\sqrt{\hat\rho}
\,.
\ea
Here operator ordering is a much more severe issue.
Note that, even though $\hat\rho$ and $\hat S$ at different spatial positions 
(but the same time) commute and thus $\na\hat\rho$ commutes with $\hat\rho$
(again at the same time), this is no longer true for the time derivatives, i.e.,
$\hat\rho$ and $\partial_t\hat\rho$ do not commute 
(same for $\hat S$ and $\partial_t\hat S$). 

\newpage
\section{Summary and outlook}\label{Summary and outlook}

Ultra-cold atoms possess many features which make them very good candidates for realizing quantum
simulators.
They can be isolated very well from the environment and cooled down (as the nomenclature already
suggests) to extremely low temperatures.
Furthermore, their characteristic energy scales are very low and thus the associated internal
frequency scales very slow such that it is possible to externally manipulate them  quite fast
in comparison, i.e., non-adiabatically (sometimes also called diabatically).
The means for external manipulation are also quite advanced, one can tune the one-particle
potential felt by the atoms to a large degree (e.g., by optical lasers) and even change their
two-particle interaction strength (e.g., by a magnetic field via Feshbach resonances).
In addition, one can exploit the various internal atomic states and induce resonant or
near-resonant transitions, e.g., via optical lasers or microwaves.
Finally, the state of the quantum simulator -- usually the atom number or density --
can be read out with high accuracy down to the single-atom level.
In contrast to all these advantages, one should also keep in mind an important drawback of
ultra-cold atoms -- the problem of three-body losses -- which places ultimate limits on
their performance.

Apart from these experimental aspects, our theoretical understanding of ultra-cold atoms is
also quite advanced, e.g., regarding Bose-Einstein condensates in the dilute-gas limit.
This combination of experimental and theoretical progress allowed us to propose and even
realize several quantum simulators based on ultra-cold atoms -- in the spirit of the famous
quote from Feynman: ``The same equations have the same solutions'' \cite{Feynman:1963uxa}.
Especially for relativistic phenomena, a selection of proposals, theoretical investigations
and experimental realizations have been discussed in the previous sections.
In chapter~\ref{Linear fields}, we discussed examples for linear fields
-- though under the influence of non-trivial external conditions.
%
%
A recurring theme was the tearing apart of quantum vacuum fluctuations,
by gravitational or electric fields, the cosmic expansion or other external influences.
Another common feature is quantum
squeezing\footnote{However, claiming the experimental verification of Hawking radiation or
the Unruh effect just because the appearance of squeezing in an experiment is not really
justified.},
which can be visualized by means of the
coupled oscillator picture sketched in Fig.~\ref{fig:hawking}.
%
%
In chapter~\ref{Non-linear fields}, we considered some cases of non-linear fields.
Here the competition between (local) minima in the potential landscape leads to
interesting phenomena beyond standard perturbation theory.



\subsection{Lessons (to be) learned}

Of course, when discussing such quantum simulators, one should also think about the question of
what we can actually learn from this endeavour.
Clearly, one of the ultimate goals of quantum simulators is to simulate a quantum system of
interest which is so complex that we cannot do these calculations on a classical computer anymore.
Even though we are (depending on who is asked that question) not quite there
yet\footnote{Note that it is a non-trivial task to define that question properly.
For example, one could argue that a system of strongly interacting electrons in a solid
(whose complex quantum dynamics cannot be simulated on available classical computers)
is a quantum simulator for itself.
Thus, one should impose certain demands (e.g., regarding state preparation and read-out)
on a quantum simulator.},
it is very likely that we can realize such a quantum simulator long before
we have a sufficiently large quantum computer that is able to solve classical problems
(e.g., factoring large numbers) better than any classical computer.
On the other hand, even before reaching the ``holy grail'' of a quantum simulation
impossible on a classical computer, the study of quantum simulators can and does
create useful insights -- as in the well-known saying attributed to Confucius:
``The journey is the reward.''
In the following, a personal and probably incomplete list is given:

{\bf Benchmarking} \quad
If we want to use a quantum simulator in order to perform a simulation which is intractable on
any available classical computer, we have to make sure that this quantum simulator is behaving
in precisely the same way we think it is, see also \cite{Lewenstein:2012afn}.
Unless we have an analytic solution to compare with (e.g., in integrable systems such as
the sine-Gordon model discussed in Sec.~\ref{sine-gordon}), the intractability on any
available classical computer prevents us from checking the obtained results.
Thus, in order to reach the ``holy grail'' mentioned above, it is crucial to benchmark the
quantum simulator -- e.g., by applying it to cases where we already know what the result should be.
This allows us to estimate the performance and accuracy 
of the quantum simulator (see also the next point).

{\bf Robustness} \quad
Turning the argument around, one can use the imperfections present in a given quantum simulator
in order to investigate the robustness of the phenomenon under consideration.
Since basically all of our descriptions are based on idealizations and/or approximations,
it is important to study how the resulting predictions ``survive'' in reality.
Taking Hawking radiation discussed in Sec.~\ref{Hawking radiation} as an example,
the theoretical calculations of this effect in the presence of the modified dispersion
relation~\eqref{BEC-disperion} of a Bose-Einstein condensate show that the particles emitted
at long wavelengths do basically not depend on these modifications at short length scales
(even though its original derivation might suggest that, cf.~the trans-Planckian problem),
see also Fig.~\ref{fig:laval-stages} and \cite{Schutzhold:2008tx}.
On the experimental side, the observation of Hawking emission at the Technion
\cite{Lahav:2009wx,Steinhauer:2015ava,Steinhauer:2015saa,MunozdeNova:2018fxv,Kolobov:2019qfs}
indicate that the main mechanism of Hawking radiation is not destroyed by the imperfections
which are undoubtedly present in the experimental apparatus.

{\bf Connecting theory and experiment} \quad
As Goethe said, ``Gray, dear friend, is all theory.''
Physics is one of the natural sciences and its goal is to describe the world around us.
Thus, theoretical physicists should leave their ``ivory tower'' from time to time and
have their theories tested in experiments (or at least compared to observations).
Real progress is only possible via an interplay of theory and experiment (or observations).
In cases where direct experimental tests are out of reach, such as Hawking radiation or
cosmological particle creation, quantum simulators are an interesting alternative
to test some of the underlying principles and mechanisms via the analogy.

{\bf Essential and inessential features} \quad
As another important point, studying the analogies between the quantum simulators and the
phenomena to be simulated helps us to distinguish their essential and inessential features,
cf.~\cite{Visser:2001kq}.
Again taking Hawking radiation discussed in Sec.~\ref{Hawking radiation} as an example,
we find that it requires an effective black-hole horizon (as in Fig.~\ref{fig:laval})
but it does not hinge on the Einstein equations.
Similarly, the Sauter-Schwinger effect discussed in Sec.~\ref{Sauter-Schwinger effect}
can be deduced from the Dirac equation or the Klein-Fock-Gordon equation describing
charged spinor or scalar particles in the presence of an external electromagnetic field,
but it does not really depend on the origin of that electromagnetic field, e.g.,
whether it is a solution of the Maxwell equations\footnote{Actually, for a given
vector potential $A_\mu$, the Maxwell equations just tell us which sources $j_\mu$
would be required to generate this field configuration.
A similar argument can be made for the Einstein equations in terms of the metric
$g_{\mu\nu}$ and the source term $T_{\mu\nu}$.}
or the Proca equation.

{\bf Physics intuition} \quad
In order to explain this point, it might be useful to make a small detour into the
field of artificial intelligence (AI).
Nowadays artificial neural networks can perform tasks such as speech recognition,
reading handwriting, or even playing chess with astonishing performances --
provided that they are trained well.
As a rule of thumb, the performance increases with the amount of data available for
training and the diversity of the data (e.g., different samples of handwriting).
Now, since such artificial neural networks are motivated by our brain, it is probably
not too much a stretch to conclude that our physics intuition can also be trained by
studying different analogies, especially those between seemingly unrelated areas
(see also the next point).

At the same time, this intuition should be substantiated by hard facts and actual calculations
in order to avoid misconceptions and confusion -- 
\rs{as in the quote attributed to Einstein:} 
``Make things as simple as possible, but no simpler.''
In the AI context, training with bad data can lead to wrong recognition and false classification.
Thus, for each quantum simulator, one must clarify to which extent the analogy applies and
where the limits of the analogy are\footnote{For example, just finding indications of
thermal behaviour and thus concluding (without further arguments) that one found signatures
of the Unruh effect or Hawking radiation is probably an exaggeration.
%
%
As another example, a system where the effective propagation speed vanishes at some position
is not necessarily a good analogue for a black hole (where everything can fall in but nothing
can escape).
As a counter-example, consider the shore line of the ocean where the speed of the water waves
goes to zero.}.

{\bf Crossing borders} \quad
In the history of physics and other sciences, transferring ideas from one sub-discipline to
another has often led to great breakthroughs.
As one example, the Nobel price in physics 2024 was awarded for the application of
methods from statistical physics to artificial neural networks.
Another example is the Anderson-Higgs mechanism bridging solid state and particle physics.
It is very likely that some of today's open problems in physics (e.g., in quantum gravity)
can also be solved by such a transfer of methods and ideas -- thereby connecting
different communities.
Actually, as became (hopefully) evident from the discussions in this article,
the study of quantum simulators is precisely doing that, i.e., transferring methods and
ideas across the borders of sub-disciplines and connecting different communities.
Thus, studying quantum simulators can help us to break through the walls which
(unfortunately) separate
the sub-disciplines  and to apply ideas which are well known and understood
in one community to another community where they are less known.
This is especially important for the training of young researchers and might help them
to ``look beyond their own nose'' and to see the bigger picture.
Last but not least, venturing into new regimes often yields totally unexpected new
discoveries and ideas, which prove very important in hindsight.

%

\subsection{Outlook}

Obviously, there are many remaining open questions, to say it with Newton:
``What we know is a drop, what we don't know is an ocean.''
For example, comparison between Figs.~\ref{fig:adiabatic} and \ref{fig:first-second}
on the one hand and Figs.~\ref{fig:kibble-zurek} and \ref{fig:false-vacuum}
on the other hand suggests an intriguing triangle of analogy including quantum information
(e.g., adiabatic quantum algorithms), relativistic quantum field theory, and
condensed matter/solid state physics (e.g., Landau-Zener tunneling, quantum phase
transitions of first and second order) which promises to yield further fascinating insights.

As another point, many of the proposals discussed above have yet to be realized experimentally
and it would be very interesting and exciting to see how far one can push the analogy and how
accurate it can be made.
Furthermore, some of the experimental realizations ``only'' apply on the classical level
-- i.e., they are not really {\em quantum} simulators yet --
and it would be great to push them into or towards the quantum regime.
Examples include super-radiance in Sec.~\ref{Super-radiance and Penrose process} and
Hubble friction as discussed in Sec.~\ref{Expanding condensates}.
For the latter example, it would be nice to observe the amplification of quantum fluctuations
in analogy to cosmic inflation via the transition from oscillation to freezing and the
resulting squeezing, see also \cite{Schutzhold:2005fh}.

On the theory side, the impact of a modified dispersion relation such as in
Eq.~\eqref{BEC-disperion} on phenomena like Hawking radiation and cosmological particle
creation (including the case of Hubble friction mentioned above) have been studied in
various works.
In contrast, the role of dissipation is far less understood, see, e.g., \cite{Lang:2019avs}
and references therein.
On the experimental side, quantum simulators could also facilitate an investigation of the
effects of dissipation and decoherence caused by the coupling of the system to some reservoir.
For example, in the Sauter-Schwinger effect or in false-vacuum decay, quantum tunneling plays
an important role.
The quantum Zeno effect \cite{Misra:1976by} shows that such tunneling processes can be influenced by
the coupling to the environment (which effectively monitors the state of the system permanently).
Usually, the quantum Zeno effect is discussed in the context of tunneling between discrete levels.
In order to apply it to the Sauter-Schwinger effect for a constant electric field, for example,
one should generalize it to a continuum of states, see also \cite{Ahmadiniaz:2022wgb}.

In addition to the interaction with an environment, non-linear fields as discussed in
chapter~\ref{Non-linear fields} can also interact with themselves and the impact of this
self-interaction on phenomena like Hawking radiation and cosmological particle creation
is far from being completely understood.
For linear fields, the main mechanism for Hawking radiation, focusing on the vicinity of
the horizon, is very similar (in suitable coordinates) to cosmological particle creation
and can also be mapped to harmonic oscillators with time-dependent potentials.
It is not clear how far this similarity extends for the case of non-linear fields --
here quantum simulators could help us understanding this point.
One could speculate that these investigations could also shed light on other open issues,
such as the the black-hole information puzzle, the origin of the black-hole entropy and its
relation to the entanglement entropy associated to the emitted Hawking radiation --
but this must be illuminated by future studies.

Of course, apart from black holes, there are many more scenarios where it would be nice to
deepen our understanding of interaction effects via suitable quantum simulators.
An obvious example is QCD as the theory of the strong interaction featuring phenomena
such as confinement (see, e.g., \cite{Zohar:2012ay}) and string breaking
(see, e.g., \cite{Surace:2019dtp}) etc.
As already explained in Sec.~\ref{Gauge fields}, external (artificial or synthetic) gauge fields
as well as dynamical gauge fields (as in full-fledged QCD) have been discussed in several works,
but actually realizing such a quantum simulator for a large enough lattice is a formidable
challenge.
As another example, neutron stars can also give rise to complex dynamics including vortices
(similar to those in Sec.~\ref{Vortices in Bose-Einstein condensates})
and it has been proposed \cite{Poli:2023vyp}
to simulate some of their properties via the analogy to super-solids
in dipolar atom systems (see also \cite{Schutzhold:supersolid}).

Of course, for each phenomenon, one should look for the best way of realizing a suitable
quantum simulator  -- or, if possible, different quantum simulators in order to benchmark and
compare their results.
Hence, one should not restrict this search {\em a priori} to ultra-cold atoms
(in spite of their advantages explained above) but also consider trapped ions
\cite{Blatt:2012chk}, for example.
However, this is then another story and beyond the scope of the present article.














\newpage
\section*{Acknowledgements}

This review is partly based on invited lectures at the 
first international school on ``Quantum Sensors for Fundamental Physics''
(6th to 10th January 2020 in Durham)
and the summer school ``Concepts and Novel Applications of Quantum Information''
(29th August to 2nd September 2022 in Vienna)
and on other occasions.
Some of the questions addressed here came up in discussions with the audiences -- 
so thank you for the feedback. 
Fruitful discussions at many workshops, conferences, schools and research stays
are also gratefully acknowledged, such as in Bad Honnef, Benasque and at Perimeter Institute
(e.g., ``Quantum Simulators of Fundamental Physics'' June 5-9, 2023).
Research at Perimeter Institute is supported in part by the Government of Canada through 
the Department of Innovation, Science and Economic Development Canada and by the Province 
of Ontario through the Ministry of Colleges and Universities.
Funded by the Deutsche Forschungsgemeinschaft (DFG, German Research Foundation)
through the Collaborative Research Center SFB~1242
``Nonequilibrium dynamics of condensed matter in the time domain'' (Project-ID 278162697).
Financial support from and collaborations with the initiative
Quantum Simulators for Fundamental Physics (QSimFP) are also gratefully acknowledged.

I thank A.~Weitzel for test reading and help with the figures 
\rs{as well as 
U.~Fischer,
T.~Jacobson,
F.~Koenig, 
S.~Liberati,  
J.~Louko, 
R.~Sauerbrey 
for helpful comments on the manuscript}.
Then I would also like to take the opportunity and thank my fellow scientists
for many inspiring discussions, where I learned at lot.
Even though we often do not agree on everything, an open and constructive scientific
debate across all borders (national and other) is extremely valuable.
%
%
%
%
%
First and foremost, I would like to thank G.~Plunien and W.~G.~Unruh as well as
my former and present colleagues from TU Dresden, HZDR, the University of Duisburg-Essen,
the University of British Columbia and from the SFB-TR12, SFB1242 and COSLAB.
Furthermore, I would like to
thank\footnote{Most probably, I forgot several important names, for which I apologize --
this is no bad intention, but just my poor memory for names.}
A. Achucarro,
I. Affleck,
Y. Aharonov,
E. Akkermans,
A. Andreev,
J. Anglin,
M. Arndt,
M. Aspelmeyer,
R. Balbinot, 
C. Barcelo
% ...
W. Zurek.

Finally, I would like to dedicate this work to the memory of R.~Parentani and G.~Soff.




\end{document}